\documentclass[article]{JHEP3}

\usepackage{amsmath,amssymb}
\usepackage[dvips]{graphics}
\usepackage{epsfig}

\newcommand{\be}{\begin{equation}}
\newcommand{\ee}{\end{equation}}
\newcommand{\bea}{\begin{eqnarray}}
\newcommand{\eea}{\end{eqnarray}}
\newcommand{\bean}{\begin{eqnarray*}}
\newcommand{\eean}{\end{eqnarray*}}

\def\beq{\begin{equation}}

\def\eeq{\end{equation}}

\relax

\def\be{\begin{equation}}
\def\bearl{\begin{array}{l}}
\def\bearll{\begin{array}{ll}}
\def\ee{\end{equation}}
\def\eear{\end{array}}
\def\a{\alpha'} 
\def\F{\mathcal{F}}

\def\X{\mathcal{X}}

\preprint{{\small LPTHE 04--20}\\ {\small \texttt{hep-th/0409017}}}

\title{Towards Quantum Dielectric Branes: Curvature Corrections in Abelian Beta Function and Nonabelian Born--Infeld Action}

\author{Pedro Bordalo$^{\dag}$, Lorenzo Cornalba$^{\ddag}$ and Ricardo
Schiappa$^{\dag\dag}$
\\
$^{\dag}$LPTHE, Universit\'e Paris VI,\\ 4 Place Jussieu, F--75252 Paris
Cedex 05, France\\
\\
$^{\ddag}$Instituut voor Theoretische Fysica, Universiteit van Amsterdam,\\
Valckenierstraat 65, 1018 XE Amsterdam, The Netherlands\\
\\
$^{\dag\dag}$CAMGSD, Departamento de Matem\'atica, Instituto Superior T\'ecnico,\\ 
Av. Rovisco Pais 1, 1049--001 Lisboa, Portugal\\
\\
$^{\dag\dag}$Faculdade de Engenharia, Universidade Cat\'olica 
Portuguesa,\\
Estrada de Tala\'\i de, 2635--631 Rio de Mouro, Lisboa, Portugal\\
\\
\email{bordalo@lpthe.jussieu.fr}, \quad
\email{lcornalb@science.uva.nl}, \quad 
\email{schiappa@math.ist.utl.pt}
}

\abstract{
We initiate a programme to compute curvature corrections to the nonabelian Born--Infeld action. This is based on the calculation of derivative corrections to the \textit{abelian} Born--Infeld action, describing a maximal brane, to \textit{all} orders in $\F = B + 2\pi\alpha' F$. An exact calculation in $\F$ allows us to apply the Seiberg--Witten map, reducing the maximal abelian brane point of view to a minimal nonabelian brane point of view (replacing $1/F$ with $[\X,\X]$ at large $F$), resulting in matrix model equations of motion in the considered background. We first study derivative corrections to the abelian Born--Infeld action and compute the two loop beta function for an open bosonic string in a WZW (parallelizable) background. This beta function is the first step in the process of computing open string equations of motion, which can be later obtained by either computing the Weyl anomaly coefficients or the partition function in the given background. The beta function for the gauge field is exact in $\F$ and computed to orders ${\mathcal{O}} (H, H^{2}, H^{3})$ (where $H=dB$ and the curvature is $R \sim H^{2}$) and ${\mathcal{O}} (\nabla\F, \nabla^{2}\F, \nabla^{3}\F)$. In order to carry out this calculation we develop a new regularization method for two loop graphs. We then relate perturbative results for abelian and nonabelian Born--Infeld actions, by showing how abelian derivative corrections yield nonabelian higher order commutators and vice--versa, at large $F$. We begin the construction of a matrix model describing $\a$ corrections to Myers' dielectric effect. This construction is carried out by first setting up a perturbative classification of the relevant nonabelian tensor structures, which can be considerably narrowed down by the physical constraint of translation invariance in the action and the possibility for generic field redefinitions. The final matrix action is not uniquely determined and depends upon two free parameters. These parameters could be computed via further calculations in the abelian theory.
}

\input{tcilatex}

\begin{document}



\vfill

\eject


\section{Introduction and Summary}


The study of $D$--branes in curved backgrounds has been the subject of a great deal of activity in recent years. $D$--branes seem to be quite interesting objects once we move away from the realm of flat space, as they generically display the ability to change their shape and their dimension, among many other phenomena. These properties arise as one embeds the branes in non--trivial closed string backgrounds, with perhaps the most renowned aspect being the dielectric effect \cite{Myers}. While there has been a great deal of research dedicated to this aspect of $D$--brane physics, it has all been performed for branes within weakly curved backgrounds (see, however, \cite{yolanda-1, yolanda-2}). It is therefore clear that a most pressing question concerns what happens as the closed string curvature is increased. An answer to this question necessarily requires a detailed analysis of curvature corrections to the open string equations of motion.

The dielectric effect is also undoubtedly associated to the fact that we are dealing with many $D$--branes, rather than a single one. The study of quantum corrections to the dielectric effect (corrections in $\a$) then seems as a great challenge: one would need to generalize the \textit{nonabelian} Born--Infeld (BI) action to curved backgrounds. Still, if one wishes to proceed along these lines, there are a couple of routes one can follow. The original approach in \cite{Myers} constrains the nonabelian BI action via an analysis of open string $T$--duality. In order to extend this work to higher order in the background curvature one would first require a set of $\a$ corrected open string $T$--duality rules, which are not yet available. Thus, an attempt along these lines would first entail a study of open string $T$--duality to higher loop order in the worldsheet. Another approach to this problem was the one followed in \cite{Kabat-Taylor, Taylor-Raamsdonk-1, Taylor-Raamsdonk-2, Taylor-Raamsdonk-3} which started with an one loop calculation in Matrix theory \cite{BFSS}. Here one can of course imagine performing higher loop calculations and, following the same approach, expect to obtain a quantum corrected nonabelian BI action. This is however a long road to follow (other approaches within the setting of Matrix theory were also attempted, \textit{e.g.} \cite{Douglas, DKO, Boer-Schalm, Raamsdonk, Hassan, BSW, BFLR}). Finally, one can of course just do the standard string theory calculation, by computing the nonabelian string effective action. This is also well known to be a hard calculation (see, \textit{e.g.}, \cite{Hashimoto-Klebanov, Garousi-Myers-1, Garousi-1, Garousi-Myers-2, Garousi-2}) .

The approach we shall use is based on recent work concerning noncommutative geometry in string theory \cite{CDS, Cheung-Krogh, Schomerus, Cornalba-Schiappa-1, Seiberg-Witten, Cornalba-2}. In this context one has the so--called Seiberg--Witten (SW) map relating commutative and noncommutative descriptions of the BI action \cite{Seiberg-Witten}, and the main idea is simple: a commutative description corresponds to a maximal brane analysis (thus, an abelian BI) while a noncommutative description corresponds to a minimal brane analysis (thus, the nonabelian BI). Knowledge of the abelian equations of motion will yield the nonabelian ones, as long as the SW map is known. That such a procedure can be used to compute curvature corrections to actions based on matrix models was first suggested in \cite{Cornalba-Schiappa-1}, and was later studied in detail in \cite{Cornalba-3, Cornalba-4, Cornalba-5} within the context of the nonabelian BI action in flat space. In here, we wish to start a programme which extends that analysis to curved backgrounds.

The starting point is the calculation of curvature corrections to the \textit{abelian} BI equations of motion, to all orders in the gauge field strength. We will work in the bosonic open string sigma model, where the beta function of the gauge field living on the $D$--brane is computed perturbatively in $\a \times$derivatives of background fields and is further expected to have corrections at every loop order\footnote{This is not the case for a supersymmetry preserving brane in a supersymmetric background.}. Observe that one key aspect of this calculation is that it needs to be \textit{exact} in $F$, the gauge field strength on the $D$--brane. Indeed, if one has any hope of performing the SW limit, open string parameters $G$ and $\Theta$ will need to be properly identified before such a limit can be taken \cite{Seiberg-Witten}. This contrasts with previous calculations performed in the literature, which were mostly perturbative in $F$. Beta functions for bosonic open string sigma models were computed up to two loops (in flat space) in \cite{DO, ACNY, CLNY, Andreev-Tseytlin}. Later, and in the superstring context, curvature corrections to the BI and Wess--Zumino (WZ) actions were computed in \cite{BBG, Scrucca-Serone, Fotopoulos, CLR}, but no $F$ or $B$ field on the brane were considered. These calculations allowed a better understanding of the physics of abelian branes in curved spaces, but are of no help on what concerns the nonabelian case. Inclusion of the $F$ (or $B$) field came later. In flat space, analysis of the BI and WZ to all orders in $F$ was carried out in \cite{Wyllard-1, Wyllard-2}, and in curved space a perturbative analysis in $F$, to first order, was done in \cite{Fotopoulos-Tseytlin, Barabanschikov}. Computations in curved backgrounds, which are exact in $F$, were performed in the context of the bosonic open string in \cite{AAGG} (but where the result turned out \textit{not} to be target space covariant), and in the context of type II string theory in \cite{Wijnholt}, where corrections to the string effective action were obtained. A different line of research concerning the computation of $\alpha'$ corrections to the nonabelian BI action, in the supersymmetric case, has been carried out in \cite{Koerber-Sevrin-1, REKS, Koerber-Sevrin-2, Sevrin-Wijns}. Let us make one more remark concerning the fact that our calculation deals with bosonic string theory. At the BI level, this will differ from the superstring calculation via the appearance of terms in $F^{n}$ with $n$ odd, where there is a proposal to sum up these odd contributions \cite{Argyres-Nappi}.

In this paper we shall compute the beta function for the gauge field on a maximal $D$--brane, at two loop order, using a short distance cutoff as a regulator. The curved closed string background is always kept on--shell. Here one could in principle choose any bosonic closed string background (without a dilaton, as while we are computing $\a$ corrections, we remain at small string coupling $g_{s}$). The sigma model analysis of closed bosonic strings in \cite{BCZ, Mukhi-1, Mukhi-2} however points out that \textit{generic} backgrounds will be extremely complicated at two loops, due to the immense possibilities in both curvature and Neveu--Schwarz--Neveu--Schwarz (NSNS) $H$--field couplings (with $H=dB$). It would thus be of some interest to restrict to some class of backgrounds which would be more tractable on a first approach. A particular class of backgrounds was pointed out precisely in \cite{BCZ}, the class of parallelizable backgrounds (this is a large class of backgrounds which includes all Wess--Zumino--Witten (WZW) group manifold models). In this regard, it is also interesting to note that the nonabelian example in \cite{Myers} of $D0$--branes expanding into a fuzzy sphere, due to the presence of Ramond--Ramond (RR) flux, is precisely matched by the example in \cite{ARS-1, ARS-2} of $D0$--branes expanding into a fuzzy sphere, this time around due to the presence of NSNS flux. Hence, one can consider the $SU(2)$ WZW model of \cite{ARS-1, ARS-2} as the prototype model of the physics of Myers' dielectric effect.

Regarding this last point, it is worth noting that the question of how will curvature corrections contribute to the dielectric effect on nonabelian branes has received some attention, with special emphasis towards possible corrections to the standard fuzzy sphere result. Indeed, the initial result in \cite{ARS-1, ARS-2} for the $SU(2)$ WZW model at level $k$ showed that at infinite level the $D0$--branes expand into a fuzzy sphere. It moreover pointed out that, at finite level, the fuzzy spheres are expected to be deformed in some manner closely related to quantum groups at a root of unity, $q = e^{\frac{i \pi}{k+2}}$, at least for integer level $k$. Another point of interest arose in later work \cite{Alekseev-Schomerus, Fredenhagen-Schomerus} where it was shown that $D$--brane charge is only well defined modulo $k+2$. This can be understood physically by the statement that $\mathbb{S}^{2}$ branes inside $\mathbb{S}^{3} \simeq SU(2)$, made out of $D0$--branes, can only grow up to a certain radius. As finite level corrections to the infinite level result amount to $\a$ corrections to the dielectric effect, this series of works seem to imply that the nonabelian branes get corrected to some sort of quantum group deformation. In \cite{BDS}, this WZW model was further investigated from the maximal brane point of view, where it was found that the semiclassical analysis yields precise agreement with the conformal field theory (CFT) analysis\footnote{These results were extended to all compact groups in \cite{pedro-1}.}, and further clues pointing towards a quantum group relation were uncovered. Still, it was not until \cite{Steinacker-1, Steinacker-2, PS-1, PS-2, PS-3} that an attempt was done in order to try to put the quantum group relation on firmer grounds. The authors of these papers started from the conjecture that the matrix model describing the full quantum corrected dielectric branes of the $SU(2)$ level $k$ WZW model was a quantum group deformed matrix model at the root of unity $q = e^{\frac{i \pi}{k+2}}$, and tried to match to both the CFT and the semiclassical results previously mentioned. A derivation of such a quantum matrix model directly from string theory remains an open problem.

Having computed the two loop beta function for an open bosonic string in a WZW (parallelizable target manifold) closed string background, to all orders in $\F = B +2\pi\alpha' F$ (which is the correct gauge invariant combination on the worldsheet boundary), one would like to relate it to the open string equations of motion (or the open string effective action). In fact, it is not the vanishing of the beta functions, $\beta=0$, which should be equivalent to the string equations of motion, but rather the vanishing of the Weyl anomaly coefficients, $\bar{\beta}=0$ (\textit{i.e.}, the ones which ensure the absence of the Weyl anomaly---the standard beta functions only ensure the absence of the scale anomaly) \cite{Tseytlin-2, Tseytlin-3, Behrndt-Dorn}. This can be accomplished in two distinct ways. On one hand, one can compute the missing terms in the relation between the beta function and the Weyl anomaly coefficient, $\bar{\beta} = \beta + \cdots$, and thus find the equations of motion. On the other hand, one may wish to compute the string effective action directly. This can be done by using the techniques of boundary string field theory \cite{BSFT-1, BSFT-2} which tell us how to relate the string effective action to the partition function, precisely via the beta function,

$$
S [A] = \left( 1 + \beta [A] \cdot \frac{\delta}{\delta A} \right) Z [A].
$$

\noindent
Because any of the aforementioned calculations is at the same level of difficulty as the two loop beta function calculation, we shall not perform them in this paper and will leave them for future work. Nevertheless, we shall carefully describe how to extract perturbative information concerning the nonabelian BI action in curved backgrounds, given higher derivative corrections to the abelian BI action. This is the first step in a programme which may eventually lead to new insight into the dielectric effect.

As we have explained, our method uses the SW map to connect commutative and noncommutative descriptions of the BI action. Derivative corrections in the abelian action can be understood, at large field strength $F$ on the abelian $D$--brane worldvolume, as corrections to the standard multiplication of functions in such a way that this multiplication actually gets deformed to a noncommutative star product. This is achieved via a particular change of variables (from commutative to noncommutative fields) which precisely corresponds to the SW map \cite{Seiberg-Witten}. The noncommutative description is then simpler to understand as a matrix model, as one represents functions with operators, star products with operator multiplication and integrals get replaced by traces \cite{Cornalba-4, Cornalba-Schiappa-2, Kontsevich}. This calculation, while naturally perturbative from the abelian BI point of view, also induces a straightforward perturbation theory for the calculation of curvature corrections to the nonabelian BI action. As it turns out, and we shall show this later on, a simple understanding of the ``nonabelian'' perturbation theory leads to a classification of all possible curvature corrections which, constrained by translation invariance and field redefinitions, \textit{are extremely few in number}. In this way, one can infer many things about the nonabelian BI action in curved backgrounds without having to carry out the (abelian) calculation to its very end. Naturally, absolute conclusions and results will only be available once all coefficients are fixed. Here, we shall content ourselves with two free parameters. Future work should then concentrate on taking the programme we set out to its full extent.

This paper is organized as follows. In section 2 we begin with a review of the perturbation theory expansion which is required in order to perform the two loop calculation we purpose to do. We explain what propagator should be used in order for the calculation to be exact in $\F$, and list all the one and two loop diagrams that arise from the possible vertices. Many details concerning the perturbative expansion of the worldsheet action, as well as technical details concerning the propagator are left to the appendices. In section 3 we shall perform the main calculation in this paper as we obtain the two loop beta function for the maximal $D$--brane gauge field, $A$, presented in section 3.9 (many technical details of the calculation are presented in the appendices). In order to compute the two loop beta function we first have to devise a new regularization scheme, which is actually valid to any loop order, that will allow us to re--sum the infinite series of powers in $\F$ we are obtaining (due to exactness of the result with respect to this field) in terms of open string tensors, $G$ and $\Theta$. After having obtained the maximal brane result, we set up its translation into a minimal brane language. This is done in section 4. This is a descriptive section where we explain how derivative corrections to the abelian BI can be used to obtain curvature couplings in the nonabelian BI action, setting up the programme of computing $\alpha'$ corrections to the dielectric effect in string theory. Mostly motivated by the abelian results we then describe a perturbative scheme within which one classifies tensor structures which may contribute (at first non--trivial order) to curvature couplings in the nonabelian BI action (we also include some appendices dealing with some known couplings of the nonabelian BI action). We list possible nonabelian tensor structures appearing at two loops in perturbation theory, and show how they correspond, at large $F$, to abelian tensor structures appearing in the abelian BI action. These structures can be further constrained by the requirement of translation invariance of the action, and by generic field redefinitions. It turns out that in the end one is left with surprisingly few allowed tensor structures. Concentrating on the $SU(2)$ model at level $k$, a prototype for Myers' dielectric effect, there are in fact only two curvature corrections to consider. This allows us to write a matrix model for quantum dielectric branes (still dependent on the two free parameters), and which we present in section 4.4. We end with some directions for future work.


\section{Setting Up the Calculation}\label{method}


We begin this section by presenting the open string sigma model, which we
shall expand using the background field method in order to yield local
covariant interaction vertices. The background field method is described at
great length in the appendix, to which we refer the reader for details and
notation. It is from this expanded sigma model action that we shall perform
the perturbation theory analysis which will allow us to compute the required
two loop beta function. We shall also discuss in this section the one and
two loop diagrams relevant for the computation. The open string sigma model
action contains a bulk contribution,

\begin{equation*}
S=\frac{1}{4\pi \alpha ^{\prime }}\int_{\Sigma }g_{\mu \nu }\left( X\right)
dX^{\mu }\wedge \ast dX^{\nu }+\frac{i}{4\pi \alpha ^{\prime }}\int_{\Sigma
}B_{\mu \nu }\left( X\right) dX^{\mu }\wedge dX^{\nu },
\end{equation*}

\noindent
as well as a boundary contribution 

\begin{equation*}
S_{B}=i\oint_{\partial \Sigma }A_{\mu }\left( X\right) dX^{\mu },
\end{equation*}

\noindent 
where $\ast$ is the worldsheet Hodge dual. $B$ is the gauge potential for the NSNS 3--form $H$ field, $H=dB$. We shall restrict our study to backgrounds which are parallelizable, in the sense that 

\begin{equation*}
R_{\mu \nu \rho \sigma }+\frac{1}{4}H_{\mu \nu \lambda }H{^{\lambda }}_{\rho
\sigma }=0, \qquad \qquad \qquad \nabla _{\mu }H_{\nu \rho
\sigma }=0.
\end{equation*}

\noindent
These manifolds represent exact closed string backgrounds, to all orders in $\alpha ^{\prime }$ \cite{BCZ} (examples of such manifolds are the group manifolds and some 
of their orbifolds, see \cite{pedro-2}). In the following, we list the
various ingredients required for the perturbation theory.


\subsection{The Propagator}\label{propagator}


We shall follow the standard background field method by choosing a classical solution to the string equations of motion $x^{\mu } \left( \sigma ^{a}\right)$, which satisfies\footnote{Let $\sigma^{a}$ be coordinates on the string worldsheet,  $\Sigma$, with disk topology and endowed with a flat fixed metric.}

\begin{eqnarray*}
\mathcal{D}^{a}\partial _{a}x^{\mu } &=& 0, \\
\left. \Big( g_{\mu \nu } \ast dx^{\nu } + i \mathcal{F}_{\mu \nu }\ dx^{\nu } \Big) \right|_{\partial\Sigma } &=& 0,
\end{eqnarray*}

\noindent
where 

\begin{equation*}
\mathcal{F}_{\mu \nu }=B_{\mu \nu }+2\pi \alpha ^{\prime }(dA)_{\mu \nu }
\end{equation*}

\noindent
and where we have defined the following covariant derivatives on sections $\zeta ^{\mu }$ of the target space tangent bundle 

\begin{eqnarray*}
D\zeta ^{\mu } &=& d\zeta ^{\mu }+dx^{\sigma }\ \Gamma^{\mu}_{\sigma\nu} \zeta ^{\nu }, \\
\mathcal{D}\zeta ^{\mu } &=& D\zeta ^{\mu }-\frac{i}{2}\ast dx^{\sigma}\ H^{\mu }{}_{\sigma \nu }\zeta ^{\nu }.
\end{eqnarray*}

\noindent
We will take the fluctuating quantum field to be a section, $\zeta^{\mu}$, of the target space tangent bundle, defining the geodesic which starts from the
classical solution and must be followed for an unit of affine parameter in
order to reach the displaced configuration of the string.

In terms of the quantum field $\zeta ^{\mu }$, the quadratic part of the action then becomes (see the appendix) 

\begin{equation*}
\frac{1}{4\pi \alpha ^{\prime }}\int_{\Sigma }g_{\mu \nu }\,\mathcal{D}\zeta
^{\mu }\wedge \ast \mathcal{D}\zeta ^{\nu }+\frac{i}{4\pi \alpha ^{\prime }} \oint_{\partial \Sigma } \Big( \nabla _{\sigma }\mathcal{F}_{\mu \nu }\zeta
^{\sigma }\zeta ^{\mu }dx^{\nu }+\mathcal{F}_{\mu \nu }\zeta ^{\mu }D\zeta
^{\nu }\Big) .
\end{equation*}

\noindent
In order to put the bulk part of the quadratic action in the standard form (from where one can easily identify the worldsheet propagator) we introduce, on the worldsheet, the spacetime vielbein $E_{\mu }^{A}$, such that (here $g_{AB}$ is constant)

\begin{equation*}
g_{\mu \nu }\left( x\right) =E_{\mu }^{A}\left( x\right) E_{\nu }^{B}\left(
x\right) g_{AB},
\end{equation*}

\noindent
and we shall reparametrize the quantum fluctuations with the tangent space
vector $\zeta ^{A}$ given by

\begin{equation*}
\zeta ^{A}=E_{\mu }^{A}\ \zeta ^{\mu }.
\end{equation*}

\noindent
It turns out that, whenever the background is parallelizable, we can always choose $E_{\mu }^{A}$ such that \cite{BCZ}

\begin{equation}
E_{\mu }^{A}\mathcal{D}\zeta ^{\mu }=d\zeta ^{A} .  \label{simpl1}
\end{equation}

\noindent
Thus, more generally, given any section $S_{\nu_{1} \cdots \nu_{n}}^{\mu_{1} \cdots \mu_{m}}$ we have that 

\begin{equation}
dS_{N_{1}\cdots N_{n}}^{M_{1}\cdots M_{m}}=E_{\mu _{1}}^{M_{1}}\cdots E_{\mu
_{m}}^{M_{m}}\,\ E_{N_{1}}^{\nu _{1}}\cdots E_{N_{n}}^{\nu _{n}}\,\,\mathcal{D} S_{\nu _{1}\cdots \nu _{n}}^{\mu _{1}\cdots \mu _{m}} .  \label{curlD}
\end{equation}

\noindent
In particular, on the boundary of the worldsheet,

\begin{equation}
D\zeta^{\mu}\ \Big|_{\partial \Sigma} = \left. E_{A}^{\mu} \left( d\zeta^{A} - \frac{1}{2} H^{ABC}{} \zeta_{B} \mathcal{F}_{CD}\,dx^{D} \right) \right|_{\partial \Sigma},  \label{simpl2}
\end{equation}

\noindent
where, clearly,

\begin{equation*}
dx^{A}\equiv E_{\mu }^{A}\,dx^{\mu }.
\end{equation*}

\noindent
We are now in a position to write the quadratic part of the action in a
final form
 
\begin{eqnarray}
&&
\frac{g_{AB}}{4\pi \alpha^{\prime}} \int_{\Sigma} d\zeta^{A} \wedge \ast
d\zeta^{B} + \frac{i}{4\pi \alpha^{\prime}} \oint_{\partial \Sigma} \mathcal{F}_{AB}\, \zeta^{A} d\zeta^{B} + \nonumber \\
&&
+ \frac{i}{4\pi \alpha^{\prime}} \oint_{\partial \Sigma} \left( \nabla_{A} \mathcal{F}_{BC} - \frac{1}{2} \mathcal{F}_{AD} H_{B}{}{}^{DE} \mathcal{F}{}{}_{CE} \right) \zeta^{A} \zeta^{B} dx^{C}. \label{kin}
\end{eqnarray}

One can now discuss the propagator for the quantum field $\zeta^{A}$. Since the tensors $\mathcal{F}_{AB}$ and $H_{ABC}$ depend on the position on the worldsheet, we shall consider only the bulk part of the action as defining the free theory, taking the 2--vertices on the boundary as interaction vertices. In the appendix we derive the propagator for a slightly more general kinetic term, corresponding to the first line of (\ref{kin}) with $\mathcal{F}_{AB}$ \textit{constant} (and when $\Sigma $ is the unit disc in the complex plane). We shall later see in the discussion that this case is actually the relevant one for the computation of the beta function. Let us quote the result for the propagator, which is written in terms of the open string tensors\footnote{We shall interchangeably use both $G^{AB}$ and $\left( \frac{1}{G} \right)^{AB}$ for the inverse open string metric.} $G$ and $\Theta$,

\begin{equation*}
\left( \frac{1}{g+\mathcal{F}} \right)^{AB} = \left( \frac{1}{G} \right)^{AB} + \Theta^{AB},
\end{equation*}

\noindent
where $G^{AB}$ and $\Theta ^{AB}$ are, respectively, symmetric and antisymmetric. Introducing the functions 

\begin{eqnarray*}
\mathcal{A} \left( z,w \right) &=& \ln \left( \frac{1-z\overline{w}}{1-\overline{z}w} \right) , \qquad \qquad \mathcal{B} \left( z,w \right) = -2 \ln \left| 1-z\overline{w} \right|, \\
\mathcal{C} \left( z,w \right) &=& - \ln \left| \frac{z-w}{1-z\overline{w}} \right| , \qquad \qquad \mathcal{D} \left( z,w \right) = \frac{1}{2} \left( z\overline{z} + w\overline{w} \right) ,
\end{eqnarray*}

\noindent
defined for $z,w\in \Sigma \subset \mathbb{C}$, one has the propagator on the disk as

\begin{equation*}
\frac{1}{\alpha^{\prime}} \left\langle \zeta^{A} \left( z \right) \zeta^{B} \left( w \right) \right\rangle = \Theta^{AB}\, \mathcal{A} (z,w) + G^{AB}\, \mathcal{B} (z,w) + g^{AB}\, \Big( \mathcal{C} (z,w) + \mathcal{D} (z,w) \Big) .
\end{equation*}

\noindent
If considering only the bulk part of (\ref{kin}) as the free action, one will simply have the above formula but with $G^{AB}=g^{AB}$ and $\Theta ^{AB}=0$.


\subsection{The Vertices}


Next we need to enumerate the relevant vertices, for the perturbative calculation we wish to perform. In the last subsection we have already discussed the 2--vertices. One can now use the results in the appendix for the action written in terms of $\zeta ^{\mu }$, together with equations (\ref{simpl1}) and (\ref{simpl2}), to quickly arrive at the results below. Recall that the vertices below have an explicit minus sign with respect to the expansion of the action, since the weight factor for the path integral is $e^{-S}$.

\paragraph{Bulk Vertices}

We begin by listing the relevant vertices arising from the bulk terms in the
perturbative expansion of the sigma model action. The resulting vertices
relevant for the beta function calculation are given by 

\begin{eqnarray*}
&&-\frac{1}{2\pi \alpha ^{\prime }}\ \frac{1}{2}\int_{\Sigma }T_{IJK}\ \zeta
^{I}d\zeta ^{J}\wedge d\zeta ^{K}, \\
&&-\frac{1}{2\pi \alpha ^{\prime }}\ \frac{1}{4}\ \int_{\Sigma }T_{IJKL}\
\zeta ^{I}\zeta ^{J}d\zeta ^{K}\wedge \ast d\zeta ^{L},
\end{eqnarray*}

\noindent
where the tensors $T_{IJK}$ and $T_{IJKL}$ are given by 

\begin{eqnarray*}
T_{IJK} &=& \frac{i}{3}\ H_{IJK}, \\
T_{IJKL} &=& \frac{1}{12} \left( H_{IKM}\, H^{M}{}_{JL} + H_{JKM}\, H^{M}{}_{IL} \right) = -\frac{1}{3} \left( R_{IKJL} + R_{JKIL} \right) .
\end{eqnarray*}

\noindent
Note that $T_{IJK}$ is totally antisymmetric, whereas $T_{IJKL}$ is
explicitly symmetric in indices $I$, $J$ and $K$, $L$.

\paragraph{Boundary Vertices}

The boundary vertices are obtained in a similar way, starting from the
expansion of the action in the appendix, and fall into two distinct classes.
First of all we have the vertices 

\begin{eqnarray*}
&&-\frac{1}{2\pi \alpha ^{\prime }}\oint_{\partial \Sigma }M_{IJ}\ \zeta
^{I}d\zeta ^{J} , \\
&&-\frac{1}{2\pi \alpha ^{\prime }}\ \frac{1}{2!}\oint_{\partial \Sigma
}M_{IJK}\ \zeta ^{I}\zeta ^{J}d\zeta ^{K} , \\
&&-\frac{1}{2\pi \alpha ^{\prime }}\ \frac{1}{3!}\oint_{\partial \Sigma }\
M_{IJKL}\ \zeta ^{I}\zeta ^{J}\zeta ^{K}d\zeta ^{L},
\end{eqnarray*}

\noindent
where the tensors $M_{A_{1} \cdots A_{n}}$ are explicitly symmetric in the
first $n-1$ indices $A_{1}, \dots, A_{n-1}$, and are given by 

\begin{eqnarray*}
M_{IJ} &=&\frac{i}{2}\ \mathcal{F}_{IJ}\,, \\
M_{IJK} &=&\frac{i}{3}\, \Big( \nabla _{I}\mathcal{F}_{JK}+\nabla _{J}
\mathcal{F}_{IK} \Big) , \\
M_{IJKL} &=&\frac{i}{8} \left( \nabla _{I}\nabla _{J}\mathcal{F}_{KL}-\frac{1}{12}\mathcal{F}_{IM}H^{M}{}_{JN}H^{N}{}_{KL}+\mathrm{Sym}_{IJK}\right) .
\end{eqnarray*}

\noindent
The second class of boundary vertices is given explicitly by 

\begin{eqnarray*}
&&-\frac{1}{2\pi \alpha ^{\prime }}\ \frac{1}{2!}\oint_{\partial \Sigma
}N_{IJ}\ \zeta ^{I}\zeta ^{J} , \\
&&-\frac{1}{2\pi \alpha ^{\prime }}\ \frac{1}{3!}\oint_{\partial \Sigma
}N_{IJK}\ \zeta ^{I}\zeta ^{J}\zeta ^{K} , \\
&&-\frac{1}{2\pi \alpha ^{\prime }}\ \frac{1}{4!}\oint_{\partial \Sigma }\
N_{IJKL}\ \zeta ^{I}\zeta ^{J}\zeta ^{K}\zeta ^{L}.
\end{eqnarray*}

\noindent
The tensors $N_{A_{1}\cdots A_{n}}$ are explicitly symmetric in all the indices $A_{1}, \dots, A_{n}$, and are worldsheet 1--forms, since they explicitly contain the 1--form $dx^{A}$. They are given by 

\begin{equation*}
N_{IJ}=\frac{i}{2!} \left( \nabla _{I}\mathcal{F}_{JK}+\nabla _{J}\mathcal{F}_{IK}-\frac{1}{2}\mathcal{F}_{IM}H_{J}{}{}^{MN}\mathcal{F}{}{}_{KN}-\frac{1}{2}\mathcal{F}_{JM}H_{I}{}{}^{MN}\mathcal{F}{}{}_{KN}\right) dx^{K},
\end{equation*}

\noindent
by 

\begin{equation*}
N_{IJK}=\frac{i}{3!} \left( \nabla _{I}\nabla _{J}\mathcal{F}_{KL}-\frac{1}{4}\mathcal{F}_{IM}H^{M}{}_{JN}H^{N}{}_{KL}-\nabla _{I}\mathcal{F}_{JM}H^{M}{}_{KN}\mathcal{F}^{N}{}_{L}+\mathrm{Sym}_{IJK}\right) dx^{L},
\end{equation*}

\noindent
and finally by 

\begin{eqnarray*}
N_{IJKL} &=&\frac{i}{4!} \left( \nabla _{I}\nabla _{J}\nabla _{K}\mathcal{F}_{LM}-\frac{3}{4}\nabla _{I}\mathcal{F}_{JA}H^{A}{}_{KB}H^{B}{}_{LM}-\frac{3}{2}\nabla _{I}\nabla _{J}\mathcal{F}_{KA}H^{A}{}_{LB}\mathcal{F}^{B}{}_{M}+\right.  \\
&&\left. +\frac{1}{8}\mathcal{F}_{IA}H^{A}{}_{JB}H^{B}{}_{KC}H^{C}{}_{LD}
\mathcal{F}^{D}{}_{M}+\mathrm{Sym}_{IJKL}\right) dx^{M}.
\end{eqnarray*}

\noindent
The above vertices, together with the free propagator described previously,
are all we shall need in order to compute the two loop beta function.


\subsection{The Diagrams}


To conclude this section we turn our attention to the diagrams contributing
to the beta function at both one and two loops, in parallelizable
backgrounds. The diagrams which contribute in this case may have either one
explicit boundary vertex of $N$--type, containing a $dx^{M}$ insertion, or
none\footnote{Of course all diagrams to consider are 1PI graphs.}. Let us comment on this further. Since we are interested in counterterms to the gauge potential, $A_{M}$, we are looking for divergences in the graphs of the form $\oint \beta _{M}dx^{M}$, which clearly contain a single $dx^{M}$ term. Diagrams
with two insertions of $dx^{M}$ would contribute to the closed string sector
beta functions, and these have already been taken into account as the closed
string physics is on--shell (new diagrams arising from the open string
sector with two insertions of $dx^{M}$ will only contribute at string loop
level as is well known \cite{CLNY}). The reason why diagrams with no insertion of $dx^{M} $ may contribute is the following: we are, in general, looking for
divergences in correlators of operators at different points on the boundary
of the string worldsheet. We can then have ultralocal divergences,
proportional to derivatives of the delta function, which will
contribute to the beta function upon integration by parts, thus 
producing the ``missing'' insertion of $dx^{M}$ \cite{Andreev-Tseytlin}. We
shall then denote the diagrams with one insertion of $dx^{M}$ as diagrams of 
$\delta $--type, and those with no $dx^{M}$ insertion as diagrams of $\delta
^{\prime }$--type (the reason for the notation $\delta $ and $\delta
^{\prime }$ type, related to the type of local divergence, will become
fully clear in the next section).

We begin by listing the one loop diagrams. There is one $\delta $--type
diagram and one $\delta ^{\prime }$--type diagram (where the thick horizontal line denotes the disk boundary):

\begin{equation}
\begin{picture}(50,50)(25,-5) \scalebox{.8}{\includegraphics{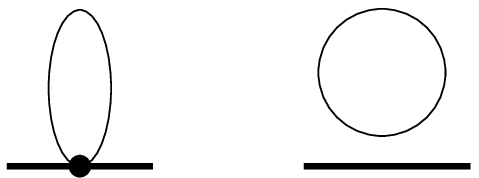}}
\put(-103,-8){ \makebox(0,0)[l]{$N_{(2)}$}} \put(-135,20){
\makebox(0,0)[l]{$\delta[1]$}} \put(-70,20){ \makebox(0,0)[l]{$\delta '
[1]$}} \end{picture}  \label{fig1loop}
\end{equation}

\noindent
Note that we are \textit{not showing} the insertions, along the propagators,
of the 2--vertex of $M$--type,

$$
-\frac{1}{2\pi \alpha ^{\prime}} \oint_{\partial \Sigma } M_{IJ}\ \zeta^{I} d\zeta^{J},
$$

\noindent
since, as will become clear in the next section, we are going to absorb its effects  to all orders in $M_{IJ}$ in the propagator.

\paragraph{Two Loop $\protect\delta$--Type Graphs}

Below are the two loop diagrams with one explicit insertion of $dx^{M}$. It
is clear from the structure of the vertices previously presented that there
are only insertions of $dx^{M}$ at the disk boundary, in this parallel
background case, due to the $N$--type vertex (in the general case, there will
also be diagrams with bulk insertions of $dx^{M}$). Of the following two
loop graphs, diagrams $3,4,6,7$ and $8$ contribute to the beta function.

\begin{equation}
\begin{picture}(220,120)(60,-15)
\scalebox{.8}{\includegraphics{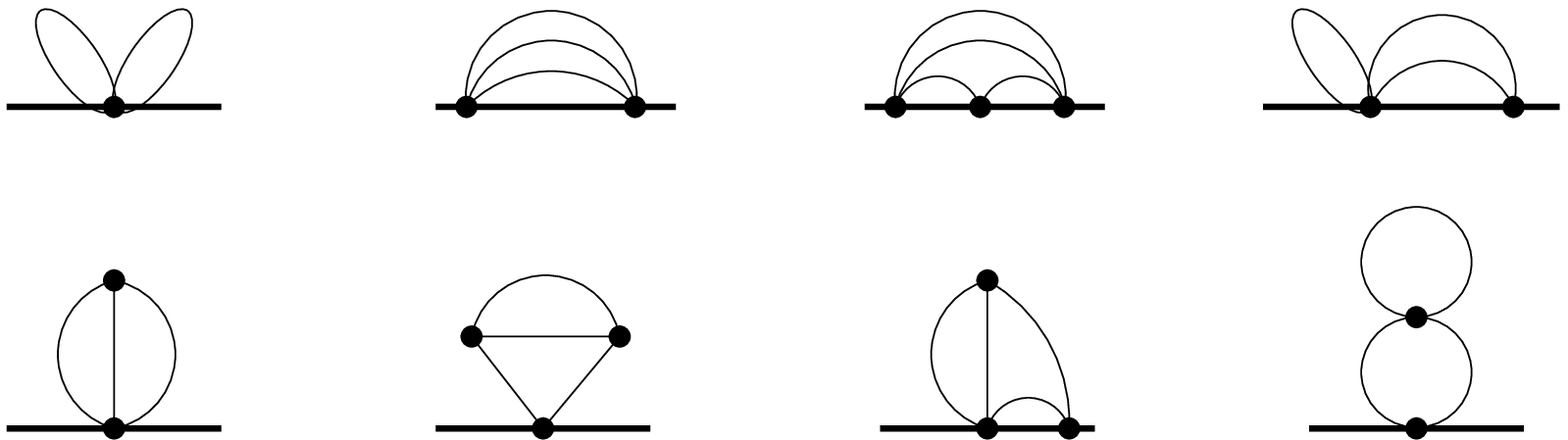}} \put(-395,90){
\makebox(0,0)[l]{$\delta[1]$}} \put(-355,70){ \makebox(0,0)[l]{$N_{(4)}$}}
\put(-290,90){ \makebox(0,0)[l]{$\delta[2]$}} \put(-275,70){
\makebox(0,0)[l]{$M_{(3)}$}} \put(-230,70){ \makebox(0,0)[l]{$N_{(3)}$}}
\put(-190,90){ \makebox(0,0)[l]{$\delta[3]$}} \put(-175,70){
\makebox(0,0)[l]{$M_{(3)}$}} \put(-125,70){ \makebox(0,0)[l]{$M_{(3)}$}}
\put(-150,70){ \makebox(0,0)[l]{$N_{(2)}$}} \put(-95,90){
\makebox(0,0)[l]{$\delta[4]$}} \put(-60,70){ \makebox(0,0)[l]{$M_{(4)}$}}
\put(-20,70){ \makebox(0,0)[l]{$N_{(2)}$}} \put(-395,8){
\makebox(0,0)[l]{$\delta[5]$}} \put(-365,-7){ \makebox(0,0)[l]{$N_{(3)}$}}
\put(-355,48){ \makebox(0,0)[l]{$T_{(3)}$}} \put(-295,8){
\makebox(0,0)[l]{$\delta[6]$}} \put(-255,-7){ \makebox(0,0)[l]{$N_{(2)}$}}
\put(-283,28){ \makebox(0,0)[l]{$T_{(3)}$}} \put(-220,28){
\makebox(0,0)[l]{$T_{(3)}$}} \put(-185,8){ \makebox(0,0)[l]{$\delta[7]$}}
\put(-150,45){ \makebox(0,0)[l]{$T_{(3)}$}} \put(-150,-7){
\makebox(0,0)[l]{$M_{(3)}$}} \put(-125,-7){ \makebox(0,0)[l]{$N_{(2)}$}}
\put(-85,8){ \makebox(0,0)[l]{$\delta[8]$}} \put(-27,32){
\makebox(0,0)[l]{$T_{(4)}$}} \put(-45,-7){ \makebox(0,0)[l]{$N_{(2)}$}}
\end{picture}  \label{fig1}
\end{equation}

\paragraph{Two Loop $\protect\delta^{\prime}$--Type Graphs}

These are the two loop diagrams with no insertion of $dx^{M}$, and in which
we shall look for ultralocal divergences. Observe that graphs $4$ and $5$,
although they look as purely bulk graphs, implicitly have an arbitrary
number of 2--vertex insertions of $M$--type on the propagators, and
therefore could have ultralocal contributions on the boundary. Nonetheless
we shall show that the diagrams contributing are $1,2$ and $3$.

\begin{equation}
\begin{picture}(220,120)(-5,-5)
\scalebox{.8}{\includegraphics{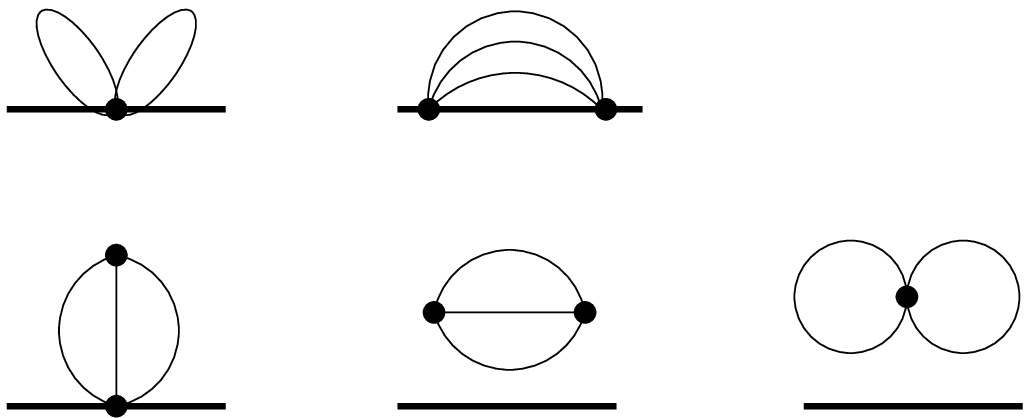}} \end{picture} 
\put(-240,80){ \makebox(0,0)[l]{$\delta'[1]$}} \put(-200,68){
\makebox(0,0)[l]{$M_{(4)}$}} \put(-150,80){ \makebox(0,0)[l]{$\delta'[2]$}}
\put(-127,70){ \makebox(0,0)[l]{$M_{(3)}$}} \put(-87,70){ \makebox(0,0)[l]
{$M_{(3)}$}} \put(-240,10){ \makebox(0,0)[l]{$\delta'[3]$}} \put(-195,0){
\makebox(0,0)[l]{$M_{(3)}$}} \put(-135,37){ \makebox(0,0)[l]{$T_{(3)}$}}
\put(-150,10){ \makebox(0,0)[l]{$\delta'[4]$}} \put(-80,37){
\makebox(0,0)[l]{$T_{(3)}$}} \put(-195,52){ \makebox(0,0)[l]{$T_{(3)}$}}
\put(-55,10){ \makebox(0,0)[l]{$\delta'[5]$}} \put(-15,50){ \makebox(0,0)[l]
{$T_{(4)}$}}  \label{fig2}
\end{equation}

Let us make one comment on counterterm graphs, which we have not considered.
When performing a two loop computation one should, in principle, include all
the one loop counterterms in the original action and further consider their
fluctuations. This yields further vertices and diagrams at two loops.
However, and for what concerns the beta function calculation we are
interested in, we may consider these terms to be zero. This reason for doing
so is that these counterterm diagrams only have divergences in $\ln^{2} \varepsilon$, and not in $\ln \varepsilon$, with $\varepsilon$ the ultraviolet  cut--off. Thus, they will not contribute to the beta function \cite{Andreev-Tseytlin}. As such, the graphs above are all the graphs one needs to take into consideration.


\section{The Two Loop Beta Function}


In order to compute the beta function associated with the gauge field, 
$A_{M} $, we must consider divergences arising in
vacuum diagrams at one and two loops built from the vertices described in
the previous section. We shall then be interested in divergences of the form 

\begin{equation*}
\frac{1}{2\pi }\ln \varepsilon \sum_{\ell}\oint \beta _{M}^{\left( \ell \right)
}\left( A,H\right) \,dx^{M}\,,
\end{equation*}

\noindent
where $\beta _{M}^{\left( \ell \right) }$ is the contribution to the beta
function at $\ell$ loops, and where $\varepsilon $ is the ultraviolet cut--off scale. The above form is dictated by dimensional analysis. As is clear, we are looking for a divergence given by the integral of a local function of the string fields, $\beta_{M}$, multiplying the 1--form $dx^{M}$. As seen in the last section, a generic vacuum graph will then fall into two large categories. It either contains a \textit{single} insertion of a $N$--type vertex, which explicitly contains a $dx^{M}$, or it will not contain an $N$--type vertex, and then the $dx^{M}$ factor will appear from differentiation in a way which we shall shortly explain. We call the two types of diagrams $\delta $--graphs and $\delta ^{\prime}$--graphs, respectively.


\subsection{Graphs of $\protect\delta$--Type}


Let us first focus on graphs of $\delta $--type---\textit{i.e.}, with a single
insertion of a vertex of $N$--type---since these graphs are the simplest ones
conceptually. We will first describe the general computational scheme and then explicitly work out two particular two loop graphs, leaving the full computation to the appendix. Also, we shall work from now on with units such that 

\begin{equation*}
\alpha ^{\prime }=1\,.
\end{equation*}

\noindent
In general, we are considering expressions of the following schematic form 

\begin{equation*}
\int \frac{d\theta}{2\pi}\, \frac{d\theta_{1}}{2\pi} \cdots \frac{d\theta_{n}}{2\pi}\ N \big[ x \left( \theta \right) \big]\ O_{1} \big[
x \left( \theta_{1} \right) \big] \cdots O_{n} \big[ x \left( \theta_{n} \right) \big]\, \times \, C \left( \theta, \theta_{1}, \cdots, \theta_{n} \right) ,
\end{equation*}

\noindent
where $\theta$, $\theta_{1}$, $\dots$, $\theta_{n}$, are the angular
positions of the $n+1$ vertices on the unit circle and where the $N$--type
vertex is located at $\theta$. We are suppressing all dependence on the radial
coordinate $\sigma $ on the disk, as well as all explicit reference to
tangent space indices on the various tensors, $N$ and $O_{i}$ (which are of $M$
or of $T$--type), which are just spectator indices as far as the arguments below are concerned. Finally, and most importantly, we are calling $C$ to the correlator of the $\zeta^{A} \left( \theta, \sigma \right)$ fields corresponding to the graph in question (again with tangent space indices understood). By standard quantum field theory arguments, the divergence in this correlation function $C$ will be local in the angular coordinates, $\theta$, $\theta_{i}$, and will be given---by dimensional arguments---by 

\begin{equation*}
c\ \ln \varepsilon\ \left( 2\pi \right) ^{n}\tprod\nolimits_{i}\,\delta
\left( \theta -\theta _{i}\right) ,
\end{equation*}

\noindent
where $c$ is a constant tensor (recall we are suppressing indices). It is for this
reason that we are dubbing these graphs as $\delta $--type graphs. Therefore,
although the vacuum graph is highly non--local in the background spacetime
fields, the divergent part is local and is given by 

\begin{equation*}
c\ \ln \varepsilon \int \frac{d\theta }{2\pi }\ N\ O_{1} \cdots O_{n} \big[
x \left( \theta \right) \big] ,
\end{equation*}

\noindent
thus yielding a contribution to the beta function of

\begin{equation*}
\Delta \beta_{M}\ dx^{M} = c\ N\ O_{1} \cdots O_{n}.
\end{equation*}

\noindent
We therefore need to find the tensor $c$. The simplest way to do this is to
consider the integral 

\begin{equation*}
\int \frac{d\theta}{2\pi}\ \frac{d\theta_{1}}{2\pi} \cdots \frac{d\theta_{n}}{2\pi}\ C \left( \theta, \theta_{1}, \dots, \theta_{n} \right) = c\ \ln \varepsilon + \mathrm{regular} .
\end{equation*}

\noindent
If we work instead in momentum space, in the angular direction $e^{in\theta
}$, where $n$ is the discrete angular momentum, we will be considering the
insertion of the $\zeta$ composite operators in the various
vertices \textit{at zero momentum}, and we will be looking at the divergent
part. Therefore, in practical terms, one must take all the vertices and
replace tensors multiplying the $\zeta$ fields, which depend on the
spacetime position $x\left( \theta ,\sigma \right) $, by \textit{constant
tensors}, and only then proceed to compute the graph (where vertices are now most easily displayed in momentum space).


\subsection{The Vertices in Momentum Space}\label{vert}


To show how this works, let us work out the form of the vertices in momentum
space. Let us first consider the $M$--type vertices, with momentum $W$
flowing \textit{into the vertex} (as discussed above, for the $\delta$--type graphs one needs $W=0$, but for the moment we will keep the discussion slightly more general, since we shall later need this extension for the $\delta^{\prime}$--type graphs). We have 

\begin{eqnarray*}
&&
-\frac{1}{\left( N-1 \right)!}\ M_{A_{1} A_{2} \cdots A_{N}} \int \frac{d\theta}{2\pi}\ e^{iW\theta}\ \zeta ^{A_{1}} \cdots \zeta^{A_{N-1}} \partial_{\theta} \zeta^{A_{N}} \left( \theta \right) \\
&&
=\frac{i}{N!}\ \big( n_{N} M_{A_{1} A_{2} \cdots A_{N}} + {\mathrm{cyc}}_{1 \cdots N} \big)\ \zeta_{n_{1}}^{A_{1}} \cdots \zeta_{n_{N}}^{A_{N}}\ \delta_{\Sigma_{i} n_{i} - W} ,
\end{eqnarray*}

\noindent
where ${\mathrm{cyc}}_{1 \cdots N}$ stands for the sum over cyclic permutations, and the sum over the outgoing momenta $n_{i}$ is implied. As we argued above, the tensor $M_{A_{1}\cdots A_{N}}$ should be thought of as \textit{constant}. We
therefore have the vertices 

\begin{equation*}
i \left( n_{N} M_{A_{1} A_{2} \cdots A_{N}} + {\mathrm{cyc}}_{1 \cdots N} \right)  \delta_{\Sigma_{i} n_{i} - W} .
\end{equation*}

\noindent
The 2--vertex is simple, since $M_{AB}=-M_{BA}$, and one has 

\begin{equation*}
i \left( n_{2} - n_{1} \right)\ M_{A_{1}A_{2}}\ \delta_{n_{1}+n_{2}-W}.
\end{equation*}

\noindent
The $N$--type vertices have no derivatives and are simply given by 

\begin{equation*}
-N_{A_{1}A_{2}\cdots A_{N}}\ \delta_{\Sigma_{i}n_{i}-W}.
\end{equation*}

\noindent
We also have two bulk $T$--type vertices. First of all, the 4--vertex
(here, we are parametrizing the unit disk with radial coordinate $\sigma \in \left[ 0,1 \right] $ and angular coordinate $\theta \in \left[ 0,2\pi \right]$) 

\begin{eqnarray*}
&&
-\frac{1}{4} T_{A_{1} \cdots A_{4}} \int \sigma d\sigma\ \frac{d\theta}{2\pi}\ e^{iW\theta}\ \zeta ^{A_{1}}\zeta ^{A_{2}} \left( \partial_{\sigma} \zeta^{A_{3}} \partial_{\sigma} \zeta^{A_{4}} + \frac{1}{\sigma^{2}}\ \partial_{\theta} \zeta^{A_{3}} \partial_{\theta} \zeta^{A_{4}} \right) \\
&=&
-\frac{1}{4} T_{A_{1} \cdots A_{4}} \int \frac{d\sigma}{\sigma}\ \zeta_{n_{1}}^{A_{1}} \zeta_{n_{2}}^{A_{2}} \zeta_{n_{3}}^{A_{3}} \zeta_{n_{4}}^{A_{4}}\ \left( \left| n_{3}n_{4} \right| - n_{3}n_{4} \right)\
\delta_{\Sigma_{i}n_{i}-W} \\
&=&
\frac{1}{4} T_{A_{1} \cdots A_{4}}\ \zeta_{n_{1}}^{A_{1}} \zeta_{n_{2}}^{A_{2}} \zeta_{n_{3}}^{A_{3}} \zeta_{n_{4}}^{A_{4}}\ \frac{\left( n_{3}n_{4} - \left| n_{3}n_{4} \right| \right)}{\Sigma_{i} \left| n_{i} \right|}\ \delta_{\Sigma_{i}n_{i}-W}.
\end{eqnarray*}

\noindent
Note that, in the above, we are able to perform the $\sigma$--integration
due to the fact that the $\zeta$--propagator factorizes as follows (more
details in the next subsection) 

\begin{equation*}
\zeta _{n}^{A}\left( \sigma \right) \zeta _{-n}^{A^{\prime }}\left( \tilde{\sigma}\right) \sim \sigma ^{\left| n\right| }\tilde{\sigma}^{\left|
n\right| } .
\end{equation*}

\noindent
Then, denoting $\left( ijkl \right) \equiv T_{A_{i}A_{j}A_{k}A_{l}} \left(
n_{k}n_{l}-\left| n_{k}n_{l} \right| \right)$, the vertex is given by 

\begin{equation*}
\frac{1}{\Sigma _{i} \left| n_{i} \right|}\ \Big[ T_{A_{1} \cdots A_{4}} \left(
n_{3}n_{4} - \left| n_{3}n_{4} \right| \right) + \left( 3412 \right) + \left(
1324 \right) + \left( 2413 \right) + \left( 1423 \right) + \left( 2314 \right) \Big]\ \delta_{\Sigma_{i}n_{i}-W}.
\end{equation*}

\noindent
Finally we have the 3--vertex 

\begin{eqnarray*}
-T_{A_{1}A_{2}A_{3}} \int d\sigma\ \frac{d\theta}{2\pi}\ e^{iW\theta}\ \zeta^{A_{1}} \partial_{x} \zeta^{A_{2}} \partial_{\theta} \zeta^{A_{3}} 
= i T_{A_{1}A_{2}A_{3}}\ \zeta_{n_{1}}^{A_{1}} \zeta_{n_{2}}^{A_{2}} \zeta_{n_{3}}^{A_{3}}\ \frac{\left| n_{2} \right| n_{3}}{\Sigma_{i} \left|
n_{i} \right|}\ \delta_{\Sigma_{i}n_{i}-W}.
\end{eqnarray*}

\noindent
Using the total antisymmetry of $T_{ABC}$, the 3--vertex finally reads 

\begin{equation*}
\frac{iT_{A_{1}A_{2}A_{3}}}{\Sigma_{i}\left| n_{i} \right|}\ \Big[ \left|
n_{1} \right| \left( n_{2}-n_{3} \right) + \left| n_{2} \right| \left(
n_{3}-n_{1} \right) + \left| n_{3} \right| \left( n_{1}-n_{2} \right) \Big]\ 
\delta_{\Sigma_{i}n_{i}-W}.
\end{equation*}


\subsection{The Dressed Propagator}


Recall that, due to the kinetic term 

\begin{equation*}
\frac{g_{AB}}{4\pi} \int d^{2}\sigma\ \delta^{ab}\ \partial_{a} \zeta^{A} \partial_{b} \zeta^{B},
\end{equation*}

\noindent
the propagator on the disk is given by 

\begin{equation*}
\left\langle \zeta^{A} \left( z \right) \zeta^{B} \left( w \right) \right\rangle = g^{AB} \Big[ \mathcal{B} \left( z,w \right) + \mathcal{C} \left( z,w \right) + \mathcal{D} \left( z,w \right) \Big] .
\end{equation*}

\noindent
The function $\mathcal{D}\left( z,w\right) $ only contributes to
the zero momentum part of the correlator, since it only depends on the
radial distances $z\bar{z}$ and $w\bar{w}$. It will therefore not contribute
in the computation of the leading divergences in the high momentum regime,
where the sums representing the various graphs diverge. The function $\mathcal{C}\left( z,w\right) $, on the other hand, vanishes if either $z$ or 
$w$ are on the boundary. It therefore only contributes in graphs where one
internal line connects two bulk vertices of $T$--type (or a $T$--vertex to
itself). We shall argue that the $\mathcal{C}$--part of the propagator does
not contribute to beta function. The reason comes from noticing that the
divergence associated to the $\beta _{M}$ function is associated to
singular behavior when all operators collide at the same point \textit{on
the boundary of the worldsheet}, exactly where the contribution of $\mathcal{C}\ $ vanishes. Therefore, for the purpose of computing the beta function, we
consider instead the propagator 

\begin{eqnarray*}
\left\langle \zeta^{A} \left( z \right) \zeta^{B} \left( w \right) \right\rangle &=& g^{AB}\ \mathcal{B} \left( z,w \right) , \\
\left\langle \zeta_{n}^{A} \zeta_{-n}^{B} \right\rangle &=& g^{AB}\ \frac{\left| zw \right|^{\left| n \right|}}{\left| n \right|},
\end{eqnarray*}

\noindent
where, in the second line, we have given the momentum space
representation, using the fact that 

\begin{equation*}
\mathcal{B}\left( xe^{i\alpha },ye^{i\beta }\right) = \sum_{n\neq 0}\frac{1}{\left| n\right| }\left( xy\right) ^{\left| n\right| }e^{-in\left( \alpha
-\beta \right) }.
\end{equation*}

\noindent
As we had previously noted, the propagator factorizes as $\left| zw \right|^{\left| n \right|}$ and the radial part has then already been absorbed in
the definition of the vertices. We are only left with the angular part,
which is simply 

\begin{equation*}
\left\langle \zeta_{n}^{A} \zeta_{-n}^{B} \right\rangle = g^{AB}\ \frac{1}{\left| n \right|}.
\end{equation*}

\noindent
Following the above discussion we are then instructed, in the computation of $\delta $--type graphs, to use the vertices in the previous subsection at zero
momentum (with $W=0$) and glue them with the above propagator.

Recall that we have two distinct 2--vertices 

\begin{equation*}
i\left( n_{2}-n_{1} \right)\ M_{A_{1}A_{2}}\ \delta _{n_{1}+n_{2}}, \qquad \qquad  \qquad - N_{A_{1}A_{2}}\ \delta_{n_{1}+n_{2}},
\end{equation*}

\noindent
and that we want to add their contribution to the propagator in order to have an exact result in $\F$. On the other hand, since in our graphs we shall have at most one insertion of the 2--vertex $N$, we shall rather find it convenient to include only the $M$--type 2--vertex in the propagator. As usual in quantum field theory, the inverse propagator sums with the 2--vertex, to give the full propagator 

\begin{equation*}
\Big( \left| n \right| g_{AB} - 2in M_{AB} \Big)^{-1} = \Big( \left| n \right|
g_{AB} + n \mathcal{F}_{AB} \Big)^{-1}.
\end{equation*}

\noindent
Introducing open string tensors as usual 

\begin{equation*}
\left( \frac{1}{g+\mathcal{F}} \right)^{AB} = \left( \frac{1}{G} \right)^{AB} + \Theta^{AB},
\end{equation*}

\noindent
we obtain the final \textit{dressed} propagator 

\begin{equation*}
\left\langle \zeta_{n}^{A} \zeta_{-n}^{B} \right\rangle = G^{AB} \frac{1}{\left| n \right|} + \Theta^{AB} \frac{1}{n} \equiv \Pi_{n}^{AB}.
\end{equation*}


\subsection{The Regularization Scheme}


The last ingredient one needs in order to compute the graphs is a regularization
scheme. Introducing a regulator $\varepsilon >0$, we consider a generic
vacuum graph contributing to a $\delta $--graph. In order to regulate the
sums defining the graph at high momenta, let us introduce a factor of 

\begin{equation*}
e^{-\left| n\right| \varepsilon }
\end{equation*}

\noindent
for every internal line connecting two vertices, where we shall consider as a 
\textit{single line any line with insertions of 2--vertices}. Note that
this is consistent due to the fact that all vertices are inserted at zero
momentum and therefore 2--vertices do not alter the momentum flowing
through a single line (which is therefore a well defined quantity, $n$).

We shall then be interested, as already advocated, in graphs which diverge \textit{primitively as} $\ln \varepsilon$, when $\varepsilon \rightarrow 0$.
In fact, graphs which diverge as $\ln ^{2}\varepsilon $ and higher (already present at two loops) will be made finite by the presence of the lower order counterterms, as described for instance in \cite{Andreev-Tseytlin}.


\subsection{One Loop Result}


As warm--up, let us rederive the standard one loop result (\ref{fig1loop}). When we later discuss $\delta ^{\prime }$--graphs, we shall see that they also contain a possible contribution at one loop---this, nonetheless, vanishes upon explicit computation. Therefore we compute the full one loop beta function by computing the $\delta$--graph.

The single graph contributing is shown in figure (\ref{fig1loop}), and is given
by 

\begin{equation*}
\frac{1}{2} N_{AB} \sum_{n\neq 0} \Pi_{n}^{AB} e^{-\left| n \right| \varepsilon}.
\end{equation*}

\noindent
The factor of $\frac{1}{2}$ comes from the symmetry factor of the graph. The
sum is explicitly symmetric in $n \rightarrow -n$, and therefore the only
contribution comes from the $G^{AB}$ term in $\Pi_{n}^{AB}$, thus yielding 

\begin{eqnarray*}
N_{AB} G^{AB} \sum_{n>0} \frac{1}{n}e^{-n\varepsilon} &=& -N_{AB} G^{AB} \ln \left( 1 - e^{-\varepsilon} \right) \\
&\sim & -N_{AB} G^{AB} \ln \left( \varepsilon \right) .
\end{eqnarray*}

\noindent
Therefore, the one loop contribution to the beta function is given
by the standard result \cite{CLNY}

\begin{equation}\label{beta1}
\beta_{M}^{\left( 1 \right)}\ dx^{M} = - N_{AB} \left( \frac{1}{G} \right)^{AB}.
\end{equation}


\subsection{Examples of Two Loop $\protect\delta$--Graphs}\label{comdelta}


We will next consider two examples of two loop diagrams, however leaving the full computation to the appendix. Let us first focus on an $\infty$--type graph, which we shall choose to be $\delta \left[ 4 \right]$ (see figure (\ref{fig1})). It is clearly given, using the Feynman rules we have just discussed and the symmetry factor of $4$, by 

\begin{eqnarray}
&&
\frac{1}{4} \sum_{n,m\neq 0} \Pi_{n}^{AR} \Pi_{n}^{SR} \Pi_{m}^{CD}\ e^{-\left| n \right| \varepsilon - \left| m \right| \varepsilon} \times \notag \\
&&
\times i \Big( m \left( M_{ABCD} - M_{DABC} \right) + n \left( M_{CDAB} - M_{BCDA} \right) \Big) N_{RS}.  \label{leq1}
\end{eqnarray}

\noindent
In order to proceed, we first quote some results on double sums (reviewed
in the appendix), which read 

\begin{equation*}
\sum_{n,m\geq 1} \frac{e^{-a\left( n+m\right)}}{n^{2}} \sim \ln \varepsilon, \qquad \qquad \qquad \sum_{n,m\geq 1} \frac{e^{-a\left( n+m\right)}}{nm}\sim 0.
\end{equation*}

\noindent
Notice that, in the first sum, we are isolating the $\ln \varepsilon$ term
part. Moreover, the second sum diverges as $\ln ^{2}\varepsilon$ and
vanishes for the purpose of computing the beta function, since we are
looking for primitive $\ln \varepsilon $ divergences. It is therefore easy
to see that the last two terms in the second line of (\ref{leq1}) vanish.
Now using that $M_{ABCD}$ is symmetric in $ABC$ and that the sum over 
$m$ is odd for the remaining terms, so that one can replace $\Pi_{m}^{CD}$
with $\frac{1}{m}\Theta ^{AB}$, we arrive at the result 

\begin{equation*}
2iM_{ABCD}\left( GNG+\Theta N\Theta \right) ^{AB}\Theta ^{CD}\,\ln
\varepsilon .
\end{equation*}

Let us next focus on the computation of a $\theta$--type graph, for instance 
$\delta \left[ 6\right]$ (see figure (\ref{fig1})). It has a symmetry factor of $4$ and is given by 

\begin{eqnarray*}
&&
- \frac{1}{4} T_{ABC} T_{A^{\prime}B^{\prime}C^{\prime}} \sum_{n,m,p} \delta_{n+m+p}\ \Pi_{n}^{AA^{\prime}} \Pi_{m}^{BB^{\prime}} \left( \Pi N \Pi \right)_{p}^{CC^{\prime}}\ e^{- \left| n \right| \varepsilon - \left| m \right| \varepsilon - \left| p \right| \varepsilon} \times \\
&&
\times \frac{\big[ \left| n \right| \left( m-p \right) + \left| m \right| \left( p-n \right) + \left| p \right| \left( n-m \right) \big]^{2}}{\big( \left| n \right| + \left| m \right| + \left| p \right| \big)^{2}} .
\end{eqnarray*}

\noindent
We must split the sum into three regions $I,II,III$, given by $n,m>0$, $n,p>0$
and $m,p>0$, alongside with the corresponding regions given by $\left( n,m,p \right) \rightarrow \left( -n,-m,-p \right)$. We are then led to consider the
function 

\begin{equation*}
F \left( n,m,p \right) = \frac{1}{nmp^{2}}\ \frac{\big[ \left| n \right| \left( m-p \right) + \left| m \right| \left( p-n \right) + \left| p \right| \left( n-m \right) \big]^{2}}{\big( \left| n \right| + \left| m \right| + \left| p \right| \big)^{2}}\ e^{- \left| n \right| \varepsilon - \left| m \right| \varepsilon - \left| p \right| \varepsilon}
\end{equation*}

\noindent
and the corresponding sums 

\begin{eqnarray*}
S_{I} &=&\sum_{n,m>0}F\left( n,m,-n-p\right) =\sum_{n,m>0}\frac{\left(
n-m\right) ^{2}}{nm\left( n+m\right) ^{2}}\ e^{-2\left( n+m\right) }, \\
S_{II} &=&\sum_{n,p>0}F\left( n,-n-p,n\right) =-\sum_{n,p>0}\frac{\left(
n-p\right) ^{2}}{p^{2}n\left( n+p\right) }\ e^{-2\left( n+p\right) }, \\
S_{III} &=&\sum_{m,p>0}F\left( -m-p,m,p\right) =-\sum_{m,p>0}\frac{\left(
m-p\right) ^{2}}{p^{2}m\left( m+p\right) }\ e^{-2\left( m+p\right) }.
\end{eqnarray*}

\noindent
The above sums can be easily computed using results from the appendix, and are given by 

\begin{equation*}
S_{I}\sim 4\ln \varepsilon, \qquad \qquad 
S_{II}\sim -\ln \varepsilon, \qquad \qquad 
S_{III}\sim -\ln\varepsilon .
\end{equation*}

\noindent
Now let us consider the propagators $\Pi_{n}^{AA^{\prime}} \Pi_{m}^{BB^{\prime}} \left( \Pi N \Pi \right)_{p}^{CC^{\prime}}$. Since the full sum must be invariant under $\left( n,m,p\right) \rightarrow \left( -n,-m,-p \right)$, we will obtain the following tensor structures 

\begin{eqnarray*}
&&\frac{1}{nmp^{2}}\ \Theta ^{AA^{\prime }}\Theta ^{BB^{\prime }}\left(
GNG+\Theta N\Theta \right) ^{CC^{\prime }}, \\
&&\frac{1}{\left| n\right| \left| m\right| p^{2}}\ G^{AA^{\prime
}}G^{BB^{\prime }}\left( GNG+\Theta N\Theta \right) ^{CC^{\prime }}, \\
&&\frac{1}{\left| n\right| mp\left| p\right| }\ G^{AA^{\prime }}\Theta
^{BB^{\prime }}\left( GN\Theta +\Theta NG\right) ^{CC^{\prime }}, \\
&&\frac{1}{n\left| m\right| p\left| p\right| }\ \Theta ^{AA^{\prime
}}G^{BB^{\prime }}\left( GN\Theta +\Theta NG\right) ^{CC^{\prime }},
\end{eqnarray*}

\noindent
which, in turn, give the following total sums 

\begin{eqnarray*}
2\left( S_{I}+S_{II}+S_{III}\right) &\sim &4\ln \varepsilon, \\
2\left( S_{I}-S_{II}-S_{III}\right) &\sim &12\ln \varepsilon, \\
2\left( -S_{I}+S_{II}-S_{III}\right) &\sim &-8\ln \varepsilon, \\
2\left( -S_{I}-S_{II}+S_{III}\right) &\sim &-8\ln \varepsilon,
\end{eqnarray*}

\noindent
where the factor of $2$ comes from regions with $\left( n,m,p\right)
\rightarrow \left( -n,-m,-p\right)$. Combining all of the above facts, we
finally have the result for the graph $\delta \left[ 6\right]$, which is: 

\begin{eqnarray*}
&&-T_{ABC}T_{A^{\prime }B^{\prime }C^{\prime }}\left( \Theta \Theta
+3GG\right) ^{AA^{\prime },BB^{\prime }}\left( GNG+\Theta N\Theta \right)
^{CC^{\prime }}\ln \varepsilon + \\
&&+2T_{ABC}T_{A^{\prime }B^{\prime }C^{\prime }}\left( \Theta G+G\Theta
\right) ^{AA^{\prime },BB^{\prime }}\left( GN\Theta +\Theta NG\right)
^{CC^{\prime }}\,\ln \varepsilon .
\end{eqnarray*}


\subsection{Graphs of $\protect\delta^{\prime}$--Type}


We now move out attention to the more complex case of the $\delta ^{\prime }$--type graphs---\textit{i.e.}, those graphs which do not contain an $N$--type vertex and which therefore do not contain an explicit $dx^{M}$ in the vertices.
Again, we are considering expressions of the following schematic form 

\begin{equation*}
\int \frac{d\theta }{2\pi}\ \frac{d\theta_{1}}{2\pi} \cdots \frac{d\theta_{n}}{2\pi}\ O \big[ x \left( \theta \right) \big]\ O_{1} \big[ x \left( \theta_{1} \right) \big] \cdots O_{n} \big[ x \left( \theta_{n} \right) \big] \times C \left( \theta, \theta _{1}, \cdots, \theta_{n} \right) ,
\end{equation*}

\noindent
where, as before, we are suppressing explicit reference to all indices and
radial dependence. To compute these graphs we again need to understand the
nature of the short--distance singularities in the correlator $C$.
Introducing the functions 

\begin{equation*}
D_{I}\left( \theta ,\theta _{i}\right) =\left( 2\pi \right) ^{n}\,\delta
^{\prime }\left( \theta -\theta _{I}\right) \tprod\nolimits_{j\neq I}\delta
\left( \theta -\theta _{j}\right) ,
\end{equation*}

\noindent
with $I=1,\cdots,n$, it is simple to see, by power counting, that the short
distance singularity of $C$ proportional to $\ln \varepsilon $ must be of
the form 

\begin{equation*}
\ln \varepsilon\ \sum_{I} c_{I} D_{I},
\end{equation*}

\noindent
where the $c_{I}$'s are constant tensors. Therefore we have a contribution
to the beta function given by 

\begin{equation*}
\Delta \beta_{M}\ dx^{M} = \sum_{J} c_{J}\ O\ O_{1} \cdots dO_{J} \cdots O_{n}.
\end{equation*}

\noindent
(note that the arbitrary choice of $O$ and $O_{J}$ is immaterial, since
different choices change the beta function by an exact 1--form). Let us
comment on the expression $dO_{J}$. Assume, for simplicity of exposition,
that the tensor $O_{J}$ is given by some section $S_{N}^{M}$. Then, using (\ref{curlD}) and (\ref{simpl2}), one concludes that 

\begin{equation*}
dS_{N}^{M} = E_{\mu}^{M} E_{N}^{\nu}\ \mathcal{D}S_{\nu}^{\mu} = 
E_{\mu}^{M} E_{N}^{\nu}\ \left[ DS_{\nu}^{\mu } + \frac{1}{2} H^{\mu}{}_{\alpha\sigma} \mathcal{F}^{\sigma}{}_{\lambda} dx^{\lambda} S_{\nu}^{\alpha} - \frac{1}{2} H^{\alpha}{}_{\nu\sigma} \mathcal{F}^{\sigma}{}_{\lambda} dx^{\lambda} S_{\alpha}^{\mu} \right] .
\end{equation*}

\noindent
Since moreover $D=dx^{\lambda} \nabla_{\lambda}$, we arrive at the final explicitly covariant result 

\begin{equation*}
dS_{N}^{M} = \left[ \nabla_{L} S_{N}^{M} + \frac{1}{2} H^{M}{}_{PA} \mathcal{F}^{A}{}_{L} S_{N}^{P} - \frac{1}{2} H^{P}{}_{NA} \mathcal{F}^{A}{}_{L} S_{P}^{M} \right] dx^{L},
\end{equation*}

\noindent
which can be easily extended to general tensors $O_{J}$. We see explicitly
that, due to the presence of $\delta ^{\prime }\left( \theta -\theta
_{J}\right) $ in the short--distance singularity of the correlator $C$, one
obtains an explicit factor of $dx^{M}$ in the expression for the graph. To
find the tensors $c_{I}$, we consider the integral 

\begin{equation*}
\frac{1}{2iW} \int \frac{d\theta}{2\pi}\ \frac{d\theta_{1}}{2\pi} \cdots 
\frac{d\theta_{n}}{2\pi}\ C \left( \theta, \theta _{1}, \cdots, \theta_{n} \right)\ \left( e^{iW \left( \theta_{J}-\theta \right)} - e^{-iW \left( \theta_{J}-\theta \right)} \right) = c_{J}\ \ln \varepsilon + \mathrm{regular}.
\end{equation*}

\noindent
The first term inserts momentum $-W$ in the vertex at $\theta$ and
momentum $+W$ at the vertex at $\theta_{J}$. Again working in momentum
space we see that one must consider the vertices of section (\ref{vert}),
with the following differences

\begin{enumerate}

\item All vertices $O_{i}$, $i\neq J$, are at zero momentum $W=0$.

\item The vertex corresponding to $O_{J}$ has momentum $W$ and has the
external tensor $M,T$, replaced by $dM,dT$.

\item The vertex corresponding to $O$ has momentum $-W$.

\end{enumerate}

\noindent
We then subtract the result with $W\rightarrow -W$, and divide by $2iW$. In
practice, we shall write down the initial expressions for arbitrary $W$
and will use the fact that the singular part is $W$--independent, in order
to simplify the results. We shall, more specifically, use the expression
above formally for $W\rightarrow 0$, by substituting it with 

\begin{equation*}
\left. \frac{\partial }{\partial W}\right| _{W=0}.
\end{equation*}

\noindent
In doing so, we shall use the following facts 

\begin{eqnarray}
\left. \frac{\partial }{\partial W}\right| _{W=0}\Pi _{n+W}^{AB} &=&-\frac{1}{n}\Pi _{n}^{AB}, \notag \\
\left. \frac{\partial }{\partial W}\right| _{W=0}e^{-\varepsilon \left|
p-W\right| } &=&\varepsilon \frac{\left| p\right| }{p}e^{-\varepsilon \left|
p\right| }. \label{Leq2}
\end{eqnarray}

The regularization scheme adopted for the $\delta $--graphs easily carries over
to the $\delta ^{\prime }$--graphs, with one minor extension. Consider an
internal line, possibly with insertions of 2--vertices. If all of
the 2--vertices are at zero momentum, the momentum flowing along the
line is just a given value $n$ and we regularize it with $e^{-\varepsilon
\left| n\right| }$. If, on the other hand, we have $K-1$ insertions of
2--vertices with non--vanishing momentum, then the momentum flowing
through the line will change, and will acquire $K$ distinct values $n_{1},\dots ,n_{K}$. Then, we shall regularize the line with the factor 

\begin{equation*}
e^{-\frac{\varepsilon}{K}\big( \left| n_{1} \right| + \cdots + \left| n_{K} \right| \big)}.
\end{equation*}

\noindent
Finally, and as far as the propagator is concerned, it is clear that whenever the 2--vertex $M_{AB}$ is inserted at zero momentum it can be re--summed as
before, to give the previously discussed dressed propagator.


\subsection{Examples of Two Loop $\protect\delta^{\prime}$--Graphs}\label{comdeltaprime}


First we want to show that the $\delta ^{\prime}$--contribution to the
one loop beta function vanishes. It is given by 

\begin{equation*}
\frac{1}{2} \sum_{n} \left( 2n-W \right)^{2}\ \Pi_{n-W}^{AA^{\prime}} \Pi_{n}^{B^{\prime}B}\ M_{AB} dM_{A^{\prime}B^{\prime}}\ e^{-\frac{\varepsilon}{2} \left( \left| n \right| + \left| n-W \right| \right)},
\end{equation*}

\noindent
on which we must act with $\left. \frac{\partial }{\partial W}\right|
_{W=0}$. Using (\ref{Leq2}), one only obtains a contribution from the $\varepsilon$--dependent part, coming from $\frac{\partial }{\partial W}$
acting on the regulator, and it is

\begin{equation*}
\varepsilon \sum_{n} n \left| n \right| \Pi_{n}^{AA^{\prime}} \Pi_{n}^{B^{\prime}B}\  M_{AB} dM_{A^{\prime}B^{\prime}}\ e^{-\varepsilon \left| n\right|}.
\end{equation*}

\noindent
Keeping only the even part of the sum under $n\rightarrow -n$, we obtain 

\begin{equation*}
2\varepsilon\ M_{AB} \left( GdM\Theta +\Theta dMG\right)
^{AB}\sum_{n>0}e^{-\varepsilon \left| n\right| }.
\end{equation*}

\noindent
The above does not contribute to the beta function for two reasons. Firstly, 
$M_{AB}$ is odd under $AB\rightarrow BA$, and $\left( GdM\Theta +\Theta
dMG\right)^{AB}$ is even. Secondly, the expression $\varepsilon
\sum_{n>0}e^{-\varepsilon \left| n\right| }=\varepsilon e^{-\varepsilon
}/\left( 1-e^{-\varepsilon}\right)$ is regular as $\varepsilon \rightarrow
0$, and therefore does not constitute a diverging term in the correlator.

Let us now compute, as an example of a two loop $\delta ^{\prime }$--graph,
the contribution to the beta function coming from $\delta ^{\prime }\left[ 1 \right]$ (see figure (\ref{fig2})). The graph, whose symmetry factor is $4$, 
has the following expression (readily obtained from the Feynman rules
previously described)

\begin{eqnarray*}
&&
\frac{1}{4} \sum_{n,m} i \Big( M_{ABCD} \left( -m \right) + M_{BCDA} \left( 
n-W \right) + M_{CDAB} \left( -n \right) + M_{DABC} \left( +m \right) \Big) \times \\
&&
\times i \left( 2n-W \right)\ dM_{RS}\ \Pi_{n-W}^{AR} \Pi_{n}^{SB} \Pi_{m}^{CD}\ e^{-\frac{\varepsilon}{2}\left( \left| n-W\right| +\left| n\right| + 2\left| m\right| \right)}.
\end{eqnarray*}

\noindent
Computing the derivative $\left. \frac{\partial }{\partial W}\right| _{W=0}$
one obtains the two terms (in the following expressions we will omit, for simplicity, the sum $\sum_{n,m}$ and the regulator $e^{-\varepsilon \left( \left| n\right| + \left| m\right| \right)}$)

\begin{equation*}
\frac{1}{4} \Big( m \left( M_{ABCD}-M_{DABC} \right) + n \left(
M_{BCDA}+M_{CDAB} \right) \Big)\ \left( \Pi dM \Pi \right)_{n}^{AB}
\Pi_{m}^{CD}
\end{equation*}

\noindent
and, from the regulator, 

\begin{equation*}
\varepsilon \frac{\left| n \right|}{4} \Big( m \left(
M_{ABCD}-M_{DABC} \right) + n \left( M_{CDAB}-M_{BCDA} \right) \Big)\ \left(
\Pi dM \Pi \right)_{n}^{AB} \Pi_{m}^{CD}.
\end{equation*}

\noindent
In the first expression, the second term vanishes, since $\sum_{n,m}\frac{1}{nm}\sim 0$. The first term gives 

\begin{equation*}
\left( M_{ABCD}-M_{DABC}\right) \left( GdMG+\Theta dM\Theta \right)
^{AB}\Theta ^{CD}\ln \varepsilon
\end{equation*}

\noindent
which itself vanishes, this time around due to symmetry of $M_{ABCD}$ and
antisymmetry of $\left( GdMG+\Theta dM\Theta \right)^{AB}$. The second
expression, coming from the regulator, gives the following two contributions 

\begin{eqnarray*}
&&\left( M_{DABC}-M_{ABCD}\right) \left( GdMG+\Theta dM\Theta \right)
^{AB}\Theta ^{CD}\ln \varepsilon , \\
&&\left( M_{BCDA}-M_{CDAB}\right) \left( GdM\Theta +\Theta dMG\right)
^{AB}G^{CD}\ln \varepsilon .
\end{eqnarray*}

\noindent
The first term vanishes as before and the remaining term yields the unique
contribution to the beta function, given by (renaming indices) 

\begin{equation*}
\left( M_{ABCD}-M_{BCDA}\right) \left( GdM\Theta +\Theta dMG\right)
^{AD}G^{BC}\ln \varepsilon .
\end{equation*}


\subsection{Final Result for the Two Loop Beta Function}


Following the techniques just described one can proceed to compute the contribution to the beta function from all of the $\delta$ and $\delta^{\prime}$--graphs. The full computation is carried out in the appendix and leads to the results we shall now present. The following graphs give a vanishing contribution:
 
\begin{eqnarray*}
&&\delta \left[ 1\right] =\delta \left[ 2\right] =\delta \left[ 5\right]
=0, \\
&&\delta ^{\prime }\left[ 4\right] =\delta ^{\prime }\left[ 5\right] =0.
\end{eqnarray*}

\noindent
The non--vanishing contributions to the beta function from the $\delta$--type graphs are: 

\begin{eqnarray*}
\delta \left[ 3 \right] &=& - M_{ABC} \left( 3M_{RST} + M_{RTS} \right) \left[
\left( \frac{1}{G} N \frac{1}{G} + \Theta N \Theta \right) \frac{1}{G} \frac{1}{G} \right]^{AR,BS,CT} + \\
&&
+ M_{ABC} \left( M_{RST} - M_{RTS} \right) \left[ \left( \frac{1}{G} N \frac{1}{G} + \Theta N \Theta \right) \Theta \Theta \right]^{AR,BS,CT} + \\
&&
+ M_{ABC} \left( M_{RST} - 3\, M_{RTS} \right) \left[ \left( \frac{1}{G} N \Theta + \Theta N \frac{1}{G} \right) \left( \frac{1}{G} \Theta + \Theta \frac{1}{G} \right) \right]^{AR,BS,CT}, \\
\delta \left[ 4 \right] &=& 2i\, M_{ABCD} \left( \frac{1}{G} N \frac{1}{G} + \Theta N \Theta \right)^{AB} \Theta^{CD}, \\
\delta \left[ 6 \right] &=& - T_{ABC} T_{A^{\prime}B^{\prime}C^{\prime}} \left[ \left( \frac{1}{G} N \frac{1}{G} + \Theta N \Theta \right) \left( 3\, \frac{1}{G} \frac{1}{G} + \Theta \Theta \right) \right]^{AA^{\prime},BB^{\prime},CC^{\prime}} +  \\
&&
+ 2\, T_{ABC} T_{A^{\prime}B^{\prime}C^{\prime}} \left[ \left( \frac{1}{G} N \Theta + \Theta N \frac{1}{G} \right) \left( \frac{1}{G} \Theta + \Theta \frac{1}{G} \right) \right]^{AA^{\prime},BB^{\prime},CC^{\prime}}, \\
\delta \left[ 7 \right] &=& 4\, T_{ABC} M_{A^{\prime}B^{\prime}C^{\prime}} \left[ 2 \left( \Theta N \Theta + \frac{1}{G} N \frac{1}{G} \right) \Theta \Theta - \left( \frac{1}{G} N \Theta + \Theta N \frac{1}{G} \right) \left( \frac{1}{G} \Theta + \Theta \frac{1}{G} \right) \right]^{AA^{\prime},BB^{\prime},CC^{\prime}}, \\
\delta \left[ 8 \right] &=& T_{ABCD} \left( \frac{1}{G} N \frac{1}{G} + \Theta N \Theta \right)^{AB} \left( \frac{1}{G} \right)^{CD}.
\end{eqnarray*}

\noindent
The non--vanishing contributions to the beta function from the $\delta^{\prime}$--type graphs are:

\begin{eqnarray*}
\delta^{\prime} \left[ 1 \right] &=& \left( M_{ABCD} - M_{BCDA} \right) \left( \frac{1}{G} dM \Theta + \Theta dM \frac{1}{G} \right)^{AD} \left( \frac{1}{G} \right)^{BC}, \\
\delta^{\prime} \left[ 2 \right] &=& \frac{1}{6}\, M_{A_{1}A_{2}A_{3}} dM_{B_{1}B_{2}B_{3}} \left[ - 2\, \Theta \Theta \Theta + 3\, \Theta \frac{1}{G} \frac{1}{G} + 3\, \frac{1}{G} \Theta \frac{1}{G} - 4\, \frac{1}{G} \frac{1}{G} \Theta \right]^{A_{1}B_{1},A_{2}B_{2},A_{3}B_{3}} + \\
&&
+ \frac{1}{3}\, M_{A_{1}A_{2}A_{3}} dM_{B_{2}B_{3}B_{1}} \left[ \Theta \Theta \Theta + 2\, \Theta \frac{1}{G} \frac{1}{G} - 5\, \frac{1}{G} \Theta \frac{1}{G} + 2\, \frac{1}{G} \frac{1}{G} \Theta \right]^{A_{1}B_{1},A_{2}B_{2},A_{3}B_{3}}, \\
\delta^{\prime} \left[ 3 \right] &=& 6i\, M_{ABC} T_{A^{\prime}B^{\prime}C^{\prime}} \left[ \left( \frac{1}{G} dM \Theta + \Theta dM \frac{1}{G} \right) \Theta \frac{1}{G} - \left( \frac{1}{G} dM \frac{1}{G} + \Theta dM \Theta \right) \Theta \Theta \right]^{AA^{\prime},BB^{\prime},CC^{\prime}}.
\end{eqnarray*}

\noindent
We then have that the two loop contribution to the beta function is given by

\begin{equation}\label{beta2}
\beta_{M}^{\left( 2 \right)}\ dx^{M} = \sum_{i} \delta \left[ i \right] + \sum_{j} \delta^{\prime} \left[ j \right].
\end{equation}

\noindent
Recall that the above terms must be summed to the one loop contribution (\ref{beta1}),

\begin{equation*}
\beta_{M}^{\left( 1 \right)}\ dx^{M} = - N_{AB} \left( \frac{1}{G} \right)^{AB}.
\end{equation*}


\section{The Nonabelian Born--Infeld Action}


We are now in possession of the abelian beta function for the open string gauge field. At this stage it is important to stress that $\beta = 0$ are \textit{not} the equations of motion for the massless open string mode \cite{Tseytlin-2}: indeed it is not the vanishing of the beta functions, $\beta = 0$, which should be equivalent to the string equations of motion, but rather the vanishing of the so--called Weyl anomaly coefficients, ${\bar{\beta}} = 0$ (\textit{i.e.}, the ones which ensure the absence of the Weyl anomaly---the standard beta functions only ensure the absence of the scale anomaly). For the closed string the Weyl anomaly coefficients have been studied in \cite{Tseytlin-2} and it is known that they are given by the beta functions plus some additive terms. These additive terms, up to two loops, are given in terms of derivatives of the dilaton and covariant derivatives of $H$ (at three and higher loops they have not been computed) so that if the dilaton is constant and $H$ is covariantly constant, as is the case for the parallel backgrounds, the Weyl anomaly coefficients will actually coincide with the beta functions. For the open string there are also some results \cite{Behrndt-Dorn}, although not as complete. Here, the Weyl anomaly coefficients are both diffeomorphism and gauge invariant \cite{Andreev-Tseytlin} (the beta functions are not) so that sigma model conformal invariance can always be rephrased as the vanishing of the beta functions \textit{up to} terms which can be generated by diffeomorphisms or gauge transformations. The general expression for the open string equations of motion is thus \cite{Andreev-Tseytlin}

\begin{equation} \label{bwac}
{\bar{\beta}}_{I}\, dx^{I} \equiv \beta_{I}\, dx^{I} + \F_{IJ} M^{J} dx^{I} + d L = 0,
\end{equation}

\noindent
where $M^{I}$ and $L$ are functions of $\F_{IJ}$ and its derivatives. Given the two loop beta function we have computed, finding the two loop open string equations of motion is thus equivalent to determining these unknown functions, $M^{I}$ and $L$, also at two loops.

An alternative way to finding the open string equations of motion is to compute the string effective action directly, although this is not necessarily simpler than computing the aforementioned Weyl anomaly coefficients. This calculation entails making use of the boundary string field theory formalism \cite{BSFT-1,BSFT-2}, where besides the beta function one is also required to compute the partition function of the sigma model theory at two loops (in the superstring case one is only required to compute the partition function). Knowledge of both these functionals then yields the open string effective action via the well known expression,

$$
S[\F_{IJ}] = \left( 1 + \beta_{M} [\F_{IJ}] \cdot \frac{\delta}{\delta A_{M}} \right) Z[\F_{IJ}].
$$

\noindent
It is important to point out that either Weyl anomaly coefficients or partition function, one must always compute to two loops and non--perturbatively in $\F$. This is to say, any of the aforementioned calculations will be non--trivial, and we shall leave them as the next step in our general programme to compute curvature couplings in the nonabelian BI action, a step to which we hope to return sometime in the future. At this stage, we want to proceed with the explanation of our strategy to compute these nonabelian couplings given the abelian equations of motion. We shall thus assume we are in possession of the abelian equations of motion and outline the general procedure from there.

These massless open string mode abelian equations of motion correspond to a commutative, or maximal brane, description of the BI action. Our goal now is to translate this abelian description of the BI action to a nonabelian description, and this is done via the SW map. The method we shall deploy has been studied at length in \cite{Cornalba-3, Cornalba-4, Cornalba-5} within the context of the nonabelian BI action in flat space, and has actually led to a detailed set of constraints that could eventually make way to a solution of the nonabelian BI. Here, we shall begin by reviewing the strategy of \cite{Cornalba-3, Cornalba-4, Cornalba-5} as it generalizes to our situation of curved backgrounds. We will spell out how one should proceed in order to move from the abelian to the nonabelian description, for the case of parallel backgrounds, and will also comment on the possibility to constraint the result to possible field redefinitions and ordering ambiguities.


\subsection{On the Seiberg--Witten Map}


The main idea behind the SW map \cite{Cornalba-Schiappa-1, Seiberg-Witten, Cornalba-2} is that derivative corrections in the abelian action can be understood, at large field strength (or $B$--field) $B$ on the abelian $D$--brane worldvolume, as corrections to the standard multiplication of functions so that multiplication actually gets deformed into a noncommutative star product. This is achieved via a particular change of variables (from commutative to noncommutative fields) which precisely corresponds to the SW map \cite{Seiberg-Witten}. The noncommutative description is then simpler to understand as a matrix model, as one represents functions with operators, star products with operator multiplication and world--volume integrals get replaced by traces \cite{Cornalba-4, Cornalba-Schiappa-2}. The shift in notation is as follows

$$
x^{A} \to {\mathcal{X}}^{A}, \qquad \int V(B)\ dx \to {\mathbf{Tr}},
$$

\noindent
where $V(B) = \sqrt{\det B}\ \left( 1 + \cdots \right)$ is a volume form which makes the integral act as a trace with respect to the star product. $B_{IJ}$ is a symplectic form in the usual case of the Moyal star product, and dots correspond to loop diagrams which contribute when the Poisson structure $\theta$ is not constant \cite{Kontsevich, Cornalba-Schiappa-2}.

Because we are going to higher order in derivatives than in \cite{Seiberg-Witten}, let us pause for a moment and explain the distinction between the several star products one can consider (see \cite{Seiberg-Witten, Kontsevich, Cornalba-4, Cornalba-Schiappa-2} for applications of the different possibilities). The standard BI situation of constant $B$ was dealt in \cite{Schomerus, Seiberg-Witten} and leads to the well known Moyal star product,

$$
f \star g\ (x) = \left. e^{\frac{i}{2} \theta^{AB} \partial_{A}^{x} \partial_{B}^{y}} f(x) g(y) \right|_{x=y}.
$$

\noindent
When $B$ is no longer constant but the target is still flat space, there are derivative corrections to the standard abelian BI action and in order for the SW map to work properly the star product should be written using Kontsevich's formula \cite{Cornalba-4, Kontsevich} (defined for nonconstant Poisson structure),

$$
f \star g = f g + \frac{i}{2} \theta^{AB} \partial_{A} f \partial_{B} g - \frac{1}{8} \theta^{AC} \theta^{BD} \partial_{A} \partial_{B} f \partial_{C} \partial_{D} g - \frac{1}{12} \theta^{AL} \partial_{L} \theta^{BC} \left( \partial_{A} \partial_{B} f \partial_{C} g - \partial_{B} f \partial_{A} \partial_{C} g \right) + {\mathcal{O}} (\theta^{3}).
$$

\noindent
Kontsevich's formula can also be of use even when $\theta$ is no longer Poisson. This corresponds to the case where the target space is no longer flat \cite{Cornalba-Schiappa-2}. We should point out that in this paper we shall \textit{not} use the point of view in \cite{Cornalba-Schiappa-2}. Even though we believe that there should be a definition of star product that will compute correlation functions at this order, just like what happened at order ${\mathcal{O}} (H)$ in \cite{Cornalba-Schiappa-2}, in here we wish to keep the background explicit in all formulas (using the approach of \cite{Cornalba-Schiappa-2} the background would be fully translated into the star product itself). Even so, there are derivative corrections to $B$ arising from our beta function calculation. This means that one should be careful when translating derivative corrections to star products as the correct star product to use could be Kontsevich's one.

Let us begin by justifying our method in general terms \cite{Cornalba-3, Cornalba-4, Cornalba-5}, explaining how the SW limit takes us from the standard BI equations of motion to the noncommutative matrix model equations of motion. Consider an abelian action $S$, written in terms of a gauge field $A_{M} ( x )$, and a SW map $\lambda$ given by (here, $*$ is the pull--back map)

$$
\lambda^{*} B = \F,
$$

\noindent
where $B_{MN}$ is constant, and $\F = B + 2\pi\alpha' dA$. Also, let the variation of $S$ under $\delta A_{M} ( x )$ be given by (defining the Weyl anomaly coefficients $\bar{\beta}$ from an effective action point of view)

$$
\delta S = -T_{p} \int d^{n}x\ {\bar{\beta}}^{M} ( x ) \delta A_{M} ( x ).
$$

\noindent
If one writes the SW map as usual \cite{Seiberg-Witten, Cornalba-2}

\begin{eqnarray*}
\X^{M} &=& x^{M} + \theta^{MN} \widehat{A}_{N} ( x ), \\
\lambda^{*} \X^{M} &=& x^{M},
\end{eqnarray*}

\noindent
then one can also write the variation of the action in terms of a variation of the coordinates $\X^{M}$, as

$$
\delta S = -T_{p} \int d^{n}x\ \sqrt{\det B}\ \ {\widehat{\bar{\beta}}}_{M}\ \delta \X^{M} = -T_{p} \int d^{n}x\ \sqrt{\det \F}\ \left( \lambda^{*} \widehat{\bar{\beta}}_{M} \right) \left( \lambda^{*} \delta \X^{M} \right).
$$

\noindent
At large $B$ field, matrix model equations of motion can be easily translated into equations depending on Poisson brackets rather than commutators (we shall see this in greater detail in the following). So, matrix model equations of motion let us know about $\lambda^{*} \widehat{\bar{\beta}}_{M}$. What is then left to understand in the above expression is $\lambda^{*} \delta \X^{M}$. We have \cite{Cornalba-2}

$$
\delta \X^{M} ( x ) = \theta^{AB} \partial_{A} \X^{M} \partial_{B} \X^{N}\ \delta A_{N} \left[ \X ( x ) \right],
$$

\noindent
so that $\lambda^{*} \delta \X^{M} ( x ) = \left( \frac{1}{B} \right)^{MN} \delta A_{N} ( x )$ and one obtains

$$
\sqrt{\det \F}\ \left( \lambda^{*} \widehat{\bar{\beta}}_{M} \right) \left( \frac{1}{B} \right)^{MN} = {\bar{\beta}}^{N} ( x ).
$$

\noindent
We have thus shown that in general it is always possible to move between abelian and nonabelian descriptions of $D$--brane physics. All one has to do is compare abelian and nonabelian equations of motion as expressed above. In practice, we shall see in the following how to implement this reasoning.

In order to establish a proper perturbative framework, the first thing one needs to do is to have a consistent set of conventions on dimensions of the various fields. We will take the target fields $g_{IJ}$ and $B_{IJ}$ to be dimensionless. In this case $F_{IJ}$ has length dimension $L^{-2}$. Next, turn to the beta function. In the standard perturbative calculation $\alpha'$ is taken as a loop counting parameter, so that different orders in perturbation theory correspond to different overall powers of $\alpha'$. In order for things to properly match later on, we shall take the following convention for the overall factors of $\alpha'$ in front of the $\ell$--loop perturbative abelian beta function (the same holds true for the Weyl anomaly coefficients),

$$
\left( \alpha' \right)^{-1} \beta^{(1)} + \left( \alpha' \right)^{0} \beta^{(2)} + \cdots + \left( \alpha' \right)^{\ell - 2} \beta^{(\ell)} + \cdots.
$$

\noindent
Now, each $\beta^{(\ell)}$ can still be expanded for large field strength $F$. This is due to the fact that we are computing the beta function with a propagator which is exact in $F$. One will thus have

$$
\beta^{(\ell)} = \sum_{n} \frac{\beta_{n}^{(\ell)}}{\left( \alpha' \right)^{n}},
$$

\noindent
where each $\beta_{n}^{(\ell)}$ does \textit{not} depend on $\alpha'$. This deals with conventions for dimensions.

Let us begin with an illustration in the case of flat space \cite{Cornalba-3, Cornalba-4, Cornalba-5}. Start with the nonabelian description and the standard quartic action,

$$
S = - \frac{1}{4 \left( \alpha' \right)^{2}}\ {\mathbf{Tr}} \Big( \left[ \X^{M}, 
\X^{N} \right] \left[ \X^{M}, \X^{N} \right] \Big),
$$

\noindent
which has the matrix equations of motion,

$$
\left( \alpha' \right)^{-2} \left[ \X^{M}, \left[ \X^{M}, \X^{N} \right] \right] = 0.
$$

\noindent
Here and in the following, repeated indices are always summed (\textit{i.e.}, contracted with the flat metric $g_{AB}$). If one now uses the star product to leading order, one has that $\left[ \X^{M}, \X^{N} \right] = i \theta^{MN}$ and $\left[ \X^{M}, \left[ \X^{M}, \X^{N} \right] \right] = i \left[ \X^{M}, \theta^{MN} \right] = - \theta^{MK} \partial_{K} \theta^{MN} + \cdots$. Observe that in here $\theta^{MN} \equiv \left( \frac{1}{F} \right)^{MN}$ and indeed $\theta$ has length dimension $L^{2}$ as expected. In terms of the Poisson structure the previous matrix equations of motion can thus be written as

\begin{equation} \label{ymnaeom}
\left( \alpha' \right)^{-2} \theta^{MK} \partial_{K} \theta^{MN} + \cdots = 0.
\end{equation}

\noindent
Let us now turn to the dual description and try to obtain the exact same result starting from the beta function for the abelian brane (at one loop there is no distinction between beta function and anomaly coefficients, for the open string). 
First, one should take notice that the open string tensors, at large $F$, behave as

\begin{eqnarray}\label{openlimits}
\lim_{F \to \infty} \left( \frac{1}{G} \right)^{MN} &=& - \frac{1}{\left( 2\pi\alpha' \right)^{2}} \left( \frac{1}{F}\ g\ \frac{1}{F} \right)^{MN} + \cdots = - \frac{1}{\left( 2\pi\alpha' \right)^{2}} \theta^{MK} \theta^{KN} + \cdots, \nonumber \\
\lim_{F \to \infty} \Theta^{MN} &=& \frac{1}{2\pi\alpha'} \left( \frac{1}{F} \right)^{MN} + \cdots = \frac{1}{2\pi\alpha'} \theta^{MN} + \cdots.
\end{eqnarray}

\noindent
Now write the contraction of the one loop beta function with the open string $\Theta$ tensor as (recall indices are raised with the open string Poisson tensor as one translates between abelian and nonabelian equations of motion)

$$
\Theta^{MN} \beta^{(1)}_{N} = \Theta^{MN} \left( \frac{1}{G} \right)^{KL} \partial_{K} \F_{LN},
$$

\noindent
and compute the large $F$ limit as

\begin{eqnarray*}
\lim_{F \to \infty} \Theta^{MN} \beta^{(1)}_{N} &=& \frac{1}{2\pi\alpha'} \theta^{MN} \left( - \frac{1}{\left( 2\pi\alpha' \right)^{2}} \theta^{JK} \theta^{KL} \right) \partial_{J} \left( 2\pi\alpha' F_{LN} \right) + \cdots \\
&=&
- \frac{1}{\left( 2\pi\alpha' \right)^{2}}\ \theta^{MN} \theta^{JK} \theta^{KL} \partial_{J} F_{LN} + \cdots.
\end{eqnarray*}

\noindent
Recalling the basic matrix fact that $\partial \left( \frac{1}{F} \right) = - \frac{1}{F} \partial F \frac{1}{F}$ it is now simple to obtain that the beta function equation of motion, at large $F$, is precisely given by

$$
\left( \alpha' \right)^{-2} \theta^{KJ} \partial_{J} \theta^{KM} + \cdots = 0,
$$

\noindent
which is the exact same expression we have previously obtained when starting from the nonabelian matrix description. This simple example thus illustrates the basics of our method.

In conclusion, at large field strength $F$ there is a map that allows us to obtain a matrix action from a loop corrected, higher derivative, beta function result. We should point out that given a generic loop corrected beta function it may be extremely hard to actually carry out this programme. An advantage in our favor is, of course, the fact that we are dealing with (WZW) parallel target space backgrounds, where the geometry simplifies immensely as compared to the generic background situation. In order to make it easier for a determination of the nonabelian equations of motion from the abelian ones, we shall now run through the expectations, based on general grounds, of which possible structures can appear for the quantum corrected matrix model. Because we are starting from loop corrections to the abelian action, it is only natural to expect that these nonabelian structures will also be organized in a consistent perturbative expansion, to which we now turn.


\subsection{Perturbative Classification of Tensor Structures}


Let us consider parallel backgrounds. Generic monomials that can appear in the action should be dimensionless and should thus have the generic form (the indices refer to powers of $H$ and $\X$)

$$
\left( \sqrt{\alpha'} \right)^{n-m} H^{n} \X^{m}.
$$

\noindent
Because we are doing a perturbative calculation, we need to understand 
how these monomials come into play as one carries forth the 
perturbation theory expansion. In order to understand this issue with 
greater clarity let us focus for the moment on a WZW model at level $k$. 
In this case, the radius (characteristic size) of the group manifold 
target space is of order $H^{-1} \sim {\mathcal{R}} \sim \ell_{s} 
\sqrt{k}$. On the other hand the radius (characteristic size) of the 
conjugacy class (the $D$--brane inside the target space) is of order 
$\X \sim n \frac{{\mathcal{R}}}{k} \sim n \frac{\ell_{s}}{\sqrt{k}}$, at 
fixed integer $n$. The above powers of $H$ and $\X$ thus behave as

$$
\left( \sqrt{\alpha'} \right)^{n-m} H^{n} \X^{m} \sim \left( \frac{1}{\sqrt{k}} \right)^{n+m}.
$$

\noindent
What we learn is that there are two distinct expansions. The 
$\alpha'$ expansion, which is organized according to powers of $n-m$; 
and the level expansion, which is organized according to powers of 
$n+m$. Indeed this is to be expected \cite{ARS-2}: one knows that in WZW 
models the relevant low energy limit is a combined limit in which 
$\alpha' \to 0$ and $\alpha' k \to \infty$. It is then only natural 
that the perturbative expansion arranges itself in this double 
expansion. We are now in a position to make a guess on the structure 
of the nonabelian monomials which will emerge from the beta function 
loop calculation. The perturbative structure we expect to find is spelled 
out in the following table.

\begin{equation} \label{tableWZW}
\begin{tabular}{|c|c|c|c|c|c|}
    \hline
    $\ $ & $\X^{2}$ & $\X^{3}$ & $\X^{4}$ & $\X^{5}$ & $\X^{6}$ \\
    \hline
    $H^{0}$ & 
    \begin{tabular}{ll}
	$\times$ \\
    \end{tabular} &
    \begin{tabular}{ll}
	$\times$ \\
    \end{tabular} &
    \begin{tabular}{ll}
	{\tiny{{\textsf{$\ell=1$ YES}}}} \\
	{\tiny{{\textsf{$\ell=2$ OK}}}} \\
    \end{tabular} &
    \begin{tabular}{ll}
	$\times$ \\
    \end{tabular} &
    \begin{tabular}{ll}
	{\tiny{{\textsf{$\ell=1$ NOT}}}} \\
	{\tiny{{\textsf{$\ell=2$ OK}}}} \\
    \end{tabular} \\
    \hline
    $H^{1}$ &
    \begin{tabular}{ll}
	$\times$ \\
    \end{tabular} &
    \begin{tabular}{ll}
	{\tiny{{\textsf{$\ell=1$ YES}}}} \\
	{\tiny{{\textsf{$\ell=2$ OK}}}} \\  
    \end{tabular} &
    \begin{tabular}{ll}
	$\times$ \\
    \end{tabular} &
    \begin{tabular}{ll}
	{\tiny{{\textsf{$\ell=1$ NOT}}}} \\
	{\tiny{{\textsf{$\ell=2$ OK}}}} \\
    \end{tabular} &
    \begin{tabular}{ll}
	$\times$ \\
    \end{tabular} \\
    \hline
    $H^{2}$ & 
    \begin{tabular}{ll}
	{\tiny{{\textsf{$\ell=1$ NOT}}}}\\
	{\tiny{{\textsf{$\ell=2$ NOT}}}} \\
    \end{tabular} &
    \begin{tabular}{ll}
	$\times$ \\
    \end{tabular} &
    \begin{tabular}{ll}
	{\tiny{{\textsf{$\ell=1$ NOT}}}} \\
	{\tiny{{\textsf{$\ell=2$ OK}}}} \\
    \end{tabular} &
    \begin{tabular}{ll}
	$\times$ \\
    \end{tabular} &
    \begin{tabular}{ll}
	$\cdots$ \\
    \end{tabular} \\
    \hline
    $H^{3}$ & 
    \begin{tabular}{ll}
	$\times$ \\
    \end{tabular} &
    \begin{tabular}{ll}
	{\tiny{{\textsf{$\ell=1$ NOT}}}} \\
	{\tiny{{\textsf{$\ell=2$ OK}}}} \\
    \end{tabular} &
    \begin{tabular}{ll}
	$\times$ \\
    \end{tabular} &
    \begin{tabular}{ll}
	$\cdots$ \\
    \end{tabular} &
    \begin{tabular}{ll}
	$\times$ \\
    \end{tabular} \\
    \hline
    $H^{4}$ & 
    \begin{tabular}{ll}
	{\tiny{{\textsf{$\ell=1$ NOT}}}} \\
	{\tiny{{\textsf{$\ell=2$ OK}}}} \\
    \end{tabular} &
    \begin{tabular}{ll}
	$\times$ \\
    \end{tabular} &
    \begin{tabular}{ll}
	$\cdots$ \\
    \end{tabular} &
    \begin{tabular}{ll}
	$\times$ \\
    \end{tabular} &
    \begin{tabular}{ll}
	$\cdots$ \\
    \end{tabular} \\
    \hline
\end{tabular}
\\
\end{equation}

\noindent
We have listed above all possible monomials which can appear in the 
\textit{action}, including all candidates at both one and two loops.
Because these are monomials in the action they must have all 
their tensorial indices contracted so that not every possible entry 
is allowed (we are not considering the $H^{0} \X^{2}$ monomial as our 
method only computes the potential terms in the action, not the kinetic 
term to which this monomial contributes). Denoted with a cross, $\times$, 
are monomials which \textit{cannot} appear in the action (it is simple 
to see that they would always have free indices). Denoted with dots, $\cdots$, 
are monomials that will only be relevant at higher loop level 
(as we will explain in the following). Each monomial has a distinct power 
of $n-m$ associated to the $\alpha'$ expansion and a distinct power of $n+m$ associated to the level expansion. For the allowed entries, one needs to 
understand at which loop level in perturbation theory will they appear and how could they contribute. Having in mind the case of WZW models, where the level expansion organizes the perturbative expansion, we can now make the following conjecture. \textit{New contributing monomials at each order, $\ell$, in perturbation theory correspond to diagonal lines of fixed $n+m$, where 
the loop order is $\ell = \frac{n+m}{2}-1$}. So, the one loop beta 
function can contribute with new monomials starting in the $n+m=4$ 
diagonal and then all entries to the right of this diagonal. Likewise, 
the two loop beta function can contribute with new monomials starting 
in the $n+m=6$ diagonal and then all entries to the right of this 
diagonal. Observe that the keyword here is \textit{new} monomials. 
What we mean by this is that, for instance, the two loop beta function 
can also contribute to monomials to the \textit{left} of the $n+m=6$ 
diagonal but it will \textit{not} generate any new monomials with 
respect to the one loop result---it will only generate derivative 
corrections to monomials which already exist from lower order results in the 
perturbative calculation. As an illustration of this point observe 
that the way the two loop results can affect the left side of the $n+m=6$ 
diagonal is via derivative corrections to the Kontsevich 
star product, in the monomials that were \textit{first} generated at one 
loop. As a consequence, if a monomial which should be first generated at 
one loop is in fact not generated (\textit{i.e.}, it has coefficient zero 
from the string theory calculation), then there can be no contribution to 
the corresponding entry arising from any higher order in perturbation theory. 
In summary, the conjecture is that the ${\bar{\beta}}^{(\ell)}$ function, at 
loop level $\ell$, contributes to the action with terms of order $k^{-N}$ with 
$N \geq 1 + \ell$, or $n+m \geq 2+2\ell$. Therefore, our computation at two 
loops can reliably fix terms in the nonabelian action with $n+m=4$ and $n+m=6$.

With these ideas in mind, let us then understand the remaining 
details in the table above. Loop monomials are separated 
by diagonal lines and we have denoted with ``{\textsf{$\ell=1$ YES}}'' 
terms which we know to exist from the one loop result (\textit{i.e.}, from the Myers action). These have been previously computed from both abelian \cite{ACNY, CLNY} and nonabelian points of view \cite{Myers, ARS-1}. In particular, we have 
denoted with ``{\textsf{$\ell=1$ NOT}}'' all the monomials which we 
know not to be generated by the one loop beta function (see the appendix). One of these lies in the $n+m=4$ diagonal and thus, according to the previous 
paragraph, there can also be no contribution to this monomial from the 
two loop beta function. This fact we have denoted with the label 
``{\textsf{$\ell=2$ NOT}}''. Finally, we have denoted with 
``{\textsf{$\ell=2$ OK}}'' all monomials which can have contributions 
from the two loop beta function. The contribution of the two loop 
beta function to the monomials in the $n+m=4$ diagonal will be only at 
the level of corrections to the star product expansion. The 
contribution of the two loop beta function to monomials in the 
$n+m=6$ diagonal will be that of generating new monomials with 
respect to the one loop result. The two loop beta function may also 
generate terms to the right of the $n+m=6$ diagonal. These terms would 
play a distinctive role in a three loop calculation, a subject on 
which we shall have nothing to say. Let us comment on the various terms 
which can be generated at one and two loops:

\begin{itemize}
    
    \item The $H^{0} \X^{6}$ term starts at two loops and corresponds
    to ${\mathrm{Tr}} F^{3}$ in the nonabelian BI action for open 
    bosonic strings. It arises from a matrix term of the type (sum over repeated 
    indices)
    
    $$
    \left( \frac{1}{\alpha'} \right)^{3} {\mathbf{Tr}} \Big( \left[ 
    \X^{M}, \X^{N} \right] \left[ \X^{N}, \X^{L} \right] 
    \left[ \X^{L}, \X^{M} \right] \Big),
    $$
    
    \noindent
    and the result one should thus expect to see from the two loop beta 
    function---\textit{i.e.}, in the equations of motion---is of the type
    (more on this later)
    
    $$
    \Theta \cdot {\bar{\beta}}^{(2)} \sim \cdots + \left( \frac{1}{\alpha'} 
    \right)^{3} \theta^{NL} \partial_{L} \left( 
    \theta^{NK} \theta^{KM} \right) + \cdots.
    $$
    
    \noindent
    Observe that the normalization of ${\mathrm{Tr}} F^{3}$ in 
    the nonabelian BI action is known, so that the full result for the two loop 
    equations of motion should reproduce the above expectations, including 
    \textit{all} factors. A detailed discussion of this term is included in the 
    appendix.
    
    \item The monomials $H^{1} \X^{5}$, $H^{2} \X^{4}$, $H^{3} \X^{3}$ and 
    $H^{4} \X^{2}$ also start at two loops. The only contribution from the 
    star product to these monomials should be the leading one, 
    \textit{i.e.}, all one should worry about is $f \star g = fg + 
    \frac{i}{2} \theta^{MN} \partial_{M} f \partial_{N} g$, 
    higher order terms not being necessary and only making an 
    appearance at three loops and higher. In particular one need not 
    worry about higher order derivative corrections to the Kontsevich star 
    products for these monomials.
    
    \item The term $H^{2} \X^{2}$ is not present at one loop (see the appendix), 
    even though dimensional analysis says it could have appeared. 
    One thus concludes that this monomial should also not be present 
    at two loops, since the two loop calculation should contribute to 
    \textit{new} monomials $H^{n} \X^{m}$ only for $n+m \geq 6$.
        
    \item The terms $H^{1} \X^{3}$ and $H^{0} \X^{4}$ start at one 
    loop. They are the well known terms in the Myers action. One thus 
    expects that the two loop result should only correct the star 
    product in these terms from Moyal to Kontsevich. We expect no new 
    monomials arising from the two loop calculation. For example, 
    the equations of motion for the term $H^{1} \X^{3}$ could have corrections 
    of the form
    
    $$
    H_{ABC}\ \X^{B} \star \X^{C} = H_{ABC} \left(
    \theta^{BC} + \kappa\ \partial_{M} \theta^{BN} 
    \partial_{N} \theta^{CM} + \cdots \right),
    $$
    
    \noindent
    where the first term arises from ${\bar{\beta}}^{(1)}$ and the second from 
    ${\bar{\beta}}^{(2)}$ with $\kappa$ some constant to be computed (and which 
    will yield information on the star product). The dots are possible higher 
    loop corrections.

\end{itemize}

This is the generic set up for obtaining the nonabelian equations of 
motion. The procedure itself is straightforward to implement: given the $\ell$ 
loop beta function result, one should start by expanding all possible
closed string background fields, $g_{IJ}$, $B_{IJ}$ and 
$\Gamma^{K}_{IJ}$, in terms of the local coordinates, $x^{A}$ (including the closed string fields which are present ``inside'' the open string tensors). This must be done in order to properly identify the matrix 
model, as it will depend on the noncommutative version of the 
coordinates. The expansion of these tensors is most simply done in 
Riemann normal coordinates (RNC) (see the appendix). Once the expansion is done, 
one should then write the beta function in powers of $\theta^{IJ}$ 
(which is to say, powers of $\frac{1}{F}$). These steps should not entail 
any difficulties. The final step is clearly the trickier as one needs 
to re--write the derivative expansions in the beta function in terms of 
star products. Here, one will need to reorganize monomials as 
$\X^{A_{1}} \star \cdots \star \X^{A_{n}}$. A point that may make this 
calculation slightly easier is the fact that, as we have just seen when 
listing the possible monomials appearing in the action, most of the time 
one is actually able to get along only with the leading terms in the star 
product.


\subsection{Field Redefinitions and Ordering Ambiguities}


One thing one should always keep present in string theory is that the coefficients of the low energy effective action, for the massless string modes, may be ambiguous.  Here, we shall have a pragmatic approach on this issue. Of all the structures which can appear in the effective action, based on the general arguments we have just seen, we wish to find out which ones can or cannot be removed via field redefinitions. The final set of structures with unambiguous coefficients will be our result for the $\a$ corrected nonabelian action.

As we have seen, there are five monomial structures arising at order $n+m=6$, which are expected to make their appearance in a two loop calculation in the abelian theory. These monomials correspond to specific tensor structures and, because they are matrix expressions, there are matrix ordering issues to be dealt with when classifying the structures that correspond to each of the two loop monomials. In other words, we would now like to have a list of \textit{all} the possible tensor structures (including \textit{all} possible orderings) that can appear from the two loop calculation. For each of the five monomials in the action we have the following classification of tensor structures:

\begin{center}
\begin{tabular}{|c|c|c|}
\hline
$H^{3} \X^{3}$ & $H^{2} \X^{4}$ & $H^{1} \X^{5}$ \\
\hline
$H_{AMN} H_{BML} H_{CNL} \X^{A} \X^{B} \X^{C}$ &
$H_{ABM} H_{MCD} \X^{A} \X^{B} \X^{C} \X^{D}$ &
$H_{ABC} \X^{A} \X^{B} \X^{C} \X^{D} \X^{D}$ \\
\hline
$H_{ABM} H_{CNL} H_{MNL} \X^{A} \X^{B} \X^{C}$ &
$H_{AMN} H_{BMN} \X^{A} \X^{C} \X^{B} \X^{C}$ &
$H_{ABC} \X^{A} \X^{B} \X^{D} \X^{C} \X^{D}$ \\
\hline
$H^{2} H_{ABC} \X^{A} \X^{B} \X^{C}$ &
$H_{AMN} H_{BMN} \X^{A} \X^{B} \X^{C} \X^{C}$ &
$\times$ \\
\hline
$\times$ &
$H^{2} \X^{A} \X^{B} \X^{A} \X^{B}$ &
$\times$ \\
\hline
$\times$ &
$H^{2} \X^{A} \X^{A} \X^{B} \X^{B}$ &
$\times$ \\
\hline
\end{tabular}
\\
\end{center}

\noindent
Repeated indices are summed and an overall trace is implied. Note that for the $H^{2} \X^{4}$ column other possible structures one could write down are not independent as the background field $H$ satisfies a Jacobi identity. Also, we have omitted the $H^{4} \X^{2}$ column, as we shall briefly see that it does not contribute to the action. One still has to classify the independent tensor structures in the monomial $H^{0} \X^{6}$. Here we will make use of a property we shall explain later: the action should be translation invariant. As we shall see shortly, this amounts to the requirement that only commutator structures can appear in the classification of this term (a fact which should not come as a great surprise since this term is the known ${\mathrm{Tr}} F^{3}$ coupling in the BI action---see the appendix). The independent structures thus are:
    
\begin{eqnarray} \label{x6}
&&
\left[ \X^{M}, \X^{N} \right] \left[ \X^{N}, \X^{S} \right] \left[ \X^{S}, \X^{M} \right], \nonumber \\
&&
\left[ \X^{A}, \left[ \X^{B}, \X^{C} \right] \right] \left[ \X^{A}, \left[ \X^{B}, \X^{C} \right] \right], \nonumber \\
&&
\left[ \X^{A}, \left[ \X^{B}, \X^{C} \right] \right] \left[ \X^{B}, \left[ \X^{A}, \X^{C} \right] \right].
\end{eqnarray}

\noindent
Let us stress once more that there are no issues concerning orderings of matrices left to resolve: we have listed all possible matrix orderings in the previous classification of tensor structures.

Having solved the issue of ordering ambiguities in the two loop monomials, we now turn to the ambiguities arising from field redefinitions. We shall begin by analyzing the constraints due to translation invariance of the matrix action. Generically speaking, an action is an integral (\textit{i.e.}, trace) over spacetime, and integrals have the property of translation invariance. This is a feature we would thus like to see made explicit in our matrix action---not all of the above tensor structures will be invariant under translations; indeed one expects that commutator structures must be present in order for such a property to hold. Let us begin with a slightly more general point of view, and let us analyze what happens to monomials $H^{n} \X^{m}$ under a change of coordinates $\X^{M}\to \X^{M} + \Xi^{M} (\X)$, where $\Xi^{M}$ is the vector generating the diffeomorphism (see \cite{Boer-Schalm, BFLR} for further discussions on these issues). The reason to start with this more general analysis is the following: the translational invariant tensor structures at one loop could, under the above field redefinition, generate new terms in the action whose role would be to affect the listed tensor structures at two loops in such a way as to make the whole action translational invariant, \textit{i.e.}, in such a way that one would no longer need the requirement that the two loop structures need to be translational invariant by themselves. We shall see that this, in fact, does not hold true---the two loop tensors must also have the property of translational invariance and thus be expressed via commutators.

Recall the matrix model we are computing is set up in RNC, so that this fact must still remain true after the above diffeomorphism. The geometric requirement associated to the coordinate change is thus that we move from RNC to RNC. If one recalls that the metric transforms as $g_{MN} \to g_{MN} - \nabla_{M} \Xi_{N} - \nabla_{N} \Xi_{M}$, and if one uses the formulas in the appendix, it is simple to check that the requirement of having RNC for the transformed metric translates to a constraint on $\Xi^{M}$ which itself translates to the fact that the field redefinition acts on the coordinate matrices as

$$
\X^{M} \to \X^{M} + \Xi^{M} (0) - \frac{1}{6} \Xi^{N} (0)\ {H_{NA}}^{L} {H_{LB}}^{M}\ \X^{A} \X^{B} + \cdots,
$$

\noindent
and on the monomials present in the matrix action the effect of the above constant and quadratic terms in the field redefinition is $H^{n} \X^{m} \to H^{n} \X^{m-1}$ and $H^{n} \X^{m} \to H^{n+2} \X^{m+1}$, respectively. There are, of course, higher order terms in the above transformation which generate many other possibilities, but for our two loop expansion this is all we care about. This action allows for two sets of ``movements'' in table (\ref{tableWZW}): moving one box to the left at fixed horizontal line as $H^{n} \X^{m} \to H^{n} \X^{m-1}$, or moving one box to the right and two boxes down as $H^{n} \X^{m} \to H^{n+2} \X^{m+1}$. As a consequence, it is simple to observe after a glance at table (\ref{tableWZW}) and immediate application of the two rules above, that none of the one loop monomials (the $n+m=4$ diagonal) can affect (under these movements) any of the two loop monomials (the $n+m=6$ diagonal). Moreover, for the two loop monomials, the $H^{n} \X^{m} \to H^{n} \X^{m-1}$ movement cannot produce any effect and this immediately implies that the full two loop diagonal must consist of monomials which are translational invariant. Thus, translation invariance requires that the $\X^{M}$ coordinates are organized into commutator structures. We shall now return to that classification and check which (if any) of the structures are translation invariant (recall that the $H^{0} \X^{6}$ term was already dealt with in a translation invariant way---thus the commutator structures):

\begin{center}
\begin{tabular}{|c|c|c|}
\hline
$H^{3} \X^{3}$ & $H^{2} \X^{4}$ & $H^{1} \X^{5}$ \\
\hline
$H_{AMN} H_{BML} H_{CNL} [ \X^{A}, \X^{B} ] \X^{C}$ &
$H_{ABM} H_{MCD} [ \X^{A}, \X^{B} ] [ \X^{C}, \X^{D} ]$ &
$H_{ABC} [ \X^{A}, \X^{B} ] [ \X^{C}, \X^{D} ] \X^{D}$ \\
\hline
$H_{ABM} H_{CNL} H_{MNL} [ \X^{A}, \X^{B} ] \X^{C}$ &
$H_{AMN} H_{BMN} [ \X^{A}, \X^{C} ] [ \X^{B}, \X^{C} ]$ &
$\times$ \\
\hline
$H^{2} H_{ABC} [ \X^{A}, \X^{B} ] \X^{C}$ &
$H^{2} [ \X^{A}, \X^{B} ] [ \X^{A}, \X^{B} ]$ &
$\times$ \\
\hline
\end{tabular}
\\
\end{center}

\noindent
The first thing one notices is that the number of independent structures has been narrowed down. For the monomial $H^{4} \X^{2}$ one can check that either the possible tensor structures are schematically of the form $\left. {\mathrm{SymTensor}} \right|_{\alpha\beta} \X^{\alpha} \X^{\beta}$ and thus vanish, or the trace ends up killing the commutator term, so that none of the tensor structures  for this monomial leads to translation invariant terms. A similar reasoning applies to other structures. Altogether, we have managed to reduce the apparently very complex problem of determining the nonabelian BI action in a curved parallel target space, at two loops, to the [still complicated but simpler] problem of determining two constants at one loop and ten constants at two loops. The constants to be determined are precisely the coefficients of the mentioned tensor structures as they appear in the action and as determined, \textit{e.g.}, via the beta function  calculation we have set up. Of course the value of the coefficients may still be zero and further simplify the action.

There is one further point to mention. We have constrained the action to field redefinitions associated to translational invariance. This is a geometric constraint based on general expectations for the form of an effective action. String theory, however, also allows for completely generic field redefinitions and this may in fact imply that some of the two loop coefficients could still be ambiguous. Let us now try to further understand this point. The most general field redefinition one can perform on the coordinate matrices $\X^{M} \to \X^{M} + \delta \X^{M}$ in target WZW parallel backgrounds consists of terms in the following monomials

\begin{center}
\begin{tabular}{cccc}
$\quad \X \quad$ & $\quad H \X^{2} \quad$ & $\quad H^{2} \X \quad$ & $\quad \cdots \quad$ \\
$\quad \X^{3} \quad$ & $\quad H \X^{4} \quad$ & $\quad H^{2} \X^{3} \quad$ & $\quad \cdots \quad$ \\
$\quad \X^{5} \quad$ & $\quad H \X^{6} \quad$ & $\quad H^{2} \X^{5} \quad$ & $\quad \cdots \quad$ \\
$\quad \vdots \quad$ & $\quad \vdots \quad$ & $\quad \vdots \quad$ & $\quad \quad$ \\
\end{tabular}
\\
\end{center}

\noindent
In this generic field redefinition, only ``movements to the right'' and ``movements down'' are allowed in table (\ref{tableWZW}). This implies that field redefinitions of two loop structures will only affect higher order structures (for which we care not). On the other hand, field redefinitions of one loop structures may affect the two loop tensors. This can happen by action of the $\X^{3}$, the $H \X^{2}$ and the $H^{2} \X$ monomials in the list above (the remaining infinite set of possibilities will only contribute at higher order). Indeed it is simple to observe that the redefinition $\X \to \X^{3}$ will produce a shift $H^{n} \X^{m} \to H^{n} \X^{m+2}$, which corresponds to moving two boxes to the right at fixed horizontal line, the redefinition $\X \to H \X^{2}$ will produce a shift $H^{n} \X^{m} \to H^{n+1} \X^{m+1}$, which corresponds to moving one box to the right and one box down, and the redefinition $\X \to H^{2} \X$ will produce a shift $H^{n} \X^{m} \to H^{n+2} \X^{m}$, which corresponds to moving two boxes down at fixed vertical line. In practical terms, some of the two loop tensor structures may thus be washed away. The redefinitions in question are properly written as

\begin{eqnarray*}
\delta \X^{M} &\sim& \left[ \X^{N}, \left[ \X^{N}, \X^{M} \right] \right], \\
\delta \X^{M} &\sim& H_{MNS} \left[ \X^{N}, \X^{S} \right], \\
\delta \X^{M} &\sim& H^{2}\ \X^{M}, \\
\delta \X^{M} &\sim& H_{MNL} H_{NLS}\ \X^{S},
\end{eqnarray*}

\noindent
their form naturally constrained by translational invariance requirements (\textit{i.e.}, while the field redefinitions by themselves need not be translation invariant, they must still be such that the action will remain translation invariant after we have performed the redefinition). 

This freedom to perform \textit{four} completely generic field redefinitions thus removes \textit{four} coefficients from the possible ten coefficients at two loops. At this stage we must pay some attention to the known results concerning the nonabelian BI action in flat backgrounds. Here we know that the $\X^{3}$ redefinition is used to remove one of the three $H^{0} \X^{6}$ structures, and we further know that once this has been done, the coefficient of the $\left[ \X, \X \right]^{3}$ term is fixed and non--zero while the coefficient of the remaining $\left[ \X, \left[ \X, \X \right] \right]^{2}$ term is fixed and zero. We will choose to do this very same field redefinition in curved space, to enjoy the fact that the $H^{0} \X^{6}$ coefficients will then be all known. We shall further make use of the $H \X^{2}$ redefinition to remove the $H^{1} \X^{5}$ one possible tensor structure. Then, we shall make use of the two $H^{2} \X$ field redefinitions to remove one of the three tensor structures at order $H^{3} \X^{3}$ and to remove one of the three tensors structures at order $H^{2} \X^{4}$. We are thus left with

\begin{center}
\begin{tabular}{|c|c|c|c|c|c|}
    \hline
    $\ $ & $\quad \X^{2} \quad$ & $\quad \X^{3} \quad$ & $\quad 
    \X^{4}\quad$ & $\quad \X^{5} \quad$ & $\quad \X^{6}\quad$ \\
    \hline
    $\quad H^{0} \quad$ & 
    $\times$ &
    $\times$ &
    1 &
    $\times$ &
    2 \\
    \hline
    $\quad H^{1} \quad$ &
    $\times$ &
    1 &
    $\times$ &
    0 &
    $\times$ \\
    \hline
    $\quad H^{2} \quad$ & 
    $\times$ &
    $\times$ &
    2 &
    $\times$ &
    $\times$ \\
    \hline
    $\quad H^{3} \quad$ & 
    $\times$ &
    2 &
    $\times$ &
    $\times$ &
    $\times$ \\
    \hline
    $\quad H^{4} \quad$ & 
    0 &
    $\times$ &
    $\times$ &
    $\times$ &
    $\times$ \\
    \hline
\end{tabular}
\\
\end{center}

\noindent
Here it is important to realize that both one loop coefficients are known as are both $H^{0} \X^{6}$ coefficients (of which one of them is zero). So, at this stage, \textit{we are only left with four unknown coefficients}. Of course an unambiguous determination of these four coefficients can only be achieved via a string theory calculation as the one we have set up in the present paper. We shall next see that, when considering Myers dielectric effect, the situation gets even better. In such a context there are only three coordinate indices, making the two $H^{3} \X^{3}$ tensor structures to be equal, and likewise for the two $H^{2} \X^{4}$ structures. Thus, \textit{when studying quantum corrections to the dielectric effect we are only left with two unknown coefficients}.


\subsection{Matrix Model for Quantum Dielectric Branes}


We shall now concentrate in the seminal example of \cite{Myers} (see also \cite{ARS-1, ARS-2}). For Myers' dielectric effect we only need to select three coordinate matrices, with indices $\{ I, J, K, \dots \} \equiv \{ i, j, k, \dots \} = \{ 1, 2, 3 \}$. This is of course in strict correspondence with the example of the $SU(2)$ WZW model at level $k$ \cite{ARS-1, ARS-2}. In this case $H_{ijk} = H\ \epsilon_{ijk}$, where $\epsilon_{ijk}$ is the totally antisymmetric Levi--Civita symbol, and where $H \sim \frac{1}{\sqrt{\a k}}$. Let us also stress once again that due to the RNC expansion of target tensors, all background tensors which appear in the matrix model are $c$--numbers.

In this three dimensional situation both the $H^{3} \X^{3}$ and the $H^{2} \X^{4}$ monomials collapse to only two independent tensor structures. It simply follows that

\begin{eqnarray*}
H_{aij} H_{bik} H_{cjk} &=& H^{3} \epsilon_{aij} \epsilon_{bik} \epsilon_{cjk} = H^{3} \epsilon_{aij} \left( \delta_{bc} \delta_{ij} - \delta_{bj} \delta_{ic} \right) = - H^{3} \epsilon_{acb} = H^{3} \epsilon_{abc}, \\
H_{abi} H_{cjk} H_{ijk} &=& H^{3} \epsilon_{abi} \epsilon_{cjk} \epsilon_{ijk} = H^{3} \epsilon_{abi} 2 \delta_{ci} = 2 H^{3} \epsilon_{abc}, \\
H^{2} H_{abc} &=& H^{3} \epsilon_{ijk} \epsilon_{ijk} \epsilon_{abc} = 3! H^{3} \epsilon_{abc}, \\
H_{abi} H_{icd} &=& H^{2} \epsilon_{abi} \epsilon_{icd} = H^{2} \left( \delta_{ac} \delta_{bd} - \delta_{ad} \delta_{bc} \right), \\
H_{aij} H_{bij} &=& H^{2} \epsilon_{aij} \epsilon_{bij} = 2 H^{2} \delta_{ab}, \\
H^{2} &=& H^{2} \epsilon_{ijk} \epsilon_{ijk} = 3! H^{2}.
\end{eqnarray*}

\noindent
This means that we are left with only two unknown coefficients for the matrix model which describes $\a$ corrections to the dielectric brane solutions of open string theory. These are the coefficients associated to the following two curvature couplings

$$
H^{3}\ \epsilon_{ijk}\ {\mathbf{Tr}} [ \X^{i}, \X^{j} ] \X^{k} \qquad {\mathrm{and}} \qquad H^{2}\ {\mathbf{Tr}} [ \X^{i}, \X^{j} ] [ \X^{i}, \X^{j} ].
$$

Let us finally, carefully understand how everything comes together in the construction of the matrix model for quantum dielectric branes. First, recall that for the one loop Myers matrix model the energy consists of two tensor couplings  with well known coefficients (we shall explicitly display powers of $\a$) \cite{Myers},

$$
V_{\mathrm{eff}}^{(1)} [\X] = - \frac{1}{4(\a)^2} {\mathbf{Tr}} [ \X^{i}, \X^{j} ] [ \X^{i}, \X^{j} ] + \frac{i}{6\a} H \epsilon_{ijk} {\mathbf{Tr}} [ \X^{i}, \X^{j} ] \X^{k}.
$$

\noindent
The equations of motion which follow are

$$
[ \X^{i}, [ \X^{i}, \X^{a} ] ] + \frac{i}{2} \a H \epsilon_{aij} [ \X^{i}, \X^{j} ]  = 0,
$$

\noindent
where an \textit{ansatz} for a fuzzy solution of the type $[ \X^i, \X^j ] = A\, \epsilon_{ijk}\, \X^k$ leads to 

$$
- 2 A^2 \delta_{ai} \X^{i} + \frac{i}{2} \a 2 A H \delta_{ai} \X^i = 0 \qquad \Leftrightarrow \qquad A = 0 \quad \vee \quad A = \frac{i}{2} \a H.
$$

\noindent
Picking the non--trivial solution, it is simple to compute its energy as $V_{\mathrm{eff}} [\X_{\mathrm{fuzzy}}] = - \frac{H^{2}}{24} {\mathbf{Tr}} \X^i \X^i < 0$, so that the stable, minimal energy, solution corresponds to the famous Myers' dielectric effect \cite{Myers}.

Proceeding at two loops, one has three tensor couplings. The $H^{0} \X^{6}$ coupling has well known coefficient, fixed via the nonabelian BI action for the bosonic string (see the appendix). The other two tensor couplings have undetermined coefficients. One can write the two loop contribution to the quantum matrix model as

$$
V^{(2)}_{\mathrm{eff}} [\X] = -\frac{1}{6\pi^{2}(\a)^3} {\mathbf{Tr}} [ \X^{i}, \X^{j} ] [ \X^{j}, \X^{k} ] [ \X^{k}, \X^{i} ] - \frac{\kappa_1}{4\a} H^{2} {\mathbf{Tr}} [ \X^{i}, \X^{j} ] [ \X^{i}, \X^{j} ] + \frac{i\kappa_2}{6} H^{3} \epsilon_{ijk} {\mathbf{Tr}} [ \X^{i}, \X^{j} ] \X^{k}
$$

\noindent
with unknown coefficients $\kappa_1$ and $\kappa_2$. These coefficients can only be obtained once the beta function programme is carried out to its very end. The energy thus becomes

\begin{eqnarray}
V_{\mathrm{eff}} [\X] &=& - \frac{1}{4(\a)^2} \left( 1 + \a \kappa_1 H^2 \right) {\mathbf{Tr}} [ \X^{i}, \X^{j} ] [ \X^{i}, \X^{j} ] + \frac{i}{6\a} \left( H + \a \kappa_2 H^3 \right) \epsilon_{ijk} {\mathbf{Tr}} [ \X^{i}, \X^{j} ] \X^{k} + \nonumber \\
&&
-\frac{1}{6\pi^{2}(\a)^3} {\mathbf{Tr}} [ \X^{i}, \X^{j} ] [ \X^{j}, \X^{k} ] [ \X^{k}, \X^{i} ].
\end{eqnarray}

\noindent
One can moreover compute equations of motion as usual and, in this case, it is not too hard to get

\begin{equation}
\left( 1 + \a \kappa_1 H^2 \right) [ \X^i, [ \X^i, \X^a ] ] + \frac{i}{2} \a \left( H + \a \kappa_2 H^3 \right) \epsilon_{aij} [ \X^i, \X^j ] - \frac{1}{2\pi^2\a} [ \X^i, [ [ \X^i, \X^j ], [ \X^j, \X^a ] ] ] = 0.
\end{equation}

\noindent
This is our final result for the quantum matrix model describing the first non--trivial curvature corrections to Myers' dielectric effect. It is definitely clear that, in the end, the answer is much simpler than we first expected when set out to do this calculation. On the other hand, the programme we have spelled in this work still needs completion, as one should soon address the question of determining the unknown coefficients $\kappa_1$ and $\kappa_2$. A parallel path of research should also try to obtain solutions to the above quantum matrix equations, possibly from a quantum group deformed fuzzy sphere point view (perhaps along the lines of \cite{PS-1, PS-2, PS-3}). The one thing which remains clear is that there is still much to learn on what concerns the physics of dielectric branes.


\subsection{Mapping Nonabelian to Abelian Branes}


In order to set the pace for the full calculation we have described, regarding two loop tensor structures in the nonabelian BI action, we shall conclude this section by exemplifying how to compute some of these structures in the flat space limit. The final calculation of all structures, thoroughly explained earlier, will be left for future work. We shall begin with the nonabelian result, at the level of the equations of motion, and obtain an abelian result which is necessarily non--perturbative in $\F$. These abelian tensor structures---derived from the nonabelian action---must be present in the two loop abelian equations of motion. Comparison of these terms with the abelian equations of motion then yields the required coefficients in the nonabelian matrix action, as we have previously described at length.

The procedure is similar to the one at the beginning of this section, where we have addressed the quartic Yang--Mills term in the nonabelian action. Indeed, it was quite simple to write the matrix equations of motion associated to this term at large $F$, as (recall (\ref{ymnaeom})),

$$
\frac{1}{\left( \alpha' \right)^{2}} \left[ \X^{M}, \left[ \X^{M}, \X^{N} \right] \right] = - \frac{1}{\left( \alpha' \right)^{2}}\ \theta^{MK} \partial_{K} \theta^{MN} + \cdots = 0.
$$

\noindent
In order to obtain the abelian results (rather than the nonabelian, as we exemplified earlier) one now needs to re--write $\theta$--derivatives as $F$--derivatives, obtaining

$$
\frac{1}{\left( \alpha' \right)^{2}}\ \theta^{NT} \theta^{RM} \theta^{MS} \partial_{R} F_{ST} + \cdots = 0,
$$

\noindent
and finally, making use of the limits (\ref{openlimits}), one can conjecture the abelian term in the equations of motion as being (this is not guaranteed to be the exact result to all orders in $F$, as our procedure only fixes the large $F$ behavior uniquely)

$$
- 4 \pi^{2}\ \Theta^{NT} \left( \frac{1}{G} \right)^{RS} \partial_{R} \F_{ST} + \cdots = 0.
$$

\noindent
Of course at one loop all is simple. At two loops the structures get more intricate, but the process of computation is the same. In flat space, the nonabelian action is known (see the appendix),

$$
V_{\mathrm{eff}} [\X] = - \frac{1}{4 \left( \a \right)^2} {\mathbf{Tr}} \left[ \X^{I}, \X^{J} \right] \left[ \X^{I}, \X^{J} \right] - \frac{1}{6 \pi^2 \left( \a \right)^3} {\mathbf{Tr}}  \left[ \X^{I}, \X^{J} \right] \left[ \X^{J}, \X^{K} \right] \left[ \X^{K}, \X^{I} \right],
$$

\noindent
with the matrix equations of motion,

$$
\left[ \X^{M}, \left[ \X^{M}, \X^{K} \right] \right] - \frac{1}{2\pi^{2}\a} \left[ \X^{M}, \left[ \left[ \X^{M}, \X^{N} \right], \left[ \X^{N}, \X^{K} \right] \right] \right] = 0.
$$

\noindent
Following the exact same procedure as before (albeit via a much cumbersome calculation), one can then conjecture the contribution of the previous two loop structure to the abelian equations of motion as being:

\begin{eqnarray*}
&&
\frac{1}{2\pi^{2}\left( \alpha' \right)^{3}} \left[ \X^{M}, \left[ \left[ \X^{M}, \X^{N} \right], \left[ \X^{N}, \X^{S} \right] \right] \right] = 8 \pi^{2} \alpha'\ \Theta^{S\bar{S}} \left( \left( \frac{1}{G} \right)^{M\bar{M}} \left( \frac{1}{G} \right)^{N\bar{N}} \Theta^{AB}\ \partial_{M} \Big( \partial_{A} \F_{\bar{M}N}\ \partial_{B} \F_{\bar{N}\bar{S}} \Big) + \right. \\
&&
- \left( \frac{1}{G} \right)^{M\bar{M}} \left( \frac{1}{G} \right)^{N\bar{N}} \Theta^{A\bar{A}}\ \Theta^{B\bar{B}}\ \Big( \partial_{M} \F_{\bar{A}B}\ \partial_{A} \F_{\bar{M}N}\ \partial_{\bar{B}} \F_{\bar{N}\bar{S}} + \partial_{\bar{A}} \F_{\bar{N}\bar{S}}\ \partial_{A} \F_{\bar{B}N}\ \partial_{M} \F_{\bar{M}B} + \\
&&
+ \partial_{\bar{A}} \F_{\bar{N}\bar{S}}\ \partial_{A} \F_{\bar{M}B}\ \partial_{M} \F_{\bar{B}N} + \partial_{A} \F_{\bar{M}N}\ \partial_{\bar{A}} \F_{\bar{B}\bar{S}}\ \partial_{M} \F_{\bar{N}B} + \partial_{A} \F_{\bar{M}N}\ \partial_{\bar{A}} \F_{\bar{N}B}\ \partial_{M} \F_{\bar{B}\bar{S}} \Big) \bigg) + \cdots.
\end{eqnarray*}

\noindent
This is an amusing expression as it can also yield a conjecture on how to obtain the Weyl anomaly coefficients at two loops, in flat space. Indeed, if one writes the expression for the two loop beta function, (\ref{beta2}), in flat space and subtracts it from the result above (which is in fact a conjecture on the final expression for the two loop Weyl anomaly coefficients), then, and according to (\ref{bwac}), whatever is left is the completion of the beta function to yield the Weyl anomaly coefficients. As we have previously explained at greater length, computing this expression from first principles in sigma model theory (or in boundary string field theory), is the most pressing problem concerning the full calculation of the nonabelian BI action.


\section{Future Directions}\label{conclusions}


Of course the most pressing question arising from our work concerns the full abelian equations of motion. The beta function we have computed is the required first step, but one will eventually have to address either the Weyl anomaly coefficients or the partition function, as we have explained in the text. Only after this is done our programme can be brought to conclusion, following the guidelines we have spelled out in the paper. Much remains to be learned from the nonabelian BI action in curved space, and in particular in what concerns $\alpha'$ corrections to the dielectric effect. Parallel to this, we believe that other lines of research can follow along the lines presented in this work, and end with a few suggestions for future research:

\begin{itemize}
    
    \item One obvious generalization concerns 
    the supersymmetric case. This is not too hard as the RNS sigma 
    model is well known to order ${\mathcal{O}}(H^{2})$ \cite{BCZ} and 
    one could easily perform the same type of calculation we perform 
    in here, for the quantum dielectric branes of type II string theory. 
    In this regard, the result in \cite{Wijnholt} may be particularly 
    relevant, as the $R^{2}$ open string effective action is 
    presented to all orders in the boundary field $\F$. It would be 
    interesting to investigate what sort of corrections one obtains in 
    superstring theory.
    
    \item Another reason why the study of superstrings is of interest 
    is, of course, the study of the quantum dielectric effect due to the 
    presence of RR background fields (this was the original example 
    in \cite{Myers}). At first sight this may seem like a hard 
    problem, as RR fields are notoriously difficult to handle. 
    However, the recent work in \cite{CCS} may be of help as it lays 
    the ground to a noncommutative analysis (as the one we perform 
    here) in the realm of RR string theory backgrounds.
    
    \item Once we are extending these results to other situations, one 
    which is of obvious interest is for the dielectric effect on the 
    fundamental string itself. This was studied in the context of 
    matrix string theory \cite{DVV} in \cite{Schiappa, BJL}. It would 
    be interesting to compute quantum corrections to those results and 
    investigate what sort of deformations they yield on $F$--strings.
    
    \item A harder problem would be to tackle the full BCFT. 
    Here we only point out that the extension of the results in 
    \cite{Cornalba-Schiappa-2} to higher orders in $H$ may provide for 
    a full algebraic description of the open string massless fields 
    embedded in curved closed string backgrounds via deformation 
    theory (\textit{i.e.}, in terms of traces and star products).
    
    \item The method described in this paper is, in a certain sense, a 
    $T$--duality short--cut, as one is moving from a maximal brane to 
    a minimal brane in one unique step. It would be 
    interesting to use the results in this paper to derive $\a$ 
    corrections to the open string $T$--duality rules. This would be 
    helpful, for instance, in generalizing the beta function we 
    have obtained for the maximal $D$--brane to arbitrary $Dp$--branes.
    
    \item If one would obtain such $\a$ corrected $T$--duality rules, 
    one could then envisage in extending this action to non--conformal 
    backgrounds. $T$--duality transformations of closed string fields 
    in non--conformal backgrounds turn out to be compatible with 
    renormalization group (RG) flows of the two dimensional sigma model 
    field theory \cite{Haagensen, Haagensen-Olsen, HOS}. Extending 
    this work to open strings is harder \cite{DO2, Olsen-Schiappa}, 
    but it could prove of interest to generalize these works to 
    higher orders: a possible solution, at all orders, for the $T$--duality 
    rules would, using the compatibility of duality and the RG flow, 
    possibly allow one to uniquely determine the solution, at all orders, for 
    the beta function equations of motion.

\end{itemize}

\section*{Acknowledgements}
PB wishes to thank S. Ribault and K. Graham for useful discussions. RS would like to thank Troels Harmark and Niclas Wyllard for useful discussions and comments at different stages of this project. PB is supported by the grant SFRH/BD/799/2000 (Portugal). LC is supported by a Marie Curie fellowship, under the European Commission's Improving Human Potential Programme (HPMF--CT--2002--02016). RS is supported in part by funds provided by the Funda\c c\~ao para a Ci\^encia e a Tecnologia, under the grant SFRH/BPD/7190/2001 (Portugal). LC and RS are supported in part by funds provided by the Funda\c c\~ao para a Ci\^encia e a Tecnologia, under the grant POCTI/FNU/38004/2001 (Portugal).

\vfill

\eject

\appendix


\section{Perturbative Expansions}


In order to set up a two dimensional sigma model perturbation 
expansion of the quantum fields in the open string action, some sort of 
geometrical expansion is required on the target manifold fields (the sigma 
model couplings), both for the closed and the open string sectors. The 
standard expansion employs the background field method \cite{BCZ, DO, 
ACNY, CLNY}. There is, however, a different method based on the radial 
gauge expansion of the target fields (the Riemann normal coordinate 
(RNC) expansion for the metric) \cite{MSV, Hatzinikitas}, which in some 
cases is better suited to the problem at hand (see, \textit{e.g}, 
\cite{Cornalba-Schiappa-2}). We wish to explain both methods in here, as 
applied to our problem. In the paper, we shall use the background field 
method in the main computation, but the RNC expansion 
also proves to be of interest for subsequent calculations.

Let us begin with a bosonic string action in a background metric 
$g_{\mu\nu} \left( X \right)$ and NSNS 2--form field $B_{\mu\nu} \left( 
X \right)$,

\begin{equation}\label{Aclosed}
S = \frac{1}{4\pi \alpha'} \int_{\Sigma} g_{\mu\nu}\left( X \right) 
dX^{\mu} \wedge \ast dX^{\nu} + \frac{i}{4\pi \alpha'} \int_{\Sigma} 
B_{\mu\nu} \left( X \right) dX^{\mu} \wedge dX^{\nu},
\end{equation}

\noindent
where $\ast$ is the Hodge dual on the worldsheet $\Sigma$. The factor 
of $i$ signals a choice of euclidean signature on the worldsheet. This 
action is to be supplemented with a boundary term as we are describing the 
open bosonic string action,

\begin{equation}\label{Aopen}
S_{B} = i \oint_{\partial\Sigma} A_{\mu} \left( X \right) dX^{\mu}.
\end{equation}

\noindent
Throughout the paper we shall consider backgrounds with constant dilaton field, $\Phi = \mathrm{constant}$, and will disregard this field. This we can do as we will be working in maximally symmetric target spaces. The background equations of motion for the closed bosonic string, to first order in $\alpha'$, are

$$
R_{\mu\nu} + \frac{1}{4} H_{\mu\lambda\sigma} 
{H^{\lambda\sigma}}_{\nu} = 0, \quad \nabla^{\lambda} 
H_{\lambda\mu\nu} = 0.
$$

\noindent
The connection which is mostly used is the one emerging from the equations
of motion of the worldsheet string action, which is $\hat{\Gamma} = \Gamma 
\pm \frac{i}{2} H$ (we shall consider the plus sign in the following). 
The curvature of the $\hat{\Gamma}$ connection follows as

$$
{\mathcal{R}}_{\mu\nu\rho\sigma} = R_{\mu\nu\rho\sigma} + \frac{i}{2} 
\nabla_{\rho} H_{\sigma\mu\nu} - \frac{i}{2} \nabla_{\sigma} 
H_{\rho\mu\nu} + \frac{1}{4} H_{\mu\rho\lambda} {H^{\lambda}}_{\nu\sigma}
- \frac{1}{4} H_{\mu\sigma\lambda} {H^{\lambda}}_{\nu\rho}. 
$$

\noindent
For parallelizable manifolds (parallelizable by inclusion of torsion, which we pick to be the field $H$), the $\hat{\Gamma}$ connection is flat (but torsionfull). In this case one has ${\mathcal{R}}_{\mu\nu\rho\sigma} = 0$, which implies

$$
R_{\mu\nu\rho\sigma} + \frac{1}{4} H_{\mu\nu\lambda} 
{H^{\lambda}}_{\rho\sigma} = 0, \quad \nabla_{\mu} H_{\nu\rho\sigma} = 0, 
\quad H_{\lambda[\mu\nu} {H^{\lambda}}_{\rho\sigma]} = 0.
$$

\noindent
Here $\nabla$ is the covariant derivative with respect to the Levi--Civita connection. This solves the closed string background equations of motion to all orders, in both the bosonic and supersymmetric cases \cite{BCZ}, so that the closed string fields will be on--shell throughout, as we analyze open strings in this class of backgrounds. Parallelizable manifolds are a large class of manifolds which include in particular all Lie groups (of course with the same choice of torsion). This means that our results will hold for generic WZW models.

While the most common expansion of the metric field in covariant tensors is probably the RNC expansion, the background field method is the one we shall employ and to which we first turn. The fourth order background field expansion for the worldsheet bulk fields was done in \cite{BCZ} and later generalized to worldsheet boundary fields in \cite{DO, Andreev-Tseytlin}, where it was precisely applied in beta function calculations. We shall briefly review this method and show how to apply it to the open string sigma model (\ref{Aclosed}) and (\ref{Aopen}) to quartic order in the fluctuations (which is the required 
expansion in order to perform the two loop calculation of the open string 
beta function in closed string background fields).

We shall take for worldsheet the disk $\Sigma=\mathbb{D}$, parameterized 
by polar coordinates and with a flat euclidean metric. Greek indices are assigned to spacetime, latin indices are for the worldsheet and $\epsilon_{01} = 1$. The aim of the background field expansion is to maintain target general coordinate covariance explicit both at the level of the sigma model local couplings and at the level of the fluctuating field (which will become a tangent vector in target space). The two dimensional quantum field $X^{\mu}$ is written as a perturbation of a fixed classical background field $x^{\mu} (\sigma)$ by a quantum fluctuating field $\xi^{\mu} (\sigma)$, such that $X^{\mu} (\sigma) = x^{\mu} (\sigma) + \xi^{\mu} (\sigma)$. The idea is then to change coordinates to the tangent vector along the geodesic linking $x$ to $\xi$: consider the geodesic from the background field $x$ to the background field plus fluctuation $x+\xi$, which will be denoted by $\rho(s)$ (with $s \in [0,1]$) such that $x=\rho(s=0)$ and $x+\xi=\rho(s=1)$. The tangent 
vector to this geodesic,

$$
\zeta^{\mu}(s) = \frac{d}{ds} \rho^{\mu}(s),
$$

\noindent
at $s=0$ is a target manifold vector field. This is the quantum field in 
powers of which we shall expand the action. Because it is a precise 
geometrical entity in target space, one is guaranteed that the 
perturbative expansion will be in terms of target tensor fields 
(evaluated at the classical background, $x$). It turns 
out that, at least up to order ${\mathcal{O}}(\zeta^{4})$, this 
expansion can be written largely in terms of the generalized Riemann 
tensor ${\mathcal{R}}_{\mu\nu\rho\sigma}$.
The procedure to carry out such an expansion is to consider the 
action (\ref{Aclosed}) as depending on the geodesic field,

\begin{equation}\label{Aclosed-parameter}
S [\rho(s)] = \frac{1}{4\pi \alpha'} \int_{\Sigma} g_{\mu\nu}\left[ 
\rho(s) \right] d\rho^{\mu}(s) \wedge \ast d\rho^{\nu}(s) + \frac{i}{4\pi 
\alpha'} \int_{\Sigma} B_{\mu\nu} \left[ \rho(s) \right] 
d\rho^{\mu}(s) \wedge d\rho^{\nu}(s),
\end{equation}

\noindent
such that the diverse terms in the perturbative expansion of the 
action (given the split of the field into background and 
fluctuation), as $S = S^{(0)} + S^{(1)} + S^{(2)} + \cdots$, will be given 
by their power series expansion coefficients

$$
S^{(n)} = \left. \frac{1}{n!} \frac{d^{n}}{ds^{n}} S[\rho(s)] 
\right|_{s=0}.
$$

\noindent
The exact same reasoning naturally applies to the boundary piece of 
the open string action,

\begin{equation}\label{Aopen-parameter}
S_{B} [\rho(s)] = i \oint_{\partial\Sigma} A_{\mu} \left[ \rho(s) \right] 
d\rho^{\mu}(s).
\end{equation}

\noindent
For the closed string worldsheet, this expansion is carefully carried 
out in the appendix of \cite{BCZ}. For the open string one needs to 
take care with partial integrations in order not to miss any boundary 
contributions. This has been carried out in \cite{DO} to quadratic 
order and later in \cite{Andreev-Tseytlin} to quartic order but in 
flat target backgrounds. Here we wish to obtain the full open string 
action to fourth order in curved backgrounds. Let us first 
concentrate on (\ref{Aclosed-parameter}), in particular in the 
boundary terms that will be generated from this worldsheet bulk 
contribution.

At zeroth order one obtains just the classical constant action (here 
and in the following, all target fields are evaluated at $x$),

$$
S^{(0)} = \frac{1}{4\pi \alpha'} \int_{\Sigma} g_{\mu\nu} dx^{\mu} 
\wedge \ast dx^{\nu} + \frac{i}{4\pi \alpha'} \int_{\Sigma} B_{\mu\nu} 
dx^{\mu} \wedge dx^{\nu}. 
$$

\noindent
At first order the action is linear in $\zeta^{\mu}$, and includes a 
boundary contribution from a partial integration,

$$
S^{(1)} [\zeta] = - \frac{1}{2\pi \alpha'} \int_{\Sigma} d\sigma d\tau\ 
g_{\mu\nu} \left( \mathcal{D}_{a} \partial^{a} x \right)^{\mu} \zeta^{\nu} + 
\frac{1}{2\pi \alpha'} \oint_{\partial\Sigma} d\tau \left( - g_{\mu\nu}
\partial_{\sigma} x^{\nu} + i B_{\mu\nu} \partial_{\tau} x^{\nu} \right) 
\zeta^{\mu},
$$

\noindent
where we have introduced worldsheet coordinates $\{\tau,\sigma\}$, 
derivatives $\mathcal{D}_{a} \zeta^{\mu} = D_{a} \zeta^{\mu} + \frac{i}{2} 
{H^{\mu}}_{\sigma\rho} \epsilon_{ab} \partial^{b} x^{\sigma} \zeta^{\rho}$ 
and $D_{a} \zeta^{\mu} = \partial_{a} \zeta^{\mu} + 
\Gamma^{\mu}_{\sigma\rho} \partial_{a} x^{\sigma} \zeta^{\rho}$, and 
where (in order to cancel tadpoles) the bulk piece yields the background 
equations of motion, $\mathcal{D}_{a} \partial^{a} x^{\mu} = 0$, while the 
boundary piece yields the boundary conditions on the background field, 
$\left. \left( - g_{\mu\nu} \partial_{\sigma} x^{\nu} + i B_{\mu\nu} 
\partial_{\tau} x^{\nu} \right) \right|_{\partial\Sigma} = 0$. 
Likewise at quadratic order one obtains the expected bulk 
contribution \cite{BCZ} with a new boundary term,

$$
S^{(2)}_{\mathrm{WZW}} [\zeta] = \frac{1}{4\pi\alpha'} \int_{\Sigma} 
d\sigma d\tau \left( g_{\mu\nu} \mathcal{D}_{a} \zeta^{\mu} \mathcal{D}^{a} 
\zeta^{\nu} \right) + \frac{i}{4\pi\alpha'} \oint_{\partial\Sigma} 
d\tau \left( \nabla_{\sigma} B_{\mu\nu} \zeta^{\sigma} \zeta^{\mu} 
\partial_{\tau} x^{\nu} + B_{\mu\nu} \zeta^{\mu} D_{\tau} 
\zeta^{\nu} \right),
$$

\noindent
In here, $\nabla$ is the target covariant derivative associated to 
the Levi--Civita connection. We have also particularized for the case 
of parallelizable backgrounds where $\mathcal{R}_{\mu\nu\rho\sigma}$
(the curvature of the $\hat{\Gamma}$ connection) vanishes. In this particular case, to third order one obtains the standard bulk term \cite{BCZ} with a new boundary contribution, which we have computed as:

\begin{eqnarray*}
S^{(3)}_{\mathrm{WZW}} [\zeta] &=& \frac{i}{12\pi\alpha'} 
\int_{\Sigma} d\sigma d\tau\ H_{\mu\nu\sigma} \epsilon^{ab} \zeta^{\mu} 
\mathcal{D}_{a} \zeta^{\nu} \mathcal{D}_{b} \zeta^{\sigma} \\
&&+ 
\frac{i}{12\pi\alpha'} \oint_{\partial\Sigma} d\tau \left( 
\nabla_{\lambda} \nabla_{\sigma} B_{\mu\nu} - \frac{1}{4} 
B_{\lambda\alpha} {H^{\alpha}}_{\sigma\rho} {H^{\rho}}_{\mu\nu} \right) 
\zeta^{\lambda} \zeta^{\sigma} \zeta^{\mu} \partial_{\tau} x^{\nu} + 
\frac{i}{6\pi\alpha'} \oint_{\partial\Sigma} d\tau\ \nabla_{\lambda} 
B_{\mu\nu} \zeta^{\lambda} \zeta^{\mu} D_{\tau} \zeta^{\nu}.
\end{eqnarray*}

\noindent
Proceeding to fourth order (which is required in order to compute the 
full two loop beta function), we have computed the following contribution 
(where again the bulk piece had been previously computed in \cite{BCZ})

\begin{eqnarray*}
S^{(4)}_{\mathrm{WZW}} [\zeta] &=& \frac{1}{48\pi\alpha'} \int_{\Sigma} 
d\sigma d\tau\ H_{\mu\nu\lambda} {H^{\lambda}}_{\rho\sigma} 
\zeta^{\mu} \zeta^{\rho} \mathcal{D}_{a} \zeta^{\nu} \mathcal{D}^{a} 
\zeta^{\sigma} \\
&&
+ \frac{i}{48\pi\alpha'} \oint_{\partial\Sigma} d\tau \left( 
\nabla_{\eta} \nabla_{\lambda} \nabla_{\sigma} B_{\mu\nu} - \frac{3}{4} 
\nabla_{\eta} B_{\lambda\beta} {H^{\beta}}_{\sigma\rho} 
{H^{\rho}}_{\mu\nu} \right) \zeta^{\eta} \zeta^{\lambda} \zeta^{\sigma} 
\zeta^{\mu} \partial_{\tau} x^{\nu} \\
&&
+ \frac{i}{16\pi\alpha'} 
\oint_{\partial\Sigma} d\tau \left( \nabla_{\lambda} \nabla_{\sigma} 
B_{\mu\nu} - \frac{1}{12} B_{\lambda\alpha} {H^{\alpha}}_{\sigma\rho} 
{H^{\rho}}_{\mu\nu} \right) \zeta^{\lambda} \zeta^{\sigma} \zeta^{\mu} 
D_{\tau} \zeta^{\nu}.
\end{eqnarray*}

\noindent
This concludes the analysis of the expansion of 
(\ref{Aclosed-parameter}) to the order we need in order to perform our 
beta function calculation. One still needs to address the open string 
action (\ref{Aopen-parameter}). This term is expanded in the exact 
same way as the previous one so that we shall limit ourselves to 
presenting the results. We have performed the expansion in a 
completely generic (parallelizable or not) background and found:

\begin{eqnarray*}
S_{B}^{(0)} &=& i \oint_{\partial\Sigma} d\tau\ A_{\mu} \partial_{\tau} 
x^{\mu}, \\
S_{B}^{(1)} [\zeta] &=& i \oint_{\partial\Sigma} d\tau\ F_{\mu\nu} 
\zeta^{\mu} \partial_{\tau} x^{\nu}, \\
S_{B}^{(2)} [\zeta] &=& \frac{i}{2} \oint_{\partial\Sigma} d\tau 
\left( \nabla_{\sigma} F_{\mu\nu} \zeta^{\sigma} \zeta^{\mu} 
\partial_{\tau} x^{\nu} + F_{\mu\nu} \zeta^{\mu} D_{\tau} \zeta^{\nu} 
\right), \\
S_{B}^{(3)} [\zeta] &=& \frac{i}{6} \oint_{\partial\Sigma} d\tau 
\left( \nabla_{\lambda} \nabla_{\sigma} F_{\mu\nu} + 
F_{\lambda\alpha} {R^{\alpha}}_{\sigma\mu\nu} \right) 
\zeta^{\lambda} \zeta^{\sigma} \zeta^{\mu} \partial_{\tau} x^{\nu} + 
\frac{i}{3} \oint_{\partial\Sigma} d\tau\ \nabla_{\lambda} F_{\mu\nu} 
\zeta^{\lambda} \zeta^{\mu} D_{\tau} \zeta^{\nu},
\end{eqnarray*}

\noindent
and

\begin{eqnarray*}
S_{B}^{(4)} [\zeta] &=& \frac{i}{24} \oint_{\partial\Sigma} d\tau 
\left( \nabla_{\eta} \nabla_{\lambda} \nabla_{\sigma} F_{\mu\nu} + 3 
\nabla_{\eta} F_{\lambda\beta} {R^{\beta}}_{\sigma\mu\nu} + 
F_{\lambda\beta} \nabla_{\eta} {R^{\beta}}_{\sigma\mu\nu} \right) 
\zeta^{\eta} \zeta^{\lambda} \zeta^{\sigma} \zeta^{\mu} \partial_{\tau} 
x^{\nu} \\
&&
+ \frac{i}{8} \oint_{\partial\Sigma} d\tau \left( \nabla_{\lambda} 
\nabla_{\sigma} F_{\mu\nu} + \frac{1}{3} F_{\lambda\alpha} 
{R^{\alpha}}_{\sigma\mu\nu} \right) \zeta^{\lambda} \zeta^{\sigma} 
\zeta^{\mu} D_{\tau} \zeta^{\nu}.
\end{eqnarray*}

\noindent
When specialized to parallelizable backgrounds and added to the 
previous bulk plus boundary expansion, it is simple to check that it 
just amounts to the replacement $B_{\mu\nu} \to B_{\mu\nu} + 
2\pi\alpha' F_{\mu\nu}$ in the WZW action, which is precisely what one 
expects from generic $D$--brane arguments. It is this resulting 
perturbative expansion that we use throughout the text for the beta function computation.

This is the essence of the background field expansion. However, once one has managed to carry out the string perturbative calculations, it is still necessary to expand the target tensors in order to obtain nonabelian matrix actions. This is now a purely target space geometric problem, where the most common expansion of the metric field in covariant tensors is probably the RNC expansion, defined via the radial gauge prescription for the metric field,

$$
X^{\mu} g_{\mu\nu} (X) = X^{\mu} g_{\mu\nu} (0),
$$

\noindent
or equivalently, $X^{\mu} X^{\nu} \Gamma^{\sigma}_{\mu\nu} (X) = 0$. We shall now turn to analyze this expansion method in the backgrounds we are interested in. This radial gauge prescription can be applied to the NSNS $B$--field as (see, \textit{e.g.}, \cite{Cornalba-Schiappa-2}),

$$
X^{\mu} B_{\mu\nu} (X) = X^{\mu} B_{\mu\nu} (0).
$$

\noindent
While for the metric field the solution of the radial gauge equation is a
bit more complicated (see, \textit{e.g.}, \cite{MSV}), for the NSNS field
one can solve in $H=dB$ as \cite{Cornalba-Schiappa-2}

$$
B_{\mu\nu} (X) = B_{\mu\nu} (0) + X^{\sigma} \int_{0}^{1} s^{2} 
H_{\mu\nu\sigma} (sX) ds.
$$

\noindent
For the metric field there is an algorithmic process that computes the RNC
expansion to any desired order, \textit{e.g.}, \cite{MSV,
Hatzinikitas}. The expansion to quartic order in the coordinates is (see 
\cite{MSV, Hatzinikitas} for higher order terms),

\begin{eqnarray}
g_{\mu\nu}(X) &=& g_{\mu\nu} + \frac{1}{3} R_{\mu\alpha\beta\nu} 
X^{\alpha} X^{\beta} + \frac{1}{6} \nabla_{\alpha} R_{\mu\beta\gamma\nu} 
X^{\alpha} X^{\beta} X^{\gamma} + \frac{1}{20} \nabla_{\alpha} 
\nabla_{\beta} R_{\mu\gamma\delta\nu} X^{\alpha} X^{\beta} X^{\gamma} 
X^{\delta} + \nonumber \\
&&
+ \frac{2}{45} R_{\mu\alpha\beta\eta} {R^{\eta}}_{\gamma\delta\nu} X^{\alpha} 
X^{\beta} X^{\gamma} X^{\delta} + \cdots. \nonumber
\end{eqnarray}

\noindent
One can likewise obtain an expansion for the Levi--Civita connection 
coefficients, which is as follows:

\begin{eqnarray*}
\Gamma^{\kappa}_{\mu\nu} (X) &=& - \frac{1}{3} \Big( 
{R^{\kappa}}_{\mu\nu\alpha} + {R^{\kappa}}_{\nu\mu\alpha} \Big) X^{\alpha} 
- \frac{1}{6} \Big( \nabla_{\alpha} {R^{\kappa}}_{\mu\nu\beta} + 
\nabla_{\alpha} {R^{\kappa}}_{\nu\mu\beta} - \frac{1}{2}\ \nabla_{\mu} 
{R^{\kappa}}_{\alpha\beta\nu} - \frac{1}{2}\ \nabla_{\nu} 
{R^{\kappa}}_{\alpha\beta\mu} + \\
&&
- \frac{1}{2}\ \nabla^{\kappa} R_{\mu\alpha\beta\nu} \Big) X^{\alpha} 
X^{\beta} + \cdots.
\end{eqnarray*}

For the special case of symmetric manifolds where $\nabla_{\lambda} 
R_{\mu\nu\rho\sigma} = 0$ the RNC expansion for the metric field simplifies 
to a closed form expression (see, \textit{e.g.}, \cite{Hatzinikitas}),

\begin{eqnarray}
g_{\mu\nu} (X) &=& g_{\mu\nu} + \frac{1}{2} \sum_{k=1}^{+\infty}\
\frac{2^{2k+2}}{(2k+2)!}\ f_{\mu\sigma_{1}} {f^{\sigma_{1}}}_{\sigma_{2}}
{f^{\sigma_{2}}}_{\sigma_{3}} \cdots {f^{\sigma_{k-1}}}_{\nu} \nonumber \\
&\equiv& g_{\mu\nu} + \frac{1}{2} \sum_{k=1}^{+\infty}\
\frac{2^{2k+2}}{(2k+2)!}\ \left( f_{\mu\nu} \right)^{k}, \nonumber
\end{eqnarray}

\noindent
where we have defined,

$$
f_{\mu\sigma_{1}} = R_{\mu\alpha\beta\sigma_{1}} X^{\alpha} X^{\beta}, 
\quad {f^{\sigma_{i}}}_{\sigma_{i+1}} = 
{R^{\sigma_{i}}}_{\gamma\delta\sigma_{i+1}} X^{\gamma} X^{\delta}, \quad 
{f^{\sigma_{k-1}}}_{\nu} = {R^{\sigma_{k-1}}}_{\lambda\rho\nu} 
X^{\lambda} X^{\rho}.
$$

\noindent
If one further deals with (WZW) parallelizable manifold backgrounds, 
where both $\nabla_{\lambda} R_{\mu\nu\rho\sigma} = 0$ and 
$R_{\mu\nu\rho\sigma} + \frac{1}{4} H_{\mu\nu\lambda} 
{H^{\lambda}}_{\rho\sigma} = 0$, the previous formula can be written as

\begin{eqnarray}
g_{\mu\nu}(X) &=& g_{\mu\nu} + 2 \sum_{k=1}^{+\infty}\ \frac{(-1)^k}{(2k+2)!}\
M_{\mu\sigma_{1}} {M^{\sigma_{1}}}_{\sigma_{2}} {M^{\sigma_{2}}}_{\sigma_{3}}
\cdots {M^{\sigma_{2k-1}}}_{\nu} \nonumber \\
&\equiv& g_{\mu\nu} + 2 \sum_{k=1}^{+\infty}\ \frac{i^{2k}}{(2k+2)!}\ \left( 
M_{\mu\nu} \right)^{2k}, \nonumber
\end{eqnarray}

\noindent
where we have defined the antisymmetric matrix $M_{\mu\nu} \equiv 
H_{\mu\rho\nu} X^{\rho}$, to appear again in the following. In this 
particular case of WZW backgrounds, the Levi--Civita connection 
coefficients can also be simplified to

$$
\Gamma^{\kappa}_{\mu\nu} (X) = \frac{1}{12} \left( 
{H^{\kappa}}_{\mu\lambda} {H^{\lambda}}_{\nu\alpha} + 
{H^{\kappa}}_{\nu\lambda} {H^{\lambda}}_{\mu\alpha} \right) 
X^{\alpha} + \cdots.
$$

\noindent
Observe that in this expression it is not very useful to use the 
matrix $M_{\mu\nu}$. This is clear also from the above expression for 
the metric: when computing the connection coefficients there will 
always be a ``free floating'' $H$ piece, due to the derivatives of 
the metric. One can develop similar methods to handle the NSNS field. The 
main concern here is to obtain a closed form expression valid for (WZW) 
parallelizable backgrounds. If one computes generic symmetrized derivatives 
of $H=dB$ to second order, and in RNC at the origin,

\begin{eqnarray}
\partial_{\rho} H_{\mu\nu\sigma} (0) &=& \nabla_{\rho} H_{\mu\nu\sigma}, 
\nonumber \\
\partial_{(\rho} \partial_{\lambda)} H_{\mu\nu\sigma} (0) &=& 
\nabla_{(\rho} \nabla_{\lambda)} H_{\mu\nu\sigma} + 3 \partial_{(\rho} 
\Gamma^{\eta}_{\lambda)[\mu} H_{\nu\sigma]\eta} = \nabla_{(\rho} 
\nabla_{\lambda)} H_{\mu\nu\sigma} + 
{R^{\eta}}_{(\rho\lambda)[\mu}H_{\nu\sigma]\eta}, \nonumber \\
&\ldots& \nonumber
\end{eqnarray}

\noindent
one can then use the previous definition of $B$ in terms of $H$ to compute 
the expansion to cubic order in the coordinates,

\begin{eqnarray}
B_{\mu\nu} (X) &=& B_{\mu\nu} + \frac{1}{3} H_{\mu\nu\sigma} X^{\sigma} + 
\frac{1}{4} \nabla_{\rho} H_{\mu\nu\sigma} X^{\rho} X^{\sigma} + 
\frac{1}{10} \nabla_{\rho} \nabla_{\lambda} H_{\mu\nu\sigma} X^{\rho} 
X^{\lambda} X^{\sigma} + \nonumber \\
&+& \frac{1}{15} {R^{\eta}}_{\rho\lambda[\mu} H_{\nu]\sigma\eta} 
X^{\rho} X^{\lambda} X^{\sigma} + \cdots.
\nonumber
\end{eqnarray}

\noindent
For the special case of WZW backgrounds, where besides the previously 
stated conditions on the curvature one has $\nabla_{\lambda} 
H_{\mu\nu\sigma} = 0$, and where one can write derivative expansions (in
RNC) for $\partial_{(\mu_{1}} \Gamma^{\eta}_{\mu_{2})\nu}$, 
$\partial_{(\mu_{1}} \partial_{\mu_{2}} \Gamma^{\eta}_{\mu_{3})\nu}$, \ldots, 
and higher symmetrized derivatives \cite{MSV, Hatzinikitas}, it follows that 
one can then compute symmetrized arbitrary derivatives of the $H$ 
field (an odd number of symmetrized derivatives yields zero),

\begin{eqnarray}
\partial_{(\alpha} \partial_{\beta)} H_{\mu\nu\sigma} &=& 
{R^{\lambda}}_{(\alpha\beta)[\mu} H_{\nu\sigma]\lambda}, \nonumber \\ 
\partial_{(\alpha} \partial_{\beta} \partial_{\gamma} \partial_{\eta)} 
H_{\mu\nu\sigma} &=& \frac{3}{5} {R^{\lambda}}_{(\alpha\beta|\rho} 
{R^{\rho}}_{|\gamma\eta)[\mu} H_{\nu\sigma]\lambda} + \frac{2}{3}
{R^{\lambda}}_{(\alpha\beta|\mu} {R^{\rho}}_{|\gamma\eta)\nu} 
H_{\lambda\rho\sigma} + \nonumber \\
&+& \frac{2}{3} {R^{\lambda}}_{(\alpha\beta|\mu} 
{R^{\rho}}_{|\gamma\eta)\sigma} H_{\lambda\nu\rho} + \frac{2}{3}
{R^{\lambda}}_{(\alpha\beta|\nu} {R^{\rho}}_{|\gamma\eta)\sigma} 
H_{\mu\lambda\rho}, \nonumber \\
&\ldots& \nonumber
\end{eqnarray}

\noindent
Finally, one can use the definition of $B$ in terms of $H$ to compute the 
expansion to fifth order in the coordinates, obtaining

$$
B_{\mu\nu} (X) = B_{\mu\nu} - \frac{2}{3!} M_{\mu\nu} - \frac{2}{5!} 
M_{\mu\lambda} {M^{\lambda}}_{\sigma} {M^{\sigma}}_{\nu} - \frac{2}{7!} 
M_{\mu\lambda} {M^{\lambda}}_{\sigma} {M^{\sigma}}_{\rho} {M^{\rho}}_{\ell}
{M^{\ell}}_{\nu} - \cdots.
$$

\noindent
For the (WZW) parallelizable backgrounds one can then guess the
following closed form expression for the expansion of the NSNS field:

$$
B_{\mu\nu}(X) = B_{\mu\nu} - 2 \sum_{k=1}^{+\infty}\ \frac{1}{(2k+1)!}\ 
\left( M_{\mu\nu} \right)^{2k-1}.
$$

\noindent
This concludes the analysis of the closed string sector 
(\ref{Aclosed}). Using these expansions and further expanding the 
quantum field $X^{\mu}$ in a zero mode and a quantum fluctuation, 
$X^{\mu} = x^{\mu} + \zeta^{\mu}$, one can devise a graphical 
procedure to perturbatively compute correlation functions in the open 
string BCFT (this procedure was done to order $\mathcal{O}(H)$ in \cite{Cornalba-Schiappa-2}, which naturally generalizes to all other orders). Observe that the matrix $M_{\mu\nu}$ will also include 
a term in the fluctuating field, $\zeta^{\mu}$, thus generating 
interacting vertices of the two dimensional bulk quantum theory. If one 
wishes to include the boundary gauge field in this expansion one further 
needs to deal with (\ref{Aopen}).

In gauge theory radial gauge is also known as Fock--Schwinger gauge, 
where one aims at explicit covariance in perturbative calculations. 
Just as before, this gauge can be defined via the condition

$$
X^{\mu} A_{\mu} (X) = X^{\mu} A_{\mu} (0),
$$

\noindent
and just like for the $B$ field one can also solve in $F=dA$ as

$$
A_{\mu} (X) = A_{\mu} (0) + X^{\nu} \int_{0}^{1} s F_{\mu\nu} (sX) 
ds. 
$$

\noindent
One now proceeds as before. In RNC we know how to compute arbitrary 
symmetrized derivatives of the metric connections, $\partial_{(\mu_{1}} 
\Gamma^{\eta}_{\mu_{2})\nu}$, $\partial_{(\mu_{1}} \partial_{\mu_{2}} 
\Gamma^{\eta}_{\mu_{3})\nu}$, \ldots, \cite{MSV, Hatzinikitas}, so 
that one can proceed to compute symmetrized derivatives of the 
covariant $F$ field, obtaining

\begin{eqnarray}
\partial_{\sigma} F_{\mu\nu} (0) &=& \nabla_{\sigma} F_{\mu\nu}, 
\nonumber \\
\partial_{(\sigma} \partial_{\rho)} F_{\mu\nu} (0) &=& \nabla_{(\sigma} 
\nabla_{\rho)} F_{\mu\nu} - \frac{2}{3} {R^{\alpha}}_{(\sigma\rho)[\mu} 
F_{\nu]\alpha}, \nonumber \\
\partial_{(\lambda} \partial_{\sigma} \partial_{\rho)} F_{\mu\nu} (0) &=& 
\nabla_{(\lambda} \nabla_{\sigma} \nabla_{\rho)} F_{\mu\nu} - 2 
{R^{\alpha}}_{(\sigma\rho[\mu} \nabla_{\lambda)} F_{\nu]\alpha}, \nonumber 
\\
&\ldots& \nonumber
\end{eqnarray}

\noindent
In this way the previous integral formula for the gauge field can be 
perturbatively expanded in terms of target tensors, thus obtaining a 
covariant perturbative expansion as

\begin{eqnarray*}
A_{\mu} (X) &=& A_{\mu} (0) + \frac{1}{2} X^{\nu} F_{\mu\nu} (0) + 
\frac{1}{3} X^{\nu} X^{\sigma} \nabla_{\sigma} F_{\mu\nu} (0) + 
\frac{1}{8} X^{\nu} X^{\sigma} X^{\rho} \nabla_{\sigma} \nabla_{\rho} 
F_{\mu\nu} (0) + \\
&&
+ \frac{1}{30} X^{\nu} X^{\sigma} X^{\rho} X^{\lambda} 
\nabla_{\sigma} \nabla_{\rho} \nabla_{\lambda} F_{\mu\nu} (0) + 
\mathcal{O} (X^{5}) - \frac{1}{6} \Big( \frac{1}{4} X^{\nu} 
X^{\sigma} X^{\rho} {R^{\alpha}}_{\sigma\rho\mu} (0) F_{\nu\alpha} 
(0) + \\
&&
+ \frac{1}{5} X^{\nu} X^{\sigma} X^{\rho} X^{\lambda} 
{R^{\alpha}}_{\sigma\rho\mu} (0) \nabla_{\lambda} F_{\nu\alpha} (0) 
+ \mathcal{O} (X^{5}) \Big) + \mathcal{O}(R^{2}).
\end{eqnarray*}

\noindent
This is the essence of the radial gauge procedure. This method was 
employed in \cite{Cornalba-Schiappa-2} for computing general 
correlation functions of open string fields in curved backgrounds. It 
is employed in this paper for expanding background fields in the 
matrix model limit (and we can also envisage that it could be of 
interest for computing correlation functions in more general 
backgrounds than the ones in \cite{Cornalba-Schiappa-2}).


\section{The Propagator on the Unit Disk}


The beta function calculation we perform in this paper is done with
worldsheet coordinates on the unit disk. Because most previous works on
these subjects deal mainly with the upper half plane computation, we address
in this appendix our conventions for the free field CFT propagator on the
disk. Let us start by considering the two point function, defined by

\begin{equation*}
\Pi^{AB} \left( z,w \right) = \frac{1}{\alpha^{\prime}} \left\langle \zeta^{A} \left( z \right) \zeta^{B} \left( w \right) \right\rangle 
\end{equation*}

\noindent 
on the unit disk $\Sigma $ in the complex plane $\mathbb{C}$. This
is a symmetric function $\Pi ^{AB}\left( z,w\right) =\Pi ^{BA}\left(
w,z\right) $ of its arguments. If one considers $\Pi ^{AB}$ as a function of 
$z$ (with $w$ fixed in the interior of the disk), one has that

\begin{equation*}
-\square\, \Pi ^{AB}=2\pi \delta \left( z-w\right) g^{AB}-2g^{AB},
\end{equation*}

\noindent 
where $\square =4\partial \overline{\partial }$, since we are
considering a sigma model with bulk kinetic term $\frac{1}{4\pi \alpha
^{\prime }}$\noindent $g_{AB}\int_{\Sigma }d\zeta ^{A}\wedge * d\zeta
^{B}$. Also, note that the second term after the equality comes from the
excluded zero mode, which acts as a background charge distribution of
uniform density. The propagator also satisfies the boundary condition
(recall that $* dz=-idz$ and $* d\overline{z}=id\overline{z}$)

\begin{equation}
\left( g_{AC} * d\Pi^{CB} + i \mathcal{F}_{AC} d\Pi^{CB} \right)
\Big|_{\partial \Sigma} = 0, \label{boundary}
\end{equation}

\noindent
where we are assuming the boundary term $\frac{i}{4\pi \alpha ^{\prime }} \mathcal{F}_{AB}\int_{\Sigma }d\zeta ^{A}\wedge d\zeta ^{B}$,
with $\mathcal{F}_{AB}$ constant. The disk is defined by $z\overline{z}=1$,
so that $zd\overline{z}+\overline{z}dz=0$, and the previous boundary
condition (\ref{boundary}) becomes

\begin{equation*}
\left( g-\mathcal{F}\right) _{AC}\,z\,\partial \Pi ^{CB}+\left( g+\mathcal{F} \right) _{AC}\,\overline{z}\,\overline{\partial }\Pi ^{CB}=0.
\end{equation*}

\noindent 
Let us now consider the modified propagator $P^{AB}$, defined by

\begin{equation*}
\Pi ^{AB}\left( z,w\right) =P^{AB}\left( z,w\right) +\frac{1}{2}\left( z \overline{z}+w\overline{w}\right) g^{AB}.
\end{equation*}

\noindent 
It satisfies the differential equation

\begin{equation}
-\square P^{AB}=2\pi \delta \left( z-w\right) g^{AB},  \label{diffeq}
\end{equation}

\noindent 
supplemented with the modified boundary condition

\begin{equation}
\left( g-\mathcal{F}\right) _{AC}\,z\,\partial P^{CB}+\left( g+\mathcal{F} \right) _{AC}\,\overline{z}\,\overline{\partial }P^{CB}=-\delta _{A}^{B}.
\label{modBC}
\end{equation}

\noindent 
The leading singularity of this modified propagator is easily seen
to be given by $P^{AB}\left( z,w\right) \sim -g^{AB}\ln \left| z-w\right| $,
so that the basic claim we now present is that

\begin{equation*}
P^{AB}=\Theta ^{AB}\,\ln \left( \frac{1-z\overline{w}}{1-\overline{z}w}
\right) -2G^{AB}\,\ln \left| 1-z\overline{w}\right| -g^{AB}\,\ln \left| 
\frac{z-w}{1-z\overline{w}}\right| ,
\end{equation*}

\noindent 
where $\Theta ^{AB}$ and $G^{AB}$ are the standard open string
tensors \cite{Schomerus, Seiberg-Witten}. One observes from this expression
that the leading singularity is correct, and therefore equation (\ref{diffeq}) is satisfied. Moreover, the derivatives of $P^{AB}$ can be computed as

\begin{eqnarray*}
\left( g-\mathcal{F}\right) _{AC}\,z\,\partial P^{CB} &=&\frac{z\overline{w}}{1-z\overline{w}}\,\delta _{A}^{B}-\frac{1}{2}\left( g-\mathcal{F}\right)
_{AC}\,g^{CB}\left( \frac{z}{z-w}+\frac{z\overline{w}}{1-z\overline{w}} 
\right) , \\
\left( g+\mathcal{F}\right) _{AC}\,\overline{z}\,\overline{\partial }P^{CB}
&=&\frac{\overline{z}w}{1-\overline{z}w}\,\delta _{A}^{B}-\frac{1}{2}\left(
g+\mathcal{F}\right) _{AC}\,g^{CB}\left( \frac{\overline{z}}{\overline{z}- 
\overline{w}}+\frac{\overline{z}w}{1-\overline{z}w}\right) .
\end{eqnarray*}

\noindent 
Using the fact that $z\overline{z}=1$ one can then sum the two
expressions above and obtain the correct equation for the modified boundary
condition (\ref{modBC}). Let us consider the functions that play a role in
the previous result for $P^{AB}\left( z,w\right) $,

\begin{eqnarray*}
\mathcal{A}\left( z,w\right) &=&\ln \left( \frac{1-z\overline{w}}{1-
\overline{z}w}\right) , \\
\mathcal{B}\left( z,w\right) &=&-2\ln \left| 1-z\overline{w}\right| , \\
\mathcal{C}\left( z,w\right) &=&-\ln \left| \frac{z-w}{1-z\overline{w}}
\right| , \\
\mathcal{D}\left( z,w\right) &=&\frac{1}{2}\left( z\overline{z}+w\overline{w}
\right) .
\end{eqnarray*}

\noindent 
Note that $\mathcal{C}$ vanishes whenever $z$ or $w$ are on the
boundary of the disk. The Fourier transforms of $\mathcal{A}$ and $\mathcal{B}$, on the unit disk $\Sigma$, are given by

\begin{eqnarray*}
\mathcal{A}\left( xe^{i\alpha },ye^{i\beta }\right)  &=&\sum_{n\neq 0}\frac{1}{n}\left( xy\right) ^{\left| n\right| }e^{-in\left( \alpha -\beta \right)
}\,, \\
\mathcal{B}\left( xe^{i\alpha },ye^{i\beta }\right)  &=&\sum_{n\neq 0}\frac{1}{\left| n\right| }\left( xy\right) ^{\left| n\right| }e^{-in\left( \alpha
-\beta \right) }\,.
\end{eqnarray*}

\noindent 
The basic result we have obtained for the propagator on the disk
is that

\begin{equation*}
\Pi ^{AB}(z,w)=\Theta ^{AB}\,\mathcal{A}(z,w)+G^{AB}\,\mathcal{B} 
(z,w)+g^{AB}\,\mathcal{C}(z,w)+\,g^{AB}\mathcal{D}(z,w).
\end{equation*}

\noindent 
We thus conclude by quoting the Fourier transform of the above
result. With the appropriate conventions of $F \left( \alpha \right) =
\sum_{n} F_{n} e^{-i n \alpha}$ and $F_{n} = \int \frac{d\alpha}{2\pi} F
\left( \alpha \right) e^{i n \alpha}$, one finally gets that

\begin{eqnarray*}
\frac{1}{\alpha ^{\prime }}\langle \zeta _{n}^{A}\left( x\right) \zeta
_{-n}^{B}\left( y\right) \rangle  &=&\left( xy\right) ^{\left| n\right|
}\left( G^{AB}\,\frac{1}{\left| n\right| }+\Theta ^{AB}\,\frac{1}{n}\right)
+\cdots, \qquad \qquad \left( n\neq 0\right) , \\
\frac{1}{\alpha ^{\prime }}\langle \zeta _{0}^{A}\left( x\right) \zeta
_{0}^{B}\left( y\right) \rangle  &=&\frac{1}{2}\left( x^{2}+y^{2}\right)
\,g^{AB}+\cdots ,
\end{eqnarray*}

\noindent 
where the terms in $\cdots$ come from the function $\mathcal{C}$
and vanish when either $x$ or $y$ are $1$ (\textit{i.e.}, when either point
is at the disk boundary).


\section{Computation of Two Loop Graphs}\label{app.graphs}


All the computations in this appendix are carried out in units where $\alpha
^{\prime }=1$. We recall the vertices used in the subsequent computation.
First we have the boundary  $M$ and $N$--type vertices  

\begin{eqnarray*}
&&
i \left( n_{N} M_{A_{1}A_{2} \cdots A_{N}} + \mathrm{cyc}_{1 \cdots N} \right) \delta_{\Sigma_{i}n_{i}-W}, \\
&&
-N_{A_{1}A_{2} \cdots A_{N}}\ \delta_{\Sigma_{i}n_{i}-W}.
\end{eqnarray*}

\noindent
Since $M_{AB}=-M_{BA}$, the two $M$--vertex reads

\begin{equation*}
i \left( n_{2}-n_{1} \right)\ M_{A_{1}A_{2}}\ \delta_{n_{1}+n_{2}-W}.
\end{equation*}

\noindent
We then have two bulk $T$--vertices 

\begin{eqnarray*}
&&
\frac{iT_{A_{1}A_{2}A_{3}}}{\Sigma_{i}\left| n_{i} \right|}\ \Big[
\left| n_{1}\right| \left( n_{2}-n_{3}\right) + \left| n_{2}\right| \left(
n_{3}-n_{1}\right) + \left| n_{3}\right| \left( n_{1}-n_{2}\right) \Big], \\
&&
\frac{1}{\Sigma_{i}\left| n_{i}\right|}\ \Big[ T_{A_{1} \cdots
A_{4}} \big( n_{3}n_{4}-\left| n_{3}n_{4}\right| \big) + \left( 3412\right)
+ \left( 1324\right) + \left( 2413\right) + \left( 1423\right) + \left(
2314\right) \Big] ,
\end{eqnarray*}

\noindent
where we are using the compact notation $\left( ijkl \right) \equiv T_{A_{i}A_{j}A_{k}A_{l}} \left( n_{k}n_{l} - \left| n_{k}n_{l} \right| \right)$.


\subsection{Computation of $\protect\delta$--Type Graphs}


\paragraph{Graph $\protect\delta{[}1{]}$:}

There is only one term 

\begin{equation*}
-\frac{1}{4}N_{ABCD} \sum_{m,n} G^{AB}G^{CD}\ \frac{1}{|n||m|}\ e^{-\epsilon
\left( \left| n\right| +\left| m\right| \right) },
\end{equation*}

\noindent
which diverges as $\ln ^{2}\epsilon $. Therefore its contribution to the
beta function is zero.

\paragraph{Graph $\protect\delta{[}2{]}$:}

The graph reads (in what follows we omit the explicit reference to the ultraviolet 
cut--off $\varepsilon$, which we consider as understood)

\begin{equation*}
\frac{1}{2}M_{ABC}N_{RST} \sum_{m,n,p} \left( -ip\right)\ \Pi_{n}^{AR}\Pi
_{m}^{BS}\Pi _{p}^{CT}\ \delta _{m+n+p}.
\end{equation*}

\noindent 
All the relevant sums are of the two forms 

\begin{equation*}
\sum_{m,n\geq 1}\frac{1}{nm}, \qquad \qquad \sum_{m,p\geq 1} \frac{1}{m(m+p)},
\end{equation*}

\noindent
which both diverge in $\ln ^{2}\epsilon $. So the contribution of this graph
to the beta function is zero.

\paragraph{Graph $\protect\delta{[}3{]}$:}

To compute this graph, we introduce the notation 

\begin{equation*}
\hat{\Pi}_{n}^{AB}=\left( \frac{1}{\left| n\right| g+n\mathcal{F}+N}\right)
^{AB}=\Pi _{n}^{AB}-\Pi _{n}^{AC}N_{CD}\Pi _{n}^{DB}+\cdots .
\end{equation*}

\noindent
The result is then compactly written as 

\begin{equation*}
\frac{1}{4}M_{ABC}M_{RST} \sum_{m,n,p} \left( p^{2}\ \hat{\Pi}_{n}^{AR} \hat{\Pi}_{m}^{BS} \hat{\Pi}_{p}^{CT}+2mp\ \hat{\Pi}_{n}^{AR} \hat{\Pi}_{m}^{BT} \hat{\Pi}_{p}^{CS} \right) \delta _{m+n+p}.
\end{equation*}

\noindent
Expanding the propagators $\hat{\Pi}$ we have, to order $N$,

\begin{eqnarray*}
&&
-\frac{1}{2}M_{ABC}M_{RST}\sum_{m,n,p}\left( p^{2}\ \left( \Pi N\Pi \right)
_{n}^{AR}\Pi _{m}^{BS}\Pi _{p}^{CT}+mp\ \left( \Pi N\Pi \right) _{n}^{AR}\Pi
_{m}^{BT}\Pi _{p}^{CS}\right) \delta _{m+n+p} + \\
&&
-\frac{1}{4}M_{ABC}M_{RST}\sum_{m,n,p}p^{2}\ \Pi _{n}^{AR}\Pi
_{m}^{BS}\left( \Pi N\Pi \right) _{p}^{CT}\delta _{m+n+p} + \\
&&
-\frac{1}{2}M_{ABC}M_{RST}\sum_{m,n,p}mp\left( \Pi _{n}^{AR}\left( \Pi
N\Pi \right) _{m}^{BT}\Pi _{p}^{CS}+\Pi _{n}^{AR}\Pi _{m}^{BT}\left( \Pi
N\Pi \right) _{p}^{CS}\right) \delta _{m+n+p}.
\end{eqnarray*}

\noindent
The second and third lines do not have a leading $\ln \epsilon $ divergence,
since the sums are of the form $\sum \frac{1}{nm}\delta _{m+n+p}$, as in the
graph $\delta \left[ 2\right] $. The non--vanishing contribution comes from
the first line. Consider first the following sums

\begin{equation*}
S\sim \sum \frac{p}{n^{2}m}\delta _{m+n+p}, \qquad \qquad R\sim \sum \frac{1}{n^{2}}\delta _{m+n+p},
\end{equation*}

\noindent
in the summation regions $I$, $II$, $III$, defined by $n,m>0$, by $n,p>0$ and
by $m,p>0$. Denote the three sums by $S_{I}$, $S_{II}$, $S_{III}$, and $R_{I}$, $R_{II}$, $R_{III}$. We have that 

\begin{eqnarray*}
S_{I} &\sim & -\ln \epsilon, \qquad \qquad S_{II} \sim
-\ln \epsilon, \qquad \qquad S_{III}\sim \ln \epsilon, \\
R_{I} &\sim & \ln \epsilon, \qquad \qquad \ \  R_{II} \sim \ln \epsilon, \qquad \qquad \ \ \, R_{III} \sim - \ln \epsilon .
\end{eqnarray*}

\noindent
The result then has the basic tensor (the $2\times $ comes from the sums
which are doubled under $\left( n,m,p\right) \rightarrow -\left(
n,m,p\right) $) 

\begin{equation*}
-2\times \frac{1}{2}M_{ABC}M_{RST}
\end{equation*}

\noindent
multiplied, for the first term ($S$ sums), by the following tensor
structures 

\begin{equation*}
\begin{array}{ccccc}
^{AR,BS,CT} & I & II & III &  \\ 
\left( GNG\right) GG & - & - & + & 3\ln \epsilon \\ 
\left( \Theta N\Theta \right) \Theta \Theta & + & + & + & -\ln \epsilon \\ 
\left( GNG\right) \Theta \Theta & + & + & + & -\ln \epsilon \\ 
\left( \Theta N\Theta \right) GG & - & - & + & 3\ln \epsilon \\ 
\left( GN\Theta \right) G\Theta & + & - & - & -\ln \epsilon \\ 
\left( \Theta NG\right) \Theta G & - & + & - & -\ln \epsilon \\ 
\left( GN\Theta \right) \Theta G & - & + & - & -\ln \epsilon \\ 
\left( \Theta NG\right) G\Theta & + & - & - & -\ln \epsilon
\end{array}
\end{equation*}

\noindent
and, for the second term ($R$ sums), by the following tensor structures 

\begin{equation*}
\begin{array}{ccccc}
^{AR,BT,CS} & I & II & III &  \\ 
\left( GNG\right) GG & + & + & + & \ln \epsilon \\ 
\left( \Theta N\Theta \right) \Theta \Theta & + & + & + & \ln \epsilon \\ 
\left( GNG\right) \Theta \Theta & + & + & + & \ln \epsilon \\ 
\left( \Theta N\Theta \right) GG & + & + & + & \ln \epsilon \\ 
\left( GN\Theta \right) G\Theta & + & + & - & 3\ln \epsilon \\ 
\left( \Theta NG\right) \Theta G & + & + & - & 3\ln \epsilon \\ 
\left( GN\Theta \right) \Theta G & + & + & - & 3\ln \epsilon \\ 
\left( \Theta NG\right) G\Theta & + & + & - & 3\ln \epsilon
\end{array}
\end{equation*}

\noindent
We arrive at the following rather lengthy result 

\begin{eqnarray*}
&&
\delta \left[ 3 \right] = - M_{ABC} \left( 3M_{RST}+M_{RTS} \right) \left[
\left( GNG + \Theta N \Theta \right) GG \right]^{AR,BS,CT} \ln \epsilon + \\
&&
+ M_{ABC} \left( M_{RST}-M_{RTS} \right) \left[ \left( GNG + \Theta N \Theta
\right) \Theta \Theta \right]^{AR,BS,CT} \ln \epsilon + \\
&&
+ M_{ABC} \left( M_{RST}-3M_{RTS} \right) \left[ \left( G N \Theta + \Theta N G \right) \left( G \Theta + \Theta G \right) \right]^{AR,BS,CT} \ln \epsilon .
\end{eqnarray*}

\paragraph{Graph $\protect\delta{[}4{]}$:}

This graph has been computed in the text, in section (\ref{comdelta}).

\paragraph{Graph $\protect\delta{[}5{]}$:}

The graph reads 

\begin{equation*}
\frac{1}{6} T_{ABC}N_{RST} \int \zeta^{A} \partial_{x} \zeta^{B} \partial_{\theta} \zeta^{C}\ dx\ \frac{d\theta}{2\pi}\ \int \zeta^{R} \zeta^{S} \zeta^{T}\ \frac{d\theta^{\prime}}{2\pi}.
\end{equation*}

\noindent
The radial part gives 

\begin{equation*}
\int_{0}^{1}dx\ x^{\left| n\right| +\left| m\right| +\left| p\right| -1}=
\frac{1}{\left| n\right| +\left| m\right| +\left| p\right|},
\end{equation*}

\noindent
whereas the angular part contributes as 

\begin{equation*}
-6i\ \left| m\right| p\ \Pi _{n}^{AR}\Pi _{m}^{BS}\Pi _{p}^{CT}\ \delta_{n+m+p}.
\end{equation*}

\noindent
The total contribution is then

\begin{equation*}
-i\ T_{ABC}N_{RST} \sum_{m,n,p} \frac{p|m|}{\left| n\right| +\left| m\right|
+\left| p\right|}\ \Pi _{n}^{AR}\Pi _{m}^{BS}\Pi _{p}^{CT}\ \delta_{n+m+p}.
\end{equation*}

\noindent 
The graph vanishes by symmetry. In fact, the most general result
for the above sum is given by 

\begin{equation*}
T_{ABC}N_{RST} \Big[ a\Theta GG+b\Theta \Theta \Theta +cGG\Theta +dG\Theta G 
\Big]^{AR,BS,CT}.
\end{equation*}

\noindent
Moreover, under the exchange $m\leftrightarrow p$ in the sum, we see that $c=d$. Recall that $N$ is symmetric in the indices $(R,S,T)$, whereas the
indices $(A,B,C)$ of $T$ are antisymmetrized. Therefore the above result
vanishes irrespective of $a,b,c=d$, due to symmetry under $S\leftrightarrow
T $ and antisymmetry under $B\leftrightarrow C$. Thus, this graph does not
contribute to the beta function.

\paragraph{Graph $\protect\delta{[}6{]}$:}

This graph has been computed in the text, in section (\ref{comdelta}).

\paragraph{Graph $\protect\delta{[}7{]}$:}

The graph reads 

\begin{equation*}
\frac{1}{2} T_{ABC} M_{A^{\prime}B^{\prime}C^{\prime}} \int \zeta^{A} \partial_{x} \zeta^{B} \partial_{\theta} \zeta^{C}\ dx\ \frac{d\theta}{2\pi} \int \zeta^{A^{\prime}} \zeta^{B^{\prime}} \partial_{\theta^{\prime}} \zeta^{C^{\prime}}\ \frac{d\theta^{\prime}}{2\pi},
\end{equation*}

\noindent
and the radial integration is readily done 

\begin{equation*}
\int dx\ x^{\left| n\right| +\left| m\right| +\left| p\right| -1}=\frac{1}{ 
\left| n\right| +\left| m\right| +\left| p\right| }.
\end{equation*}

\noindent
After doing the angular integrations, one gets 

\begin{eqnarray*}
&&
T_{ABC} M_{A^{\prime}B^{\prime}C^{\prime}} \sum_{m,n,p} \frac{1}{\left|
n\right| +\left| m\right| +\left| p\right|}\ \delta_{n+m+p} \times \\
&&
\times \left( p^{2}\left| m\right|\ \hat{\Pi}_{n}^{AA^{\prime }}\hat{\Pi}_{m}^{BB^{\prime }} \hat{\Pi}_{p}^{CC^{\prime }}+mp\left| m\right|\ \hat{\Pi}_{n}^{AB^{\prime }}\hat{\Pi}_{m}^{BC^{\prime }}\hat{\Pi}_{p}^{CA^{\prime
}}+np\left| m\right|\ \hat{\Pi}_{n}^{AC^{\prime }}\hat{\Pi}_{m}^{BA^{\prime }} 
\hat{\Pi}_{p}^{CB^{\prime }}\right) .
\end{eqnarray*}

\noindent
Using the total antisymmetry of $T_{ABC}$, and the symmetry of $M_{A^{\prime
}B^{\prime }C^{\prime }}$ in $A^{\prime }B^{\prime }$, we obtain 

\begin{equation*}
T_{ABC}M_{A^{\prime }B^{\prime }C^{\prime }}\sum_{m,n,p}F\left( n,m,p\right)\ \hat{\Pi}_{n}^{AA^{\prime }}\hat{\Pi}_{m}^{BB^{\prime }} \hat{\Pi}_{p}^{CC^{\prime }}\ \delta_{n+m+p},
\end{equation*}

\noindent
where 

\begin{equation*}
F\left( n,m,p\right) =\frac{1}{2}\left[ \frac{p^{2}\left| m\right| +np\left|
p\right| +mp\left| n\right| }{\left| n\right| +\left| m\right| +\left|
p\right| }-\left( m\leftrightarrow n\right) \right] .
\end{equation*}

\noindent
We must then consider the two following sums 

\begin{equation*}
S \sim \sum_{m,n,p}\delta _{n+m+p}\ \frac{F\left( n,m,p\right) }{n^{2}mp}, \qquad \qquad \qquad R\sim \sum_{m,n,p}\delta_{n+m+p}\ \frac{F\left( n,m,p\right) }{nmp^{2}},
\end{equation*}

\noindent
in the three regions $I, II, III$ (recall, with $n,m>0$, $n,p>0$ and $m,p>0$), where we obtain 

\begin{eqnarray*}
S_{I} &=&\frac{1}{2}\sum_{m,n\geq 1}\frac{n-m}{n^{2}m}\sim -\frac{1}{2}\ln
\epsilon , \\
S_{II} &=&\frac{1}{2}\sum_{n,p\geq 1}\frac{n-p}{n^{2}\left( n+p\right) }\sim
-\frac{1}{2}\ln \epsilon , \\
S_{III} &=&\frac{1}{2}\sum_{m,p\geq 1}\frac{m-p}{\left( m+p\right) ^{2}m} 
\sim -\ln \epsilon ,
\end{eqnarray*}

\noindent
and 

\begin{eqnarray*}
R_{I} &=&\frac{1}{2}\sum_{m,n\geq 1}\frac{m-n}{nm\left( n+m\right) }=0, \\
R_{II} &=&\frac{1}{2}\sum_{n,p\geq 1}\frac{n-p}{np\left( n+p\right) }=0, \\
R_{III} &=&\frac{1}{2}\sum_{m,p\geq 1}\frac{p-m}{\left( m+p\right) mp}=0.
\end{eqnarray*}

\noindent
Finally we can compute the only non--vanishing contribution to the graph
itself, given by 

\begin{equation*}
-2T_{ABC}M_{A^{\prime }B^{\prime }C^{\prime }}\sum_{m,n,p}F\left(
n,m,p\right) \ \left( \Pi N\Pi \right) _{n}^{AA^{\prime }}\Pi
_{m}^{BB^{\prime }}\Pi _{p}^{CC^{\prime }}\ \delta _{n+m+p}.
\end{equation*}

\noindent
This is given by 

\begin{equation*}
-4T_{ABC}M_{A^{\prime }B^{\prime }C^{\prime }}
\end{equation*}

\noindent
multiplied by the following tensor structures 

\begin{equation*}
\begin{array}{ccccc}
^{AA^{\prime },BB^{\prime },CC^{\prime }} & I & II & III &  \\ 
\left( GNG\right) GG & - & - & + & 0 \\ 
\left( \Theta N\Theta \right) \Theta \Theta & + & + & + & -2\ln \epsilon \\ 
\left( GNG\right) \Theta \Theta & + & + & + & -2\ln \epsilon \\ 
\left( \Theta N\Theta \right) GG & - & - & + & 0 \\ 
\left( GN\Theta \right) G\Theta & + & - & - & \ln \epsilon \\ 
\left( \Theta NG\right) \Theta G & - & + & - & \ln \epsilon \\ 
\left( GN\Theta \right) \Theta G & - & + & - & \ln \epsilon \\ 
\left( \Theta NG\right) G\Theta & + & - & - & \ln \epsilon
\end{array}
\end{equation*}

\noindent
We finally obtain 

\begin{equation*}
\delta \left[ 7\right] =4T_{ABC}M_{A^{\prime }B^{\prime }C^{\prime }}\left[
2\left( \Theta N\Theta +GNG\right) \Theta \Theta -\left( GN\Theta +\Theta
NG\right) \left( G\Theta +\Theta G\right) \right] ^{AA^{\prime },BB^{\prime
},CC^{\prime }}\ln \epsilon.
\end{equation*}

\paragraph{Graph $\protect\delta{[}8{]}$:}

This graph is computed from the expectation value of

\begin{equation*}
\frac{1}{8} T_{ABCD}N_{RS} \int \zeta^{A} \zeta ^{B} \left( \partial_{x} \zeta^{C} \partial_{x} \zeta^{D} + \frac{1}{x^{2}}\ \partial_{\theta} \zeta^{C} \partial_{\theta} \zeta^{D} \right)\ x\ dx\ \frac{d\theta}{2\pi} \int
\zeta^{R} \zeta^{S}\ \frac{d\theta^{\prime}}{2\pi}.
\end{equation*}

\noindent
In what follows we concentrate on the expectation values of the $\zeta $'s,
omitting the term $\frac{1}{8}T_{ABCD}N_{RS}$. First of all, the radial part
always contributes as 

\begin{equation*}
\int_{0}^{1}x\ dx\ x^{2\left| n\right| +2\left| m\right| -2}=\frac{1}{2}\ \frac{1}{\left| m\right| +\left| n\right| }.
\end{equation*}

\noindent
The angular part then gives three contributions

\begin{eqnarray*}
&&2\left( 2m^{2}\right)\ \Pi _{n}^{AR}\Pi _{-n}^{BS}\Pi _{m}^{CD}, \\
&&2\left( 2n^{2}\right)\ \Pi _{m}^{AB}\Pi _{n}^{CR}\Pi _{-n}^{DS}, \\
&&8\left( \left| m\right| \left| n\right| -mn\right)\ \Pi _{m}^{BD}\Pi
_{n}^{AR}\Pi _{-n}^{CS}.
\end{eqnarray*}

\noindent
The combined result is then 

\begin{eqnarray*}
&&
\frac{1}{4} T_{ABCD} N_{RS} \sum_{n,m} \frac{1}{\left| m\right| +\left|
n\right| }\times \\
&&
\times \Big[ m^{2}\ \Pi_{n}^{AR} \Pi_{-n}^{BS} \Pi_{m}^{CD} + n^{2}\ \Pi_{m}^{AB} \Pi_{n}^{CR} \Pi_{-n}^{DS} + 2 \big( \left| m \right| \left| n \right| - m n \big)\ 
\Pi_{m}^{BD} \Pi_{n}^{AR} \Pi_{-n}^{CS} \Big].
\end{eqnarray*}

\noindent 
The second and the third terms diverge in $\ln ^{2}\epsilon $, so
their contribution to the beta function is zero. The first term, on the
other hand, contributes as 

\begin{equation*}
\delta \left[ 8\right] =T_{ABCD}\left( GNG+\Theta N\Theta \right)
^{AB}G^{CD}\,\ln \epsilon .
\end{equation*}


\subsection{Computation of $\protect\delta^{\prime}$--Type Graphs}


\paragraph{Graph $\protect\delta^{\prime}{[}1{]}$:}

This graph has been computed in the text, in section (\ref{comdeltaprime}).

\paragraph{Graph $\protect\delta^{\prime}{[}2{]}$:}

We look at the graph, with symmetry factor of $6$, given by 

\begin{eqnarray*}
&&
\left. \frac{\partial}{\partial W} \right|_{W=0} \frac{i^{2}}{6}\ \Big( \left( p-W \right)\ M_{A_{1}A_{2}A_{3}} + n\ M_{A_{2}A_{3}A_{1}} + m\ M_{A_{3}A_{1}A_{2}} \Big)\ \Pi_{n}^{A_{1}B_{1}} \Pi_{m}^{A_{2}B_{2}} \Pi_{p-W}^{A_{3}B_{3}} \times \\
&&
\times \Big( - \left( p-W \right)\ dM_{B_{1}B_{2}B_{3}} - n\ dM_{B_{2}B_{3}B_{1}} - m\ dM_{B_{3}B_{1}B_{2}} \Big)\ \delta_{n+m+p}.
\end{eqnarray*}

\noindent
We then obtain the following terms (a factor of $\delta_{n+m+p}$ is understood) 

\begin{eqnarray*}
&&
\frac{1}{6}\ \frac{1}{p}\ \Big[ -p^{2}\ M_{A_{1}A_{2}A_{3}} dM_{B_{1}B_{2}B_{3}} + \big( n\ M_{A_{2}A_{3}A_{1}} + m\ M_{A_{3}A_{1}A_{2}} \big) \big( n\ dM_{B_{2}B_{3}B_{1}} + m\ dM_{B_{3}B_{1}B_{2}} \big) \Big] \times \\
&&
\times \Pi_{n}^{A_{1}B_{1}} \Pi_{m}^{A_{2}B_{2}} \Pi_{p}^{A_{3}B_{3}},
\end{eqnarray*}

\noindent
and, from the regulator 

\begin{eqnarray*}
&&
\varepsilon\ \frac{1}{12}\ \frac{\left| p \right|}{p}\ \Big( p\ M_{A_{1}A_{2}A_{3}} + n\ M_{A_{2}A_{3}A_{1}} + m\ M_{A_{3}A_{1}A_{2}} \Big)\ \Pi_{n}^{A_{1}B_{1}} \Pi_{m}^{A_{2}B_{2}} \Pi_{p}^{A_{3}B_{3}} \times \\
&&
\times \Big( p\ dM_{B_{1}B_{2}B_{3}} + n\ dM_{B_{2}B_{3}B_{1}} + m\ dM_{B_{3}B_{1}B_{2}} \Big) .
\end{eqnarray*}

\noindent
Using the symmetry of $M_{ABC}$ and $dM_{ABC}$ in the first two indices and
combining terms, we get 

\begin{eqnarray*}
&&
\frac{1}{12}\ \Pi _{n}^{A_{1}B_{1}}\Pi _{m}^{A_{2}B_{2}}\Pi
_{p}^{A_{3}B_{3}} \times \\
&&
\times \left[ M_{A_{1}A_{2}A_{3}}dM_{B_{1}B_{2}B_{3}}\left( \frac{-2p^{2}}{p}+
\frac{2p^{2}}{m}+\frac{2p^{2}}{n}+\varepsilon \frac{\left| p\right| }{p}
p^{2}+\varepsilon \frac{\left| m\right| }{m}p^{2}+\varepsilon \frac{\left|
n\right| }{n}p^{2}\right) \right. + \\
&&
\left. +2\,M_{A_{1}A_{2}A_{3}}dM_{B_{2}B_{3}B_{1}}\left( \frac{2pn}{m}
+\varepsilon \frac{\left| p\right| pn}{p}+\varepsilon \frac{\left| m\right|
pn}{m}+\varepsilon \frac{\left| n\right| pn}{n}\right) \right] .
\end{eqnarray*}

\noindent
Doing the sums, we obtain 

\begin{eqnarray*}
\delta ^{\prime }\left[ 2\right] &=&\frac{1}{6} 
M_{A_{1}A_{2}A_{3}}dM_{B_{1}B_{2}B_{3}}\left[ -2\Theta \Theta \Theta
+3\Theta GG+3G\Theta G-4GG\Theta \right] ^{A_{1}B_{1},A_{2}B_{2},A_{3}B_{3}}+
\\
&&+\frac{1}{3}M_{A_{1}A_{2}A_{3}}dM_{B_{2}B_{3}B_{1}}\left[ \Theta \Theta
\Theta +2\Theta GG-5G\Theta G+2GG\Theta \right]
^{A_{1}B_{1},A_{2}B_{2},A_{3}B_{3}}.
\end{eqnarray*}

\noindent
The second contribution is given by 

\begin{eqnarray*}
&&
\left. \frac{\partial}{\partial W}\right|_{W=0} \frac{i^{3}}{4}\ \Big(
\left( p-W \right)\ M_{A_{1}A_{2}A_{3}} + n\ M_{A_{2}A_{3}A_{1}} + m\ M_{A_{3}A_{1}A_{2}} \Big)\ \Pi_{n}^{A_{1}B_{1}} \Pi_{m}^{A_{2}B_{2}} \Pi_{pW}^{A_{3}C} \Pi_{p}^{DB_{3}} \times \\
&&
\times \left( 2p-W\right)\ dM_{CD}\ \Big( - p\ M_{B_{1}B_{2}B_{3}} - n\ M_{B_{2}B_{3}B_{1}} - m\ M_{B_{3}B_{1}B_{2}} \Big)\ \delta_{n+m+p}.
\end{eqnarray*}

\noindent
It leads to the following terms 

\begin{equation} \label{DP1}
- \frac{i}{2}\ M_{A_{1}A_{2}A_{3}}\ \Pi_{n}^{A_{1}B_{1}} \Pi_{m}^{A_{2}B_{2}} \left( \Pi dM \Pi \right)_{p}^{A_{3}B_{3}}\ p\ \Big( p\ M_{B_{1}B_{2}B_{3}} + n\ M_{B_{2}B_{3}B_{1}} + m\ M_{B_{3}B_{1}B_{2}} \Big).  
\end{equation}

\noindent
and 

\begin{eqnarray*}
&&
\frac{i}{4}\ \left( 1+\left| p\right| \varepsilon \right)\ \Big(
p\ M_{A_{1}A_{2}A_{3}} + n\ M_{A_{2}A_{3}A_{1}} + m\ M_{A_{3}A_{1}A_{2}} \Big) 
\Pi_{n}^{A_{1}B_{1}} \Pi_{m}^{A_{2}B_{2}} \left( \Pi dM\Pi \right)_{p}^{A_{3}B_{3}} \times \\
&&
\times \Big( p\ M_{B_{1}B_{2}B_{3}} + n\ M_{B_{2}B_{3}B_{1}} + m\ M_{B_{3}B_{1}B_{2}} \Big) .
\end{eqnarray*}

\noindent
The first expression, (\ref{DP1}), gives a vanishing contribution. The second
also gives a vanishing contribution, due to the following symmetry argument.
Notice that the possible contractions are between the tensor 

\begin{equation*}
\Big[ \left( aGG+b\Theta \Theta \right) \left( \Theta dM\Theta +GdMG\right)
+\left( cG\Theta +d\Theta G\right) \left( GdM\Theta +\Theta dMG\right) 
\Big] ^{A_{1}B_{1},A_{2}B_{2},A_{3}B_{3}}
\end{equation*}

\noindent
(with constants $a,b,c,d$) and an expression which is given by a product of
two tensors $M_{ABC}$, and which is symmetric under the interchange $A_{i}\leftrightarrow B_{i}$. On the other hand, the tensor above is
antisymmetric in the same interchange, and we therefore get a vanishing result.

\paragraph{Graph $\protect\delta ^{\prime}{[}3{]}$:}

The graph has a symmetry factor of $2$, and reads 

\begin{eqnarray*}
&&
- \frac{i}{2}\ \Big( \left( p - W \right) M_{ABC} + n\ M_{BCA} + m\ M_{CAB} \Big)\  T_{A^{\prime}B^{\prime}C^{\prime}} dM_{RS} \times \\
&&
\times \frac{1}{\left| n \right| + \left| m \right| + \left| p \right|}\ \Big[ \left| n \right| \left( m - p \right) + \left| m \right| \left( p - n \right) + \left| p \right| \left( n - m \right) \Big]\ \left( W - 2 p \right)\ \Pi_{n}^{AA^{\prime}} \Pi_{m}^{BB^{\prime}} \Pi_{p-W}^{CR} \Pi_{p}^{SC^{\prime}}.
\end{eqnarray*}

\noindent
Take the derivative $\left. \frac{\partial}{\partial W}\right|_{W=0}$ to get 

\begin{eqnarray*}
&&
\frac{i}{2}\ \Big( - p\ M_{ABC} + n\ M_{BCA} + m\ M_{CAB} \Big)\ T_{A^{\prime}B^{\prime}C^{\prime}} \times \\
&&
\times \frac{1}{\left| n \right| + \left| m \right| + \left| p \right| }\ \Big[ \left| n \right| \left( m - p \right) + \left| m \right| \left( p - n \right) + \left| p \right| \left( n - m \right) \Big]\ \Pi_{n}^{AA^{\prime}} \Pi_{m}^{BB^{\prime}} \left( \Pi dM \Pi \right)_{p}^{CC^{\prime}},
\end{eqnarray*}

\noindent
and also, from the regularization, 

\begin{eqnarray*}
&&
\varepsilon\ \frac{i}{2}\ \Big( p\ M_{ABC} + n\ M_{BCA} + m\ M_{CAB} \Big)\ T_{A^{\prime}B^{\prime}C^{\prime}} \times \\
&&
\times \frac{\left| p \right|}{\left| n \right| + \left| m \right| + \left| p \right|}\ \Big[ \left| n \right| \left( m - p \right) + \left| m \right| \left( p - n \right) + \left| p \right| \left( n - m \right) \Big]\ \Pi_{n}^{AA^{\prime}} \Pi_{m}^{BB^{\prime}} \left( \Pi dM \Pi \right)_{p}^{CC^{\prime}}.
\end{eqnarray*}

\noindent
We must consider the sums $\sum \frac{1}{nmp^{2}}pR$, $\sum \frac{1}{nmp^{2}} 
nS$ and $\sum \frac{1}{nmp^{2}}mS$, where 

\begin{eqnarray*}
R &\sim &\frac{\left( -1+\varepsilon \left| p\right| \right) }{\left|
n\right| +\left| m\right| +\left| p\right| }\ \Big[ \left| n\right| \left(
m-p\right) +\left| m\right| \left( p-n\right) +\left| p\right| \left(
n-m\right) \Big], \\
S &\sim &\frac{\left( 1+\varepsilon \left| p\right| \right) }{\left|
n\right| +\left| m\right| +\left| p\right| }\ \Big[ \left| n\right| \left(
m-p\right) +\left| m\right| \left( p-n\right) +\left| p\right| \left(
n-m\right) \Big].
\end{eqnarray*}

\noindent
We obtain the following three tables. Firstly 

\begin{eqnarray*}
pR_{I} &\sim &\sum_{n,m\geq 1}\frac{n-m}{mn\left( n+m\right) }-\varepsilon 
\frac{n-m}{nm}\sim 0, \\
pR_{II} &\sim &\sum_{n,p\geq 1}\frac{p-n}{np\left( p+n\right) }+\varepsilon 
\frac{n-p}{n\left( n+p\right) }\sim \frac{1}{2}\ln \varepsilon, \\
pR_{III} &\sim &\sum_{m,p\geq 1}\frac{m-p}{mp\left( m+p\right) }-\varepsilon 
\frac{m-p}{\left( m+p\right) m}\sim -\frac{1}{2}\ln \varepsilon,
\end{eqnarray*}

\noindent
where we use the fact that $\sum_{n,m\geq 1}\frac{\varepsilon }{n} 
e^{-\varepsilon \left( an-bm\right) }\sim -\frac{1}{b}\ln \varepsilon $, and
that, within our regularization scheme, in $\theta$--type graphs $a=b=2$,
whereas in $\infty$--type graphs $a=b=1$. Secondly 

\begin{eqnarray*}
nS_{I} &\sim &\sum_{n,m\geq 1}\frac{n-m}{m\left( n+m\right) ^{2}} 
+\varepsilon \frac{n-m}{m\left( n+m\right) }\sim \frac{3}{2}\ln \varepsilon,
\\
nS_{II} &\sim &\sum_{n,p\geq 1}\frac{n-p}{p^{2}\left( n+p\right) } 
+\varepsilon \frac{n-p}{\left( n+p\right) p}\sim \frac{1}{2}\ln \varepsilon,
\\
nS_{III} &\sim &\sum_{m,p\geq 1}\frac{m-p}{p^{2}m}+\varepsilon \frac{m-p}{mp} 
\sim \ln \varepsilon,
\end{eqnarray*}

\noindent
and finally 

\begin{eqnarray*}
mS_{I} &\sim &\sum_{n,m\geq 1}\frac{n-m}{n\left( n+m\right) ^{2}} 
+\varepsilon \frac{n-m}{n\left( n+m\right) }\sim -\frac{3}{2}\ln \varepsilon,
\\
mS_{II} &\sim &\sum_{n,p\geq 1}\frac{p-n}{p^{2}n}+\varepsilon \frac{p-n}{np} 
\sim -\ln \varepsilon, \\
mS_{III} &\sim &\sum_{m,p\geq 1}\frac{p-m}{p^{2}\left( m+p\right) } 
+\varepsilon \frac{p-m}{\left( m+p\right) p}\sim -\frac{1}{2}\ln \varepsilon.
\end{eqnarray*}

\noindent
Using the table for the $R$--sums 

\begin{equation*}
\begin{array}{ccccc}
^{AA^{\prime },BB^{\prime },CC^{\prime }} & I & II & III &  \\ 
GG\left( GdMG\right) & + & - & - & 0 \\ 
\Theta \Theta \left( \Theta dM\Theta \right) & + & + & + & 0 \\ 
\Theta \Theta \left( GdMG\right) & + & + & + & 0 \\ 
GG\left( \Theta dM\Theta \right) & + & - & - & 0 \\ 
G\Theta \left( GdM\Theta \right) & - & + & - & \ln \epsilon \\ 
\Theta G\left( \Theta dMG\right) & - & - & + & -\ln \epsilon \\ 
\Theta G\left( GdM\Theta \right) & - & - & + & -\ln \epsilon \\ 
G\Theta \left( \Theta dMG\right) & - & + & - & \ln \epsilon
\end{array}
\end{equation*}

\noindent
and that for the $S$ sums 

\begin{equation*}
\begin{array}{ccccc}
^{AA^{\prime },BB^{\prime },CC^{\prime }} & I & II & III &  \\ 
GG\left( GdMG\right) & + & - & - & 0 \\ 
\Theta \Theta \left( \Theta dM\Theta \right) & + & + & + & 3\ln \varepsilon
\\ 
\Theta \Theta \left( GdMG\right) & + & + & + & 3\ln \varepsilon \\ 
GG\left( \Theta dM\Theta \right) & + & - & - & 0 \\ 
G\Theta \left( GdM\Theta \right) & - & + & - & -2\ln \epsilon \\ 
\Theta G\left( \Theta dMG\right) & - & - & + & -\ln \epsilon \\ 
\Theta G\left( GdM\Theta \right) & - & - & + & -\ln \epsilon \\ 
G\Theta \left( \Theta dMG\right) & - & + & - & -2\ln \epsilon
\end{array}
\,\ \ \ \ \ \ \ \ \ \ \ \ \ 
\begin{array}{ccccc}
^{AA^{\prime },BB^{\prime },CC^{\prime }} & I & II & III &  \\ 
GG\left( GdMG\right) & + & - & - & 0 \\ 
\Theta \Theta \left( \Theta dM\Theta \right) & + & + & + & -3\ln \varepsilon
\\ 
\Theta \Theta \left( GdMG\right) & + & + & + & -3\ln \varepsilon \\ 
GG\left( \Theta dM\Theta \right) & + & - & - & 0 \\ 
G\Theta \left( GdM\Theta \right) & - & + & - & \ln \epsilon \\ 
\Theta G\left( \Theta dMG\right) & - & - & + & 2\ln \epsilon \\ 
\Theta G\left( GdM\Theta \right) & - & - & + & 2\ln \epsilon \\ 
G\Theta \left( \Theta dMG\right) & - & + & - & \ln \epsilon
\end{array}
\end{equation*}

\noindent
the various terms combine as follows 

\begin{equation*}
3i\left( M_{BCA}-M_{CAB}\right) T_{A^{\prime }B^{\prime }C^{\prime }}\left(
\Theta \Theta \right) ^{AA,BB^{\prime }}\left( GdMG+\Theta dM\Theta \right)
^{CC^{\prime }}\ln \varepsilon,
\end{equation*}

\noindent
and 

\begin{eqnarray*}
&&i\left( M_{ABC}\right) T_{A^{\prime }B^{\prime }C^{\prime }}\left( G\Theta
-\Theta G\right) ^{AA^{\prime },BB^{\prime }}\left( GdM\Theta +\Theta
dMG\right) ^{CC^{\prime }}\ln \varepsilon + \\
&&+i\left( M_{BCA}\right) T_{A^{\prime }B^{\prime }C^{\prime }}\left(
-2G\Theta -\Theta G\right) ^{AA^{\prime },BB^{\prime }}\left( GdM\Theta
+\Theta dMG\right) ^{CC^{\prime }}\ln \varepsilon + \\
&&+i\left( M_{CAB}\right) T_{A^{\prime }B^{\prime }C^{\prime }}\left(
G\Theta +2\Theta G\right) ^{AA^{\prime },BB^{\prime }}\left( GdM\Theta
+\Theta dMG\right) ^{CC^{\prime }}\,\ln \varepsilon .
\end{eqnarray*}

\noindent
Using the fact that $M_{ABC}+M_{BCA}+M_{CAB}=0$, that $T_{ABC}$ is totally
antisymmetric, and that $M_{ABC}$ is symmetric in $AB$, we can combine the
various terms and obtain, after reshuffling indices, 

\begin{equation*}
\delta^{\prime} \left[ 3 \right] = 6i M_{ABC} T_{A^{\prime }B^{\prime
}C^{\prime }} \left[ \left( GdM\Theta +\Theta dMG\right) \Theta G-\left( GdMG+\Theta
dM\Theta \right) \Theta \Theta \right] ^{AA^{\prime },BB^{\prime
},CC^{\prime }} \ln \varepsilon.
\end{equation*}

\paragraph{Graph $\protect\delta^{\prime}{[}4{]}$:}

This graph has symmetry factor $4$, and is given by 

\begin{eqnarray*}
&&
- \frac{i}{4}\ T_{ABC} T_{A^{\prime}B^{\prime}C^{\prime}} dM_{RS}\ \left( W- 2 p \right)\ \Pi_{n}^{AA^{\prime}} \Pi_{m}^{BB^{\prime}} \Pi_{p-W}^{CR} \Pi_{p}^{SC^{\prime}} \times \\
&&
\times \frac{1}{\left| n\right| +\left| m\right| +\left| p-W\right|}\ \Big[
\left| n\right| \left( m-p+W\right) +\left| m\right| \left( p-n-W\right)
+\left| p-W\right| \left( n-m\right) \Big] \times \\
&&
\times \frac{1}{\left| n\right| +\left| m\right| +\left| p\right| }\ \Big[
\left| n\right| \left( m-p\right) +\left| m\right| \left( p-n\right) +\left|
p\right| \left( n-m\right) \Big] .
\end{eqnarray*}

\noindent
Taking the derivative $\left. \frac{\partial }{\partial W}\right| _{W=0}$
one gets (excluding terms from the regularization) $-\frac{i}{4} 
T_{ABC}T_{A^{\prime }B^{\prime }C^{\prime }}$ multiplying by

\begin{equation*}
\frac{-2\left| p\right| }{\left( \left| n\right| +\left| m\right| +\left|
p\right| \right) ^{3}}\ \Pi _{n}^{AA^{\prime }}\Pi _{m}^{BB^{\prime
}}\left( \Pi dM\Pi \right) _{p}^{CC^{\prime }}\ \Big[ \left| n\right| \left( m-p\right) +\left| m\right| \left(
p-n\right) +\left| p\right| \left( n-m\right) \Big] ^{2},
\end{equation*}

\noindent
and by

\begin{eqnarray*}
&& \frac{-2}{\left( \left| n\right| +\left| m\right| +\left| p\right| \right)
^{2}}\ \Pi _{n}^{AA^{\prime }}\Pi _{m}^{BB^{\prime }}\left( \Pi dM\Pi
\right) _{p}^{CC^{\prime }} \times \\
&&
\times \Big[ \left| n\right| \left( m+p\right) -\left| m\right| \left(
p+n\right) -\left| p\right| \left( n-m\right) \Big]\ 
\Big[ \left| n\right| \left( m-p\right) +\left| m\right| \left(
p-n\right) +\left| p\right| \left( n-m\right) \Big] .
\end{eqnarray*}

\noindent
Notice that the factors multiplying the propagators are even under the
exchange $\left( n,m,p\right) \rightarrow \left( -n,-m,-p\right) $, and
therefore, one concludes that the general form of the tensor multiplying $T_{ABC}T_{A^{\prime }B^{\prime }C^{\prime }}$ will be 

\begin{equation*}
\left[ \left( aGG+b\Theta \Theta \right) \left( \Theta dM\Theta +GdMG\right)
+\left( cG\Theta +d\Theta G\right) \left( GdM\Theta +\Theta dMG\right) 
\right] ^{AA^{\prime },BB^{\prime },CC^{\prime }},
\end{equation*}

\noindent
for some constants $a,b,c,d$. The expression $T_{ABC}T_{A^{\prime }B^{\prime
}C^{\prime }}$ is symmetric when interchanging primed and unprimed indices,
whereas the above tensor is antisymmetric. Therefore the contribution above
vanishes.

Consider finally the terms coming from the regulator. They read 

\begin{equation*}
\frac{i}{4}\ T_{ABC} T_{A^{\prime}B^{\prime}C^{\prime}}\ \Pi_{n}^{AA^{\prime}} \Pi_{m}^{BB^{\prime}} \left( \Pi dM \Pi \right)_{p}^{CC^{\prime}}\ \frac{\varepsilon \left| p \right|}{\left( \left| n \right| + \left| m \right| + \left| p \right| \right)^{2}}\ \Big[ \left| n \right| \left( m - p \right) + \left| m \right| \left( p - n \right) + \left| p \right| \left( n - m \right) \Big]^{2}.
\end{equation*}

\noindent
This vanishes for the same reasons as above.

\paragraph{Graph $\protect\delta^{\prime}{[}5{]}$:}

The graph, which has symmetry factor $4$, is given by 

\begin{eqnarray*}
&&
\frac{1}{4}\ \frac{1}{\Sigma_{i}\left| n_{i} \right|}\ \Big[ T_{A_{1} \cdots A_{4}} \left( n_{3} n_{4} - \left| n_{3} n_{4} \right| \right) + \left( 3412\right) + \left( 1324\right) + \left( 2413\right) + \left( 1423\right) + \left( 2314\right) \Big] \times \\
&&
\times i \left( n_{1}-n_{2} \right)\ dM_{B_{1}B_{2}}\ \Pi_{n_{1}}^{A_{1}B_{1}} \Pi_{n_{2}}^{A_{2}B_{2}} \Pi_{n_{3}}^{A_{3}A_{4}},
\end{eqnarray*}

\noindent
where 

\begin{eqnarray*}
n_{1} &=&n, \qquad \qquad \, n_{2}=-n-W,\\
n_{3} &=&m, \qquad \qquad n_{4}=-m.
\end{eqnarray*}

\noindent
Let us now compute the derivative $\left. \frac{\partial }{\partial W}\right|
_{W=0}$ and get, excluding---for the moment---terms coming from the
regularization, 

\begin{eqnarray*}
&&
-\frac{i}{4}\ \frac{1}{\Sigma_{i} \left| n_{i} \right|}\ \Big[ T_{A_{1} \cdots A_{4}} \left( n_{3}n_{4} - \left| n_{3}n_{4} \right| \right) +\left( 3412\right)
+\left( 1324\right) +\left( 2413\right) +\left( 1423\right) +\left(
2314\right) \Big] \times \\
&&
\times dM_{B_{1}B_{2}}\ \Pi _{n_{1}}^{A_{1}B_{1}}\Pi
_{n_{2}}^{A_{2}B_{2}}\Pi _{n_{3}}^{A_{3}A_{4}},
\end{eqnarray*}

\noindent
and 

\begin{eqnarray*}
&&
-\frac{i}{2}\ \frac{\left| n_{2}\right| }{\left( \Sigma _{i}\left|
n_{i}\right| \right)^{2}}\ \Big[ T_{A_{1}\cdots A_{4}}\left(
n_{3}n_{4}-\left| n_{3}n_{4}\right| \right) +\left( 3412\right) +\left(
1324\right) +\left( 2413\right) +\left( 1423\right) +\left( 2314\right) 
\Big] \times \\
&&
\times dM_{B_{1}B_{2}}\ \Pi _{n_{1}}^{A_{1}B_{1}}\Pi
_{n_{2}}^{A_{2}B_{2}}\Pi _{n_{3}}^{A_{3}A_{4}},
\end{eqnarray*}

\noindent
and finally 

\begin{equation*}
\frac{i}{2}\ \frac{1}{\Sigma_{i} \left| n_{i} \right|}\ \Big[ \left(
3412\right) +\left( 1324\right) +\left( 1423\right) \Big]\ 
dM_{B_{1}B_{2}}\ \Pi _{n_{1}}^{A_{1}B_{1}}\Pi
_{n_{2}}^{A_{2}B_{2}}\Pi _{n_{3}}^{A_{3}A_{4}}.
\end{equation*}

\noindent
We then have the following six terms 

\begin{equation*}
-\frac{i}{8}\ \frac{1}{\left| n\right| +\left| m\right| }\ \left( \Pi dM\Pi
\right) _{n}^{A_{1}A_{2}}\Pi _{m}^{A_{3}A_{4}}
\end{equation*}

\noindent
times ($A_{i}\rightarrow i$) 

\begin{eqnarray*}
&&T_{1234}\,\left( -2m^{2}\right) \left( \frac{\left| n\right| }{\left|
n\right| +\left| m\right| }+1\right), \\
&&T_{3412}\left( 2n^{2}\right) \left( \frac{\left| m\right| }{\left|
n\right| +\left| m\right| }\right), \\
&&T_{1324}\left( -nm+\left| nm\right| \right) \left( \frac{\left| m\right| }{ 
\left| n\right| +\left| m\right| }\right), \\
&&T_{2413}\left( nm-\left| nm\right| \right) \left( \frac{\left| n\right| }{ 
\left| n\right| +\left| m\right| }+1\right), \\
&&T_{1423}\left( nm+\left| nm\right| \right) \left( \frac{\left| m\right| }{ 
\left| n\right| +\left| m\right| }\right), \\
&&T_{2314}\left( -nm-\left| nm\right| \right) \left( \frac{\left| n\right| }{ 
\left| n\right| +\left| m\right| }+1\right).
\end{eqnarray*}

\noindent
The first and second line vanish due to symmetry. In fact $dM_{AB}$ is
antisymmetric, and therefore the symmetric part of $\Pi dM\Pi $ is $\Theta
dMG+GdM\Theta $, but it comes with a $\left( n\left| n\right| \right) ^{-1}$, which is odd under $n\rightarrow -n$, whereas the rest of the sum is even.
The other four term can be rewritten as 

\begin{eqnarray*}
&&SS_{1234}\left( \frac{-4n^{2}\left| m\right| }{\left| n\right| +\left|
m\right| }\right), \\
&&SA_{1234}\left( -4nm\right), \\
&&AS_{1234}\left( 4\left| nm\right| \right), \\
&&AA_{1234}\left( \frac{4nm\left| n\right| }{\left| n\right| +\left|
m\right| }\right),
\end{eqnarray*}

\noindent
where $SS$ is symmetric in both pairs $12$ and $34$, $SA$ is symmetric in pair $12$ and antisymmetric in pair $34$, $AS$ is antisymmetric in pair $12$ and symmetric in pair $34$, etc. The only surviving terms (due to symmetry)\ are the second and third, but they both vanish as $\sum_{n,m}\frac{1}{n\left( n+m\right) }\sim 0$.

Finally let us compute the terms coming from the regularization 

\begin{eqnarray*}
&&-\varepsilon\ \frac{i}{4}\ \frac{\left| n_{1}\right| }{\Sigma _{i}\left|
n_{i}\right| }\ \Big[ T_{A_{1}\cdots A_{4}}\left( n_{3}n_{4}-\left|
n_{3}n_{4}\right| \right) +\left( 3412\right) +\left( 1324\right) +\left(
2413\right) +\left( 1423\right) +\left( 2314\right) \Big] \times \\
&&\times \left( \Pi dM\Pi \right) _{n}^{A_{1}A_{2}}\Pi _{m}^{A_{3}A_{4}},
\end{eqnarray*}

\noindent
where 

\begin{eqnarray*}
&&
-\varepsilon\ \frac{i}{8}\ \frac{\left| n\right| }{\left| n\right| +\left|
m\right| }\ \left( \Pi dM\Pi \right) _{n}^{A_{1}A_{2}}\Pi
_{m}^{A_{3}A_{4}} \times \\
&&\times \Big[ -2m^{2}\ T_{1234}-2n^{2}\ T_{3412}+\left( nm-\left| nm\right|
\right) \left( T_{1324}+T_{2413}\right) -\left( nm+\left| nm\right| \right)
\left( T_{1423}+T_{2314}\right) \Big] .
\end{eqnarray*}

\noindent
The first two terms vanish for the same reasons as before. The other four
terms re--sum to 

\begin{equation*}
-\varepsilon\ \frac{i}{8}\ \frac{\left| n\right| }{\left| n\right| +\left|
m\right| }\ \left( \Pi dM\Pi \right) _{n}^{A_{1}A_{2}}\Pi _{m}^{A_{3}A_{4}}
\Big[ -4\left| nm\right| SS_{1324}+4nm\ AA_{1234}\Big] ,
\end{equation*}

\noindent
which also vanishes by symmetry.


\section{Formul\ae\ for Useful Sums}


In the computation of the beta function, the only results we shall need are
the $\ln \varepsilon $ contributions in the following sums 

\begin{eqnarray}
\sum_{n,m\geq 1}\frac{1}{n^{2}} &\sim &\ln \varepsilon, \notag \\
\sum_{n,m\geq 1}\frac{1}{\left( n+m\right) ^{2}} &\sim &-\ln \varepsilon, \notag \\
\varepsilon \sum_{n,m\geq 1}\frac{1}{n} &\sim &-\frac{1}{a}\ln \varepsilon, \label{sum1}
\end{eqnarray}

\noindent
together with the vanishing contributions of the following ones 

\begin{eqnarray}
\sum_{n,m\geq 1}\frac{1}{nm} &\sim &0, \notag \\
\varepsilon \sum_{n,m\geq 1}\frac{1}{n+m} &\sim &0, \notag \\
\sum_{n,m\geq 1}\frac{1}{n\left( n+m\right) } &\sim &0. \label{sum2}
\end{eqnarray}

\noindent
Note that all the sums are regulated using the factor 

\begin{equation*}
e^{-\varepsilon a\left( n+m\right) }.
\end{equation*}

\noindent
Denoting with 

\begin{equation*}
x=e^{-\varepsilon a},
\end{equation*}

\noindent
the three sums in (\ref{sum1}) read 

\begin{eqnarray*}
\mathrm{DiLn}\left( 1-x\right)\ \frac{x}{1-x} &\sim &\frac{\pi ^{2}}{6a}\ \frac{1}{\varepsilon }+\ln \varepsilon , \\
-\ln \left( 1-x\right) -\mathrm{DiLn}\left( 1-x\right)  &\sim &\,-\ln
\varepsilon , \\
-\varepsilon \ln \left( 1-x\right)\ \frac{x}{1-x} &\sim &-\frac{1}{a}\ \ln
\varepsilon ,
\end{eqnarray*}

\noindent
where we have used that 

\begin{equation*}
\mathrm{DiLn}\left( 1-x\right) =\sum_{n\geq 1}\frac{x^n}{n^{2}} \sim \frac{\pi ^{2}}{6}+a\varepsilon\ \ln \varepsilon + {\mathcal{O}} \left( \varepsilon \right).
\end{equation*}

\noindent
The first and second sum in (\ref{sum2}) read, on the other hand, 

\begin{eqnarray*}
\ln ^{2}\left( 1-x\right)  &\sim &\ln ^{2}\varepsilon \sim 0, \\
\varepsilon\ \frac{x}{1-x}+\varepsilon\ \ln \left( 1-x\right)  &\sim &0.
\end{eqnarray*}

\noindent
To evaluate the third sum in (\ref{sum2}), we observe that

\begin{equation*}
x\ \frac{\partial}{\partial x} \sum_{n,m\geq 1} \frac{1}{n\left( n+m\right)} = \sum_{n,m\geq 1} \frac{1}{n} = - \frac{x\ \ln \left( 1-x\right)}{1-x},
\end{equation*}

\noindent
so that the sum is given, after integration, by

\begin{equation*}
\frac{1}{2}\ln ^{2}\left( 1-x\right) + {\mathrm{constant}} \sim \ln ^{2}\varepsilon \sim 0.
\end{equation*}


\section{Born--Infeld Action at One Loop}


In the main text we describe a general procedure to extract curvature couplings for the nonabelian BI action, given the two loop abelian BI equations of motion computed non--perturbatively in $\F$. This procedure is implemented at the level of the equations of motion, and when the curved background is described in terms of a RNC expansion. Here, and as a complement to what is done in the main text, let us briefly analyze what are the general expectations from the [one loop] abelian BI action. For the WZW parallelizable target manifolds we are working with, one can define $\mathfrak{M}_{AB} (x) \equiv g_{AB} (x) + B_{AB} (x)$ in a RNC expansion as

$$
\mathfrak{M}_{AB} (x) = g_{AB} + \frac{1}{3} H_{ABC}\ x^{C} - \frac{1}{12} H_{ACK} {H^{K}}_{DB}\ x^{C} x^{D} + \cdots,
$$

\noindent
where we extracted the constant part of the NSNS 2--form field, and where one raises and lowers indices with the constant metric $g_{AB}$ (which can be chosen to be $\delta_{AB}$). In this case the abelian BI action is written

$$
S = -T_{p} \int d^{n}x\ \sqrt{\det \Big( \mathfrak{M}_{IJ} (x) + B_{IJ} + 2\pi\alpha' F_{IJ} (x) \Big)}.
$$

\noindent
Next, one applies the SW map to the previous expression to learn which nonabelian couplings are present in this one loop result. First, change coordinates to $\sigma^{i}$ such that

$$
\Big( B_{IJ} + 2\pi\alpha' F_{IJ} (x) \Big)\ dx^{I} \wedge dx^{J} = B_{ij}\ d\sigma^{i} \wedge d\sigma^{j},
$$

\noindent
so that the dynamical variables become the functions $x^{I} (\sigma^{i})$. In these new coordinates the BI action becomes (we have denoted $\theta \equiv \frac{1}{B}$)

\begin{eqnarray*}
S &=& -T_{p} \int d^{n} \sigma\ \sqrt{\det \Big( B_{ij} + \partial_{i} x^{I} \partial_{j} x^{J}\ \mathfrak{M}_{IJ} (x) \Big)} \\
&=&
-T_{p} \int d^{n} \sigma\ \sqrt{\det B}\ \sqrt{\det \Big( \delta^{i}_{j} + \theta^{ik} \partial_{k} x^{I} \partial_{j} x^{J}\ \mathfrak{M}_{IJ} (x) \Big)}.
\end{eqnarray*}

\noindent
We are now in a position to obtain a matrix model from the previous expression. Use the Poisson bracket, $\left\{ f,g \right\} = \theta^{ij} \partial_{i} f \partial_{j} g$, and the fact that (recall that the determinant can be expanded in powers of the trace)

$$
\left( \mathrm{Tr}_{ij} \left( \theta^{ik} \partial_{k} x^{I} \partial_{j} x^{J}\ \mathfrak{M}_{IJ} (x) \right) \right)^{n} = \Big( \mathrm{Tr}_{IJ} \Big( \{ x^{I}, x^{K} \}\ \mathfrak{M}_{KJ} (x) \Big) \Big)^{n},
$$

\noindent
to conclude that the abelian action can be finally written as

$$
S = -T_{p} \int d^{n} \sigma\ \sqrt{\det B}\ \sqrt{\det \Big( \delta^{I}_{J} + \{ x^{I}, x^{K} \}\ \mathfrak{M}_{KJ} (x) \Big)}.
$$

\noindent
Our point here concerns the power expansion of this action: it is simple to see that it contains the two known terms which are fixed at one loop, it contains none of the terms which are fixed at two loops, and it contains infinite other terms which can only be fixed at higher loop order (see the main text). Indeed, one obtains upon expansion

$$
S = -T_{p} \int d^{n} \sigma\ \sqrt{\det B}\ \Big( 1 + \frac{1}{4} \{ x^{I}, x^{J} \} \{ x^{I}, x^{J} \} - \frac{1}{6} H_{IJK}\ \{ x^{I}, x^{J} \} x^{K} + \cdots \Big).
$$

\noindent
A ``would--be'' nonabelian BI action can be obtained from the previous expression upon the familiar deformation quantization map of $\left\{ x^{I}, x^{J} \right\} \rightarrow -i \left[ \X^{I}, \X^{J} \right]$ and also $\int d^{n} \sigma \sqrt{\det B} \rightarrow \mathbf{Tr}$ (where the trace is now over the matrices $\X$). It includes the terms in the one loop diagonal, but no terms in the two loop diagonal (see the main text)

$$
S = -T_{p}\ \mathbf{Tr} \left( 1 - \frac{1}{4} \left[ \X^{I}, \X^{J} \right] \left[ \X^{I}, \X^{J} \right] + \frac{i}{6} H_{IJK} \left[ \X^{I}, \X^{J} \right] \X^{K} + \cdots \right).
$$


\section{Two Loop Matrix Model in a Flat Background}


In this appendix we analyze the two loop matrix model we have found, when the background is flat (which is basically the flat space nonabelian BI action for a bosonic string). A question we want to address is whether or not the standard solution of commuting $D$--branes can get replaced with some other, possibly fuzzy, solution. Let us begin precisely with the flat space nonabelian BI action for a bosonic string (see, \textit{e.g.}, Polchinski's volume 1, page 188, for details and notation)

$$
S[A_{I}^{a}] = \frac{2\a}{g_{o}^2} \int d^{26}x \left( -\frac{1}{4} {\mathrm{Tr}} \left( F_{IJ} F^{IJ} \right) - \frac{2i\a}{3} {\mathrm{Tr}} \left( {F_{I}}^{J} {F_{J}}^{K} {F_{K}}^{I} \right) \right).
$$

\noindent
We have disregarded the open string tachyon above. As we compactify from maximal abelian brane to minimal nonabelian brane, one has \cite{Taylor}

$$
A_{I} \to \frac{\X_{I}}{2\pi\a}, \qquad
F_{IJ} \to -i \left[ \frac{\X_{I}}{2\pi\a}, \frac{\X_{J}}{2\pi\a} \right] = - \frac{i}{\left( 2\pi\a \right)^2} [ \X_{I}, \X_{J} ].
$$

\noindent
After factoring out the volume of spacetime, and disregarding the kinetic term, the matrix action follows

\begin{eqnarray*}
S[\X] &=& \frac{2\a}{g_{o}^2} \left( - \frac{1}{4} \left( - \frac{i}{\left( 2\pi\a \right)^2} \right)^2 {\mathbf{Tr}} \left( \big[ \X_{I}, \X_{J} \big] \left[ \X^{I}, \X^{J} \right] \right) + \right. \\
&&
\left. - \frac{2i\a}{3} \left( - \frac{i}{\left( 2\pi\a \right)^2} \right)^3 {\mathbf{Tr}} \left( \left[ \X_{I}, \X^{J} \right] \left[ \X_{J}, \X^{K} \right] \left[ \X_{K}, \X^{I} \right] \right) \right).
\end{eqnarray*}

\noindent
One should now recall that the effective potential relates to the action as $V_{\mathrm{eff}} [\X] = - S [\X]$, so that by factoring out a (strictly positive) common term in the expression above one obtains the final expression we use in this paper

$$
V_{\mathrm{eff}} [\X] = - \frac{1}{4 \left( \a \right)^2} {\mathbf{Tr}} \left[ \X^{I}, \X^{J} \right] \left[ \X^{I}, \X^{J} \right] - \frac{1}{6 \pi^2 \left( \a \right)^3} {\mathbf{Tr}}  \left[ \X^{I}, \X^{J} \right] \left[ \X^{J}, \X^{K} \right] \left[ \X^{K}, \X^{I} \right],
$$

\noindent
where repeated indices are summed (contracted with the flat metric $\delta_{IJ}$). It is straightforward to obtain the equations of motion which follow from this potential, and they are

$$
\left[ \X^{M}, \left[ \X^{M}, \X^{K} \right] \right] - \frac{1}{2\pi^{2}\a} \left[ \X^{M}, \left[ \left[ \X^{M}, \X^{N} \right], \left[ \X^{N}, \X^{K} \right] \right] \right] = 0.
$$

\noindent
One immediately observes that commuting branes, where $\left[ \X^{M}, \X^{N} \right] = 0$, satisfy the equations of motion with energy $V_{\mathrm{eff}} [ \X_{\mathrm{commuting}} ] = 0$. On the other hand, one can also try a non--commuting fuzzy solution, where $\left[ \X^{M}, \X^{N} \right] = A\ \epsilon_{MNK}\, \X^{K}$ (let us consider only three coordinate matrices). In this case, the equations of motion tell us that

$$
- 2 A^{2} \delta_{KM} \X^{M} - \frac{1}{2\pi^{2}\a} 2 A^{4} \delta_{KM} \X^{M} = 0 \qquad \Leftrightarrow \qquad A = 0 \quad \vee \quad A^{2} = - 2 \pi^{2} \a,
$$

\noindent
where the $A=0$ case corresponds to the previously analyzed commuting solution and one should pick non--zero $A$ for the fuzzy solution. The fuzzy branes satisfy the equations of motion with energy $V_{\mathrm{eff}} [ \X_{\mathrm{fuzzy}} ] = \frac{\pi^{2}}{3\a} {\mathbf{Tr}} \X^{M} \X^{M} > 0$. It thus follows that the stable, minimal energy solution, corresponds to the standard solution of commuting $D$--branes, as should have been expected from scratch in a flat space background.


\vfill

\eject

\bibliographystyle{plain}

\end{document}